\documentclass{fic-l}

\usepackage{amssymb}
\usepackage{eucal}
\usepackage{amsmath}
\usepackage{amscd}
\usepackage[all]{xy}           
\usepackage{graphicx}
\usepackage{xspace}
\usepackage{axodraw}

\begin{document}

\newcommand{\nc}{\newcommand}
\newcommand{\delete}[1]{}

\newtheorem{theorem}{Theorem}[section]
\newtheorem{prop}[theorem]{Proposition}
\newtheorem{defn}[theorem]{Definition}
\newtheorem{lemma}[theorem]{Lemma}
\newtheorem{coro}[theorem]{Corollary}
\newtheorem{prop-def}{Proposition-Definition}[section]
\newtheorem{claim}{Claim}[section]
\newtheorem{remark}[theorem]{Remark}
\newtheorem{propprop}{Proposed Proposition}[section]
\newtheorem{conjecture}{Conjecture}
\newtheorem{exam}{Example}[section]
\newtheorem{assumption}{Assumption}
\newtheorem{condition}[theorem]{Assumption}

\nc{\grad}[1]{^{({#1})}} \nc{\fil}[1]{_{#1}}

\nc{\Hom}{\mathrm{Hom}} \nc{\id}{\mathrm{id}}
\nc{\mchar}{\rm char}

\nc{\BA}{{\mathbb A}} \nc{\CC}{{\mathbb C}} \nc{\DD}{{\mathbb D}}
\nc{\EE}{{\mathbb E}} \nc{\FF}{{\mathbb F}} \nc{\GG}{{\mathbb G}}
\nc{\HH}{{\mathbb H}} \nc{\LL}{{\mathbb L}} \nc{\NN}{{\mathbb N}}
\nc{\QQ}{{\mathbb Q}} \nc{\RR}{{\mathbb R}} \nc{\TT}{{\mathbb T}}
\nc{\VV}{{\mathbb V}} \nc{\ZZ}{{\mathbb Z}}


\nc{\cala}{{\mathcal A}} \nc{\calb}{{\mathcal B}}
\nc{\calc}{{\mathcal C}} \nc{\cald}{{\mathcal D}}
\nc{\cale}{{\mathcal E}} \nc{\calf}{{\mathcal F}}
\nc{\calg}{{\mathcal G}} \nc{\calh}{{\mathcal H}}
\nc{\cali}{{\mathcal I}} \nc{\calj}{{\mathcal J}}
\nc{\call}{{\mathcal L}} \nc{\calm}{{\mathcal M}}
\nc{\caln}{{\mathcal N}} \nc{\calo}{{\mathcal O}}
\nc{\calp}{{\mathcal P}} \nc{\calr}{{\mathcal R}}
\nc{\calt}{{\mathcal T}} \nc{\calu}{{\mathcal U}}
\nc{\calv}{{\mathcal V}} \nc{\calw}{{\mathcal W}}
\nc{\calx}{{\mathcal X}}


\font \eightrm=cmr8
\def\BCH{{\rm{BCH}}}
\font\cyr=wncyr10

\nc{\redtext}[1]{\textcolor{red}{#1}}


\newcommand{\newfish}{\parbox{1.5pc}{\begin{picture}(10,10) 
\put(1,5){\qbezier(0,-2)(8,10)(16,-2)}
\put(1,5){\qbezier(0,2)(8,-10)(16,2)}
\end{picture}}}

\newcommand{\Anewfish}{\parbox{1pc}{\begin{picture}(10,10)
\put(1,5){\qbezier(-2,-8)(10,0)(-2,8)}
\put(1,5){\qbezier(2,-8)(-10,0)(2,8)}
\end{picture}}}

\newcommand{\bikini}{\parbox{2pc}{\begin{picture}(10,5) 
\put(0,2){\qbezier(0,-1.5)(6,7.5)(11.4,0)}
\put(0,2){\qbezier(0,1.5)(6,-7.5)(11.4,0)}
\put(10.8,2){\qbezier(0.6,0)(6,7.5)(12,-1.5)}
\put(10.8,2){\qbezier(0.6,0)(6,-7.5)(12,1.5)}
\end{picture}}}

\newcommand{\winecup}{\parbox{1.4pc}{\begin{picture}(10,10) 
\put(5.9,-1.4){\line(2,3){9}} \put(11.1,-1.4){\line(-2,3){9}}
\put(4,7.5){\qbezier(0,1.8)(4.5,5)(9,1.8)}
\put(4,7.5){\qbezier(0,1.8)(4.5,-1.25)(9,1.8)}
\end{picture}}}

\newcommand{\shark}{\parbox{2pc}{\begin{picture}(10,10) 
\put(0,6){\qbezier(0,-1.5)(6,7.5)(11.4,0)}
\put(0,6){\qbezier(0,1.5)(6,-7.5)(11.4,0)}
\put(11.4,6){\line(2,1){10}} \put(11.4,6){\line(2,-1){10}}
\put(19.4,2){\qbezier(0,0)(-4,4)(0,8)}
\put(19.4,2){\qbezier(0,0)(4,4)(0,8)}
\end{picture}}}

\newcommand{\roll}{\parbox{1.5pc}{\begin{picture}(10,10) 
\put(8,8){\circle{16}} \put(8,8){\qbezier(8,8)(0,0)(-8,8)}
\put(8,8){\qbezier(8,-8)(0,0)(-8,-8)}
\end{picture}}}

\newcommand{\kite}{\parbox{1.5pc}{\begin{picture}(20,15) 
\put(6.4,0){\line(-1,3){6.4}} \put(6.4,0){\line(1,3){4.8}}
\put(4.4,-1.5){\line(4,3){14}} \put(11.2,14.4){\line(2,-3){7.2}}
\put(1.6,14.4){\qbezier(0,0)(4.8,4.8)(9.6,0)}
\put(1.6,14.4){\qbezier(0,0)(4.8,-4.8)(9.6,0)}
\end{picture}}}


\def\coprod2{{\scalebox{0.13}{ 
\begin{picture}(420,150)(30,-135)
\SetWidth{3.0} \SetColor{Black} \Line(105,-60)(240,-135)
\Line(375,-60)(450,-60) \Line(240,-135)(240,15)
\Line(375,-60)(240,-135) \Line(105,-60)(240,15)
\Line(30,-60)(105,-60) \Line(240,15)(375,-60)
\end{picture}}}}

\def\l2cop{{\scalebox{0.13}{ 
\begin{picture}(195,150)(60,-135)
\SetWidth{3.0} \SetColor{Black} \Line(105,-60)(255,-135)
\Line(105,-60)(255,15) \Line(225,-115)(225,-4)
\Line(60,-60)(105,-60)
\end{picture}}}}

\def\c2cop{{\scalebox{0.13}{ 
\begin{picture}(270,90)(15,-165)
\SetWidth{3.0} \SetColor{Black} \GOval(150,-120)(45,75)(0){1.0}
\Line(75,-120)(15,-120) \Line(285,-120)(225,-120)
\end{picture}}}}

\def\r2cop{{\scalebox{0.13}{ 
\begin{picture}(195,150)(135,-135)
\SetWidth{3.0} \SetColor{Black} \Line(135,-135)(285,-60)
\Line(135,15)(285,-60) \Line(330,-60)(285,-60)
\Line(165,-115)(165,-1)
\end{picture}}}}

\def\Fra1coprod2{{\scalebox{0.13}{
 \begin{picture}(420,180) (30,-120)
    \SetWidth{0.8}
    \SetColor{Black}
    \GBox(90,-120)(255,60){0.823}
    \SetWidth{3.0}
    \Line(105,-30)(240,-105)
    \Line(375,-30)(450,-30)
    \Line(240,-105)(240,45)
    \Line(375,-30)(240,-105)
    \Line(105,-30)(240,45)
    \Line(30,-30)(105,-30)
    \Line(240,45)(375,-30)
  \end{picture}}}}

\def\fra2coprod2{{\scalebox{0.13}{
   \begin{picture}(420,180) (30,-120)
    \SetWidth{0.8}
    \SetColor{Black}
    \GBox(225,-120)(390,60){0.823}
    \SetWidth{3.0}
    \Line(105,-30)(240,-105)
    \Line(375,-30)(450,-30)
    \Line(240,-105)(240,45)
    \Line(375,-30)(240,-105)
    \Line(105,-30)(240,45)
    \Line(30,-30)(105,-30)
    \Line(240,45)(375,-30)
  \end{picture}}}}

\def\vertex{{\scalebox{0.13}{ 
\begin{picture}(75,60)(45,-30)
\SetWidth{2.0} \SetColor{Black} \Line(45,0)(90,0)
\Line(90,0)(120,30) \Line(90,0)(120,-30)
\end{picture}}}}

\def\FERMprop{{\scalebox{0.4}{ 
\begin{picture}(90,0)(30,-45)
\SetWidth{1.5} \SetColor{Black} \ArrowLine(30,-45)(120,-45)
\end{picture}}}}

\def\BOSONprop{{\scalebox{0.4}{ 
\begin{picture}(90,24)(30,-33)
\SetWidth{1.5} \SetColor{Black} \Photon(30,-21)(120,-21){6}{5}
\end{picture}}}}

\def\QEDvertex{{\scalebox{0.4}{ 
\begin{picture}(105,90)(15,-30)
\SetWidth{1.5} \SetColor{Black} \ArrowLine(120,60)(75,15)
\ArrowLine(75,15)(120,-30) \SetWidth{0.5} \Vertex(75,15){5.66}
\SetWidth{1.5} \Photon(75,15)(15,15){6}{5}
\end{picture}}}}

\def\overQED{{\scalebox{0.2}{
\begin{picture}(390,150) (60,-75)
    \SetWidth{0.8}
    \SetColor{Black}
    \Photon(60,0)(150,0){7.5}{4}
    \SetWidth{1.0}
    \ArrowLine(225,-75)(150,0)
    \ArrowLine(285,75)(360,0)
    \ArrowLine(225,75)(285,75)
    \ArrowLine(285,-75)(225,-75)
    \ArrowLine(360,0)(285,-75)
    \ArrowLine(150,0)(225,75)
    \SetWidth{0.8}
    \Photon(60,0)(150,0){7.5}{4}
    \Photon(360,0)(450,0){7.5}{4}
    \Photon(225,75)(225,-75){7.5}{7}
    \Photon(285,75)(285,-75){7.5}{7}
  \end{picture}}}}

\def\RRoverQED{{\scalebox{0.2}{
 \begin{picture}(270,150) (180,-75)
    \SetWidth{1.0}
    \SetColor{Black}
    \ArrowLine(285,75)(360,0)
    \ArrowLine(225,75)(285,75)
    \ArrowLine(285,-75)(225,-75)
    \ArrowLine(360,0)(285,-75)
    \SetWidth{0.8}
    \Photon(360,0)(450,0){7.5}{4}
    \Photon(225,75)(225,-75){7.5}{7}
    \Photon(285,75)(285,-75){7.5}{7}
    \SetWidth{1.0}
    \ArrowLine(225,-75)(180,-75)
    \ArrowLine(180,75)(225,75)
  \end{picture}}}}

\def\LLoverQED{{\scalebox{0.2}{
\begin{picture}(255,150) (60,-75)
    \SetWidth{0.8}
    \SetColor{Black}
    \Photon(60,0)(150,0){7.5}{4}
    \SetWidth{1.0}
    \ArrowLine(225,-75)(150,0)
    \ArrowLine(225,75)(285,75)
    \ArrowLine(285,-75)(225,-75)
    \ArrowLine(150,0)(225,75)
    \SetWidth{0.8}
    \Photon(60,0)(150,0){7.5}{4}
    \Photon(225,75)(225,-75){7.5}{7}
    \Photon(285,75)(285,-75){7.5}{7}
    \SetWidth{1.0}
    \ArrowLine(315,-75)(285,-75)
    \ArrowLine(285,75)(315,75)
  \end{picture}}}}

\def\LoverQED{{\scalebox{0.2}{
  \begin{picture}(195,150) (255,-75)
    \SetWidth{1.0}
    \SetColor{Black}
    \ArrowLine(285,75)(360,0)
    \ArrowLine(360,0)(285,-75)
    \SetWidth{0.8}
    \Photon(360,0)(450,0){7.5}{4}
    \Photon(285,75)(285,-75){7.5}{7}
    \SetWidth{1.0}
    \ArrowLine(285,-75)(255,-75)
    \ArrowLine(255,75)(285,75)
  \end{picture}}}}

\def\RoverQED{{\scalebox{0.2}{
\begin{picture}(195,150) (60,-75)
    \SetWidth{0.8}
    \SetColor{Black}
    \Photon(60,0)(150,0){7.5}{4}
    \SetWidth{1.0}
    \ArrowLine(225,-75)(150,0)
    \ArrowLine(150,0)(225,75)
    \SetWidth{0.8}
    \Photon(60,0)(150,0){7.5}{4}
    \Photon(225,75)(225,-75){7.5}{7}
    \SetWidth{1.0}
    \ArrowLine(255,-75)(225,-75)
    \ArrowLine(225,75)(255,75)
  \end{picture}}}}

\def\loopQED{{\scalebox{0.2}{
\begin{picture}(300,90) (60,-105)
    \SetWidth{0.8}
    \SetColor{Black}
    \Photon(60,-60)(150,-60){7.5}{4}
    \Photon(60,-60)(150,-60){7.5}{4}
    \SetWidth{1.0}
    \ArrowArc(210,-77.5)(62.5,16.26,163.74)
    \ArrowArc(210,-42.5)(62.5,-163.74,-16.26)
    \SetWidth{0.8}
    \Photon(270,-60)(360,-60){7.5}{4}
  \end{picture}}}}

\def\over1QED{{\scalebox{0.2}{
\begin{picture}(330,150) (60,-75)
    \SetWidth{0.8}
    \SetColor{Black}
    \Photon(60,0)(150,0){7.5}{4}
    \SetWidth{1.0}
    \ArrowLine(225,-75)(150,0)
    \ArrowLine(150,0)(225,75)
    \SetWidth{0.8}
    \Photon(60,0)(150,0){7.5}{4}
    \Photon(225,75)(225,-75){7.5}{7}
    \SetWidth{1.0}
    \ArrowLine(225,75)(300,0)
    \ArrowLine(300,0)(225,-75)
    \SetWidth{0.8}
    \Photon(300,0)(390,0){7.5}{4}
  \end{picture}}}}


\def\ladtr1{\scalebox{0.8}{
 \begin{picture}(81,111) (113,-60)
    \SetWidth{0.5}
    \SetColor{Black}
    \Vertex(120,-9){5.66}
    \Vertex(120,-54){5.66}
    \Vertex(120,36){5.66}
    \SetWidth{1.0}
    \Line(120,-9)(120,-54)
    \Line(120,36)(120,-24)
    \SetWidth{0.9}
    \CArc(165,-31.5)(37.5,36.87,143.13)
    \CArc(165,28.5)(37.5,-143.13,-36.87)
    \CArc(165,73.5)(37.5,-143.13,-36.87)
    \CArc(165,13.5)(37.5,36.87,143.13)
    \CArc(165,-76.5)(37.5,36.87,143.13)
    \CArc(165,-16.5)(37.5,-143.13,-36.87)
  \end{picture}}}

\def\chertr1{\scalebox{0.8}{
 \begin{picture}(150,105) (15,-90)
    \SetWidth{0.5}
    \SetColor{Black}
    \Vertex(120,-60){5.66}
    \Vertex(60,-60){5.66}
    \Vertex(90,-15){5.66}
    \SetWidth{1.0}
    \Line(90,-15)(60,-60)
    \Line(90,-15)(120,-60)
    \SetWidth{0.9}
    \CArc(45,-112.5)(37.5,36.87,143.13)
    \CArc(45,-52.5)(37.5,-143.13,-36.87)
    \CArc(135,-112.5)(37.5,36.87,143.13)
    \CArc(135,-52.5)(37.5,-143.13,-36.87)
    \CArc(105,37.5)(37.5,-143.13,-36.87)
    \CArc(105,-22.5)(37.5,36.87,143.13)
  \end{picture}}}

\def\tw1{{\scalebox{0.2}{ 
 \begin{picture}(150,49) (45,-55)
    \SetWidth{0.8}
    \SetColor{Black}
    \PhotonArc(120,-56)(45,-0,180){-5}{10.5}
    \SetWidth{1.0}
    \ArrowLine(75,-56)(165,-56)
    \ArrowLine(45,-56)(75,-56)
    \ArrowLine(165,-56)(195,-56)
  \end{picture}}}}

\def\sw1{{\scalebox{0.2}{ 
 \begin{picture}(210,65) (45,-69)
    \SetWidth{1.0}
    \SetColor{Black}
    \ArrowLine(105,-70)(195,-70)
    \SetWidth{0.8}
    \PhotonArc(150,-86.88)(76.88,12.68,167.32){-5}{15.5}
    \PhotonArc(150,-88.75)(48.75,22.62,157.38){-5}{8.5}
    \SetWidth{1.0}
    \ArrowLine(75,-70)(105,-70)
    \ArrowLine(195,-70)(225,-70)
    \ArrowLine(45,-70)(75,-70)
    \ArrowLine(225,-70)(255,-70)
  \end{picture}}}}

\def\uw1{{\scalebox{0.2}{ 
\begin{picture}(240,94) (30,-40)
    \SetWidth{0.8}
    \SetColor{Black}
    \PhotonArc(150,-41)(60,-0,180){-5}{13.5}
    \PhotonArc(150,-41)(30,-0,180){-5}{7.5}
    \PhotonArc(150,-41)(90,-0,180){-5}{20.5}
    \SetWidth{1.0}
    \ArrowLine(30,-41)(60,-41)
    \ArrowLine(60,-41)(90,-41)
    \ArrowLine(90,-41)(120,-41)
    \ArrowLine(240,-41)(270,-41)
    \ArrowLine(210,-41)(240,-41)
    \ArrowLine(180,-41)(210,-41)
    \ArrowLine(180,-41)(120,-41)
  \end{picture}}}}

\def\vw1{{\scalebox{0.2}{ 
\begin{picture}(300,80) (30,-54)
    \SetWidth{1.0}
    \SetColor{Black}
    \ArrowLine(30,-55)(60,-55)
    \SetWidth{0.8}
    \PhotonArc(120,-55)(30,-0,180){-5}{7.5}
    \PhotonArc(180,-113.5)(133.5,25.99,154.01){-5}{21.5}
    \PhotonArc(240,-55)(30,-0,180){-5}{7.5}
    \SetWidth{1.0}
    \ArrowLine(60,-55)(90,-55)
    \ArrowLine(270,-55)(300,-55)
    \ArrowLine(300,-55)(330,-55)
    \ArrowLine(210,-55)(270,-55)
    \ArrowLine(210,-55)(150,-55)
    \ArrowLine(90,-55)(150,-55)
  \end{picture}}}}

\def\ab1{{\scalebox{0.2}{
\begin{picture}(240,165) (30,-135)
    \SetWidth{0.8}
    \SetColor{Black}
    \Photon(225,15)(270,15){7.5}{2}
    \SetWidth{1.0}
    \ArrowArc(195,-7.5)(37.5,36.87,143.13)
    \ArrowArc(195,37.5)(37.5,-143.13,-36.87)
    \SetWidth{0.8}
    \Photon(165,15)(120,15){7.5}{2}
    \SetWidth{1.0}
    \Line(120,0)(150,-60)
    \SetWidth{0.5}
    \Vertex(120,0){5.66}
    \SetWidth{1.0}
    \Line(120,0)(90,-60)
    \SetWidth{0.5}
    \Vertex(90,-60){5.66}
    \Vertex(150,-60){5.66}
    \SetWidth{0.8}
    \Photon(60,-105)(30,-105){6.5}{2}
    \SetWidth{1.0}
    \ArrowLine(105,-75)(60,-105)
    \ArrowLine(60,-105)(105,-135)
    \SetWidth{0.8}
    \Photon(91,-85)(89,-122){6.5}{3.5}
    \Photon(150,-105)(120,-105){6.5}{2}
    \SetWidth{1.0}
    \ArrowLine(195,-75)(150,-105)
    \ArrowLine(150,-105)(195,-135)
    \SetWidth{0.8}
    \Photon(183,-124)(185,-84){6.5}{3.5}
  \end{picture}}}}

\def\bb1{{\scalebox{0.2}{
 \begin{picture}(171,120) (113,-135)
    \SetWidth{1.0}
    \SetColor{Black}
    \ArrowArc(210,-7.5)(37.5,-143.13,-36.87)
    \ArrowArc(210,-52.5)(37.5,36.87,143.13)
    \SetWidth{0.8}
    \Photon(180,-30)(135,-30){7.5}{2}
    \Photon(240,-30)(285,-30){7.5}{2}
    \SetWidth{0.5}
    \Vertex(120,-45){5.66}
    \Vertex(120,-105){5.66}
    \SetWidth{1.0}
    \Line(120,-45)(120,-105)
    \SetWidth{0.8}
    \Photon(165,-105)(135,-105){6.5}{2}
    \SetWidth{1.0}
    \ArrowLine(165,-105)(210,-135)
    \ArrowLine(210,-75)(165,-105)
    \SetWidth{0.8}
    \Photon(197,-85)(195,-125){6.5}{3.5}
  \end{picture}}}}

\def\cb1{{\scalebox{0.2}{
 \begin{picture}(171,180) (113,-135)
    \SetWidth{1.0}
    \SetColor{Black}
    \ArrowArc(210,52.5)(37.5,-143.13,-36.87)
    \ArrowArc(210,7.5)(37.5,36.87,143.13)
    \SetWidth{0.8}
    \Photon(180,30)(135,30){7.5}{2}
    \Photon(240,30)(285,30){7.5}{2}
    \SetWidth{0.5}
    \Vertex(120,15){5.66}
    \SetWidth{0.8}
    \Photon(165,-105)(135,-105){6.5}{2}
    \SetWidth{1.0}
    \ArrowLine(165,-105)(210,-135)
    \ArrowLine(210,-75)(165,-105)
    \SetWidth{0.8}
    \Photon(165,-30)(135,-30){6.5}{2}
    \SetWidth{1.0}
    \ArrowLine(165,-30)(210,-60)
    \ArrowLine(210,0)(165,-30)
    \SetWidth{0.5}
    \Vertex(120,-45){5.66}
    \Vertex(120,-105){5.66}
    \SetWidth{1.0}
    \Line(120,-45)(120,-105)
    \Line(120,15)(120,-45)
    \SetWidth{0.8}
    \Photon(204,-6)(203,-54){6.5}{3.5}
    \Photon(202,-81)(201,-129){6.5}{3.5}
  \end{picture}}}}


\def\ta1{{\scalebox{0.25}{ 
\begin{picture}(12,12)(38,-38)
\SetWidth{0.5} \SetColor{Black} \Vertex(45,-33){5.66}
\end{picture}}}}

\def\tb2{{\scalebox{0.25}{ 
\begin{picture}(12,42)(38,-38)
\SetWidth{0.5} \SetColor{Black} \Vertex(45,-3){5.66}
\SetWidth{1.0} \Line(45,-3)(45,-33) \SetWidth{0.5}
\Vertex(45,-33){5.66}
\end{picture}}}}

\def\tc3{{\scalebox{0.25}{ 
\begin{picture}(12,72)(38,-38)
\SetWidth{0.5} \SetColor{Black} \Vertex(45,27){5.66}
\SetWidth{1.0} \Line(45,27)(45,-3) \SetWidth{0.5}
\Vertex(45,-33){5.66} \SetWidth{1.0} \Line(45,-3)(45,-33)
\SetWidth{0.5} \Vertex(45,-3){5.66}
\end{picture}}}}

\def\td31{{\scalebox{0.25}{ 
\begin{picture}(42,42)(23,-38)
\SetWidth{0.5} \SetColor{Black} \Vertex(45,-3){5.66}
\Vertex(30,-33){5.66} \Vertex(60,-33){5.66} \SetWidth{1.0}
\Line(45,-3)(30,-33) \Line(60,-33)(45,-3)
\end{picture}}}}

\def\te4{{\scalebox{0.25}{ 
\begin{picture}(12,102)(38,-8)
\SetWidth{0.5} \SetColor{Black} \Vertex(45,57){5.66}
\Vertex(45,-3){5.66} \Vertex(45,27){5.66} \Vertex(45,87){5.66}
\SetWidth{1.0} \Line(45,57)(45,27) \Line(45,-3)(45,27)
\Line(45,57)(45,87)
\end{picture}}}}

\def\tf41{{\scalebox{0.25}{ 
\begin{picture}(42,72)(38,-8)
\SetWidth{0.5} \SetColor{Black} \Vertex(45,27){5.66}
\Vertex(45,-3){5.66} \SetWidth{1.0} \Line(45,27)(45,-3)
\SetWidth{0.5} \Vertex(60,57){5.66} \SetWidth{1.0}
\Line(45,27)(60,57) \SetWidth{0.5} \Vertex(75,27){5.66}
\SetWidth{1.0} \Line(75,27)(60,57)
\end{picture}}}}

\def\tg42{{\scalebox{0.25}{ 
\begin{picture}(42,72)(8,-8)
\SetWidth{0.5} \SetColor{Black} \Vertex(45,27){5.66}
\Vertex(45,-3){5.66} \SetWidth{1.0} \Line(45,27)(45,-3)
\SetWidth{0.5} \Vertex(15,27){5.66} \Vertex(30,57){5.66}
\SetWidth{1.0} \Line(15,27)(30,57) \Line(45,27)(30,57)
\end{picture}}}}

\def\th43{{\scalebox{0.25}{ 
\begin{picture}(42,42)(8,-8)
\SetWidth{0.5} \SetColor{Black} \Vertex(45,-3){5.66}
\Vertex(15,-3){5.66} \Vertex(30,27){5.66} \SetWidth{1.0}
\Line(15,-3)(30,27) \Line(45,-3)(30,27) \Line(30,27)(30,-3)
\SetWidth{0.5} \Vertex(30,-3){5.66}
\end{picture}}}}

\def\thj44{{\scalebox{0.25}{ 
\begin{picture}(42,72)(8,-8)
\SetWidth{0.5} \SetColor{Black} \Vertex(30,57){5.66}
\SetWidth{1.0} \Line(30,57)(30,27) \SetWidth{0.5}
\Vertex(30,27){5.66} \SetWidth{1.0} \Line(45,-3)(30,27)
\SetWidth{0.5} \Vertex(45,-3){5.66} \Vertex(15,-3){5.66}
\SetWidth{1.0} \Line(15,-3)(30,27)
\end{picture}}}}

\def\ti5{{\scalebox{0.25}{ 
\begin{picture}(12,132)(23,-8)
\SetWidth{0.5} \SetColor{Black} \Vertex(30,117){5.66}
\SetWidth{1.0} \Line(30,117)(30,87) \SetWidth{0.5}
\Vertex(30,87){5.66} \Vertex(30,57){5.66} \Vertex(30,27){5.66}
\Vertex(30,-3){5.66} \SetWidth{1.0} \Line(30,-3)(30,27)
\Line(30,27)(30,57) \Line(30,87)(30,57)
\end{picture}}}}

\def\tj51{{\scalebox{0.25}{ 
\begin{picture}(42,102)(53,-38)
\SetWidth{0.5} \SetColor{Black} \Vertex(61,27){4.24}
\SetWidth{1.0} \Line(75,57)(90,27) \Line(60,27)(75,57)
\SetWidth{0.5} \Vertex(90,-3){5.66} \Vertex(60,27){5.66}
\Vertex(75,57){5.66} \Vertex(90,-33){5.66} \SetWidth{1.0}
\Line(90,-33)(90,-3) \Line(90,-3)(90,27) \SetWidth{0.5}
\Vertex(90,27){5.66}
\end{picture}}}}

\def\tk52{{\scalebox{0.25}{ 
\begin{picture}(42,102)(23,-8)
\SetWidth{0.5} \SetColor{Black} \Vertex(60,57){5.66}
\Vertex(45,87){5.66} \SetWidth{1.0} \Line(45,87)(60,57)
\SetWidth{0.5} \Vertex(30,57){5.66} \SetWidth{1.0}
\Line(30,57)(45,87) \SetWidth{0.5} \Vertex(30,-3){5.66}
\SetWidth{1.0} \Line(30,-3)(30,27) \SetWidth{0.5}
\Vertex(30,27){5.66} \SetWidth{1.0} \Line(30,57)(30,27)
\end{picture}}}}

\def\tl53{{\scalebox{0.25}{ 
\begin{picture}(42,102)(8,-8)
\SetWidth{0.5} \SetColor{Black} \Vertex(30,57){5.66}
\Vertex(30,27){5.66} \SetWidth{1.0} \Line(30,57)(30,27)
\SetWidth{0.5} \Vertex(30,87){5.66} \SetWidth{1.0}
\Line(30,27)(45,-3) \SetWidth{0.5} \Vertex(15,-3){5.66}
\SetWidth{1.0} \Line(15,-3)(30,27) \Line(30,57)(30,87)
\SetWidth{0.5} \Vertex(45,-3){5.66}
\end{picture}}}}

\def\tm54{{\scalebox{0.25}{ 
\begin{picture}(42,72)(8,-38)
\SetWidth{0.5} \SetColor{Black} \Vertex(30,-3){5.66}
\SetWidth{1.0} \Line(30,27)(30,-3) \Line(30,-3)(45,-33)
\SetWidth{0.5} \Vertex(15,-33){5.66} \SetWidth{1.0}
\Line(15,-33)(30,-3) \SetWidth{0.5} \Vertex(45,-33){5.66}
\SetWidth{1.0} \Line(30,-33)(30,-3) \SetWidth{0.5}
\Vertex(30,-33){5.66} \Vertex(30,27){5.66}
\end{picture}}}}

\def\tn55{{\scalebox{0.25}{ 
\begin{picture}(42,72)(8,-38)
\SetWidth{0.5} \SetColor{Black} \Vertex(15,-33){5.66}
\Vertex(45,-33){5.66} \Vertex(30,27){5.66} \SetWidth{1.0}
\Line(45,-33)(45,-3) \SetWidth{0.5} \Vertex(45,-3){5.66}
\Vertex(15,-3){5.66} \SetWidth{1.0} \Line(30,27)(45,-3)
\Line(15,-3)(30,27) \Line(15,-3)(15,-33)
\end{picture}}}}

\def\tp56{{\scalebox{0.25}{ 
\begin{picture}(66,111)(0,0)
\SetWidth{0.5} \SetColor{Black} \Vertex(30,66){5.66}
\Vertex(45,36){5.66} \SetWidth{1.0} \Line(30,66)(45,36)
\Line(15,36)(30,66) \SetWidth{0.5} \Vertex(30,6){5.66}
\Vertex(60,6){5.66} \SetWidth{1.0} \Line(60,6)(45,36)
\SetWidth{0.5}
\SetWidth{1.0} \Line(45,36)(30,6) \SetWidth{0.5}
\Vertex(15,36){5.66}
\end{picture}}}}

\def\tq57{{\scalebox{0.25}{ 
\begin{picture}(81,111)(0,0)
\SetWidth{0.5} \SetColor{Black} \Vertex(45,36){5.66}
\Vertex(30,6){5.66} \Vertex(60,6){5.66} \SetWidth{1.0}
\Line(60,6)(45,36) \SetWidth{0.5}
\SetWidth{1.0} \Line(45,36)(30,6) \SetWidth{0.5}
\Vertex(75,36){5.66} \SetWidth{1.0} \Line(45,36)(60,66)
\Line(60,66)(75,36) \SetWidth{0.5} \Vertex(60,66){5.66}
\end{picture}}}}

\def\tr58{{\scalebox{0.25}{ 
\begin{picture}(81,111)(0,0)
\SetWidth{0.5} \SetColor{Black} \Vertex(60,6){5.66}
\Vertex(75,36){5.66} \SetWidth{1.0} \Line(60,66)(75,36)
\SetWidth{0.5} \Vertex(60,66){5.66}
\SetWidth{1.0} \Line(60,36)(60,66) \Line(60,6)(60,36)
\SetWidth{0.5} \Vertex(60,36){5.66} \Vertex(45,36){5.66}
\SetWidth{1.0} \Line(60,66)(45,36)
\end{picture}}}}

\def\ts59{{\scalebox{0.25}{ 
\begin{picture}(81,111)(0,0)
\SetWidth{0.5} \SetColor{Black}
\Vertex(75,36){5.66} \SetWidth{1.0} \Line(60,66)(75,36)
\SetWidth{0.5} \Vertex(60,66){5.66}
\SetWidth{1.0} \Line(60,36)(60,66) \SetWidth{0.5}
\Vertex(60,36){5.66} \Vertex(45,36){5.66} \SetWidth{1.0}
\Line(60,66)(45,36) \Line(75,6)(75,36) \SetWidth{0.5}
\Vertex(75,6){5.66}
\end{picture}}}}

\def\tt591{{\scalebox{0.25}{ 
\begin{picture}(81,111)(0,0)
\SetWidth{0.5} \SetColor{Black}
\Vertex(75,36){5.66} \SetWidth{1.0} \Line(60,66)(75,36)
\SetWidth{0.5} \Vertex(60,66){5.66}
\SetWidth{1.0} \Line(60,36)(60,66) \SetWidth{0.5}
\Vertex(60,36){5.66} \Vertex(45,36){5.66} \SetWidth{1.0}
\Line(60,66)(45,36) \SetWidth{0.5} \Vertex(45,6){5.66}
\SetWidth{1.0} \Line(45,6)(45,36)
\end{picture}}}}

\nc{\li}[1]{\textcolor{red}{#1}}
\nc{\bluetext}[1]{\textcolor{blue}{#1}}


\hyphenation{equi-va-lence equi-va-lent equi-vari-ant ge-ne-ral
ge-ne-rate ge-ne-ra-ted ge-o-des-ic geo-met-ric geo-met-ries
geo-met-ry Hamil-ton-ian Her-mit-ian ma-ni-fold ma-ni-folds
neigh-bour-hood ope-ra-tor ope-ra-tors or-tho-go-nal pro-duct
qua-drat-ic re-nor-ma-li-za-tion Rie-mann-ian semi-def-i-nite
skew-ad-joint sum-ma-bi-li-ty sum-ma-ble to-po-lo-gi-cal to-po-lo-gy
va-cuum}


\title[Rota--Baxter Algebras in Renormalization of Perturbative QFT]
{Rota--Baxter Algebras in Renormalization\\ of Perturbative
Quantum Field Theory}

\author{Kurusch Ebrahimi-Fard}
\address{I.H.\'E.S.
         Le Bois-Marie,
         35, Route de Chartres,\\
         F-91440 Bures-sur-Yvette,
         France}
\email{kurusch@ihes.fr}

\author{Li Guo}
\address{Department of Mathematics and Computer Science,
         Rutgers University,\\
         Newark, NJ 07102}
\email{liguo@newark.rutgers.edu}

\maketitle


\begin{abstract}
Recently, the theory of renormalization in perturbative quantum
field theory underwent some exciting new developments. Kreimer
discovered an organization of Feynman graphs into combinatorial
Hopf algebras. The process of renormalization is captured by a
factorization theorem for regularized Hopf algebra characters. In
this context the notion of Rota--Baxter algebras enters the scene.
We review several aspects of Rota--Baxter algebras as they appear
in other sectors also relevant to perturbative renormalization,
for instance multiple-zeta-values and matrix differential
equations.
\end{abstract}


\noindent {\footnotesize{${}\phantom{a}$ 2001 PACS Classification:
03.70.+k, 11.10.Gh, 02.10.Hh, 02.10.Ox}}




\tableofcontents


\section{Introduction}
\label{sect:intro}

In this paper we extend the talk given by one of us (L.~G.) at the
workshop {\textit{Renormalization and Universality in Mathematical
Physics}} held at the Fields Institute for Research in
Mathematical Sciences, October 18-22, 2005. This talk, as well as
the following presentation, is based on our results in
\cite{E-G-K2,E-G-K3,EGGV,EG} and explores in detail the algebraic
structure underlying the Birkhoff decomposition in the theory of
renormalization of perturbative quantum field theory. This
decomposition result was discovered by Alain Connes and Dirk
Kreimer using the minimal subtraction scheme in dimensional
regularization in the context of their Hopf algebraic approach to
renormalization \cite{CK2,CK3,KreimerHopf,KreimerChen}. Recently,
we extended our findings in joint work with D.~Manchon in
\cite{EGMbch06}, presenting a unified picture of several
apparently unrelated factorizations arisen from quantum field
theory, vertex operator algebras, combinatorics and numerical
methods in differential equations.

With a broad range of audience in mind, we summarize briefly some
background in perturbative quantum field theory (\textsf{pQFT}) as
well as in Hopf algebra and Rota--Baxter algebra, before
presenting some of the recent developments. In doing so, we hope
to reach a middle ground between mathematics and physics where the
algebraic structures which emerged in the work of Connes and
Kreimer can be appreciated and further studied by students and
researchers in pure and applied areas. For other aspects and more
details on the work of Kreimer and collaborators on the Hopf
algebra approach to renormalization in \textsf{pQFT} we refer the
reader to \cite{EK,FG05,KreimerRev,Ma01}, as well as the
contributions of D.~Kreimer and S.~Weinzierl in this volume. For
readable and brief sources on the basics of renormalization in
\textsf{pQFT} using the standard terminology, that is, before the
appearance of the more recent Hopf algebra approach, the reader
might wish to take a look into J.~Gracey's conference
contribution, as well as \cite{Callan75,CaswellKennedy82}, and
more recently \cite{Collins06,Delamotte04}. For a more complete
picture on renormalization theory in the context of applications,
including calculational details, one should consult the standard
textbooks in the field, e.g.
\cite{BS59,Collins84,IZ80,Muta,Vasilev04}.
\medskip

{\bf{The new picture of renormalization in \textsf{pQFT}}} The
very concept of renormalization has a long history \cite{Brown93}
going beyond its most famous appearance in perturbative quantum
field theory, which started with a short paper on the Lamb shift
in perturbative quantum electrodynamics (\textsf{QED}) by H.~Bethe
\cite{Bethe47}. Eventually, in the context of \textsf{pQFT},
renormalization together with the gauge principle reached the
status of a fundamental concept. Its development is marked by
contributions by some of the most important figures in 20 century
theoretical physics including the founding fathers of
\textsf{pQFT}, to wit, Feynman, Dyson, Schwinger, and Tomonaga,
see \cite{Kaiser05,Schweber94} for more details. Renormalization
theory in \textsf{pQFT} reached a satisfying form, from the
physics' point of view, with the formulation known as BPHZ
renormalization prescription due to Bogoliubov and Parasiuk
\cite{BoPa57}, further improved by Hepp \cite{Hepp66} and
Zimmermann \cite{Zimmermann69}.

As a matter of fact, in most of the interesting and relevant
$4$-dimensional quantum field theories, even simple perturbative
calculations are plagued with ill-defined, i.e., ultraviolet
divergent multidimensional integrals. The removal of these
so-called short-distance singularities in a physically and
mathematically sound way is the process of renormalization.
Although the work of the aforementioned authors played a decisive
role in the acceptance of the theory of renormalization in
theoretical physics, the conceptual foundation as well as
application of renormalization was nevertheless plagued by an
ambivalent reputation. Let us cite the following statement from
\cite{Delamotte04}, {\it{Removing [these] divergencies has been
during the decades, the nightmare and the delight of many
physicists working in particle physics. [$\dots$] This matter of
fact even participated to the nobility of the subject}}. Moreover,
renormalization theory suffered from its lack of a concise
mathematical formulation. It was only recently that the
foundational, i.e., mathematical structures of renormalization
theory in \textsf{pQFT} experienced new and enlightening
contributions. These developments started out with the Hopf
algebraic formulation of its combinatorial-algebraic structures
discovered by Kreimer in \cite{KreimerHopf} --50 years after
Bethe's paper. This new approach found its satisfying formulation
in the work of Connes and Kreimer \cite{CK1,CK2,CK3}, and
Broadhurst and Kreimer \cite{BK98,BK99,BK00a,BK00b,BK01}.

In fact, in the Hopf algebra picture the analytic and algebraic
aspects of perturbative renormalization nicely decouple, hence
making its structure more transparent. The process of
renormalization in \textsf{pQFT} finds a surprisingly compact
formulation in terms of a classical group theoretical
factorization problem. This is one of the main results discovered
by Connes and Kreimer in \cite{CK2,CK3} using the minimal
subtraction scheme in dimensional regularization. It was achieved
by first organizing the set $F$ of one-particle irreducible
Feynman graphs coming from a perturbative renormalizable
\textsf{QFT} (e.g. $\phi^3_6$, $\phi^4_4$, \textsf{QED},...) into
a commutative graded connected Hopf algebra $\calh_\calf$, where
$\calf:=\mathbb{K}F$ denotes the vector space freely generated by
$F$. Its non-cocommutative coproduct map essentially encodes the
disentanglement of Feynman graphs into collections of ultraviolet
divergent proper one-particle irreducible subgraphs and the
corresponding cograph, just as it appears in the original BPHZ
prescription, more precisely, Bogoliubov's
${\bar{\mathrm{R}}}$-map.

Furthermore, regularized Feynman rules seen as linear maps $\phi:
\calf \to A$ from the vector space of graphs into a commutative
unital algebra $A$ are extended to $\phi: \calh_\calf \to A$ in
the group $G_A$ of algebra homomorphisms from $\calh_\calf$ to
$A$. This setting, as we will see, already gives rise to a
factorization of such generalized, that is, $A$-valued algebra
homomorphisms in the group $G_A$. In addition we will observe that
when the target space algebra $A$ carries a Rota--Baxter algebra
structure, Connes--Kreimer's Birkhoff decomposition for $G_A$ in
terms of Bogoliubov's $\bar{\rm{R}}$-map follows immediately. The
last ingredient, i.e. the Rota--Baxter structure is naturally
provided by the choice of a renormalization scheme on the
regularization space $A$, e.g. minimal subtraction in dimensional
regularization.\smallskip

It was the goal of this talk to present the algebraic underpinning
for the result of Connes--Kreimer in terms of a general
factorization theorem for complete filtered associative, not
necessarily commutative algebras combined with other key results,
such as Spitzer's identity and Atkinson's theorem for such
algebras in the presence of a Rota--Baxter structure.

In fact, we would like to show how the above factorization problem
for renormalization may be formulated more transparently as a
classical decomposition problem for matrix Lie groups. This is
achieved by establishing a representation of the group of
regularized Hopf algebra characters by --infinite dimensional--
unipotent lower triangular matrices with entries in a
commutative unital Rota--Baxter algebra. The representation space
is given by --a well-chosen vector subspace of-- the vector space
$\calf$ of 1PI Feynman graphs corresponding to the \textsf{pQFT}.
The demand for entries in a commutative Rota--Baxter algebra
induces a non-commutative Rota--Baxter structure on the underlying
matrix algebra such that Spitzer's theorem applies, and indeed, as
we will see, this paves smoothly the way to Bogoliubov's
$\bar{\mathrm{R}}$-map, though in matrix form.\medskip

Before giving a more detailed account on this purely algebraic
structure we should underline that the initial proof of the
Connes--Kreimer factorization of regularized Feynman rules was
achieved using dimensional regularization, i.e., assuming $A$ to
be the field of Laurent series, hereby opening a hitherto hidden
geometric viewpoint on perturbative renormalization in terms of a
correspondence to the Riemann--Hilbert problem. For the minimal
subtraction scheme it amounts to the multiplicative decomposition
of the Laurent series associated to a Feynman graph using
dimensionally regularized Feynman rules. Such a Laurent series has
poles of finite order in the regulator parameter and the
decomposition happens to be the one into a part holomorphic at the
origin and a part holomorphic at the complex infinity. The latter
moreover distinguishes itself by the particular property of
locality. This has far reaching geometric implications, starting
with the interpretation upon considering the Birkhoff
decomposition of a loop around the origin, providing the clutching
data for the two half-spheres defined by that very loop. Connes
and Kreimer found a scattering type formula for the part
holomorphic at complex infinity, that is, the counterterm, giving
rise to its description in terms of iterated integrals.

Moreover, by some recent progress, which has been made in the
mathematical context of number theory, and motivic aspects of
Feynman integrals, this geometric point of view leads to motivic
Galois theory upon studying the notion of equisingular connections
in the related Riemann--Hilbert correspondence. This was used to
explore Tannakian categories and Galois symmetries in the spirit
of differential Galois theory in~\cite{CM1,CM2,CM3}. Underlying
the notion of an equisingular connection is the locality of
counterterms, which itself results from Hochschild cohomology. The
resulting Dyson--Schwinger equation allows for gradings similar to
the weight- and Hodge filtrations for the polylogarithm
\cite{KreimerLesHouches,Kreimer8}. More concretely, the motivic
nature of primitive graphs has been established very recently by
Bloch, Esnault and Kreimer in~\cite{BlochEK}.
\medskip

We now outline the organization of the paper. In
Section~\ref{sect:renormalization} we review some pieces from
renormalization theory in \textsf{pQFT} up to the introduction of
Bogoliubov's $\bar{\mathrm{R}}$-map. We use
Section~\ref{sect:HopfCK} for introducing most of the basic
mathematical notions needed in the sequel, especially with respect
to Hopf algebras. A general factorization theorem for complete
filtered algebras is presented, based on a new type of recursion
equation defined in terms of the Baker--Campbell--Hausdorff
($BC\!H$) formula. We establish Connes' and Kreimer's Hopf
respectively Lie algebra of Feynman graphs as well as their
Birkhoff decomposition of regularized Hopf algebra characters.
Section~\ref{sect:RBA} is devoted to a not-so-short review of the
concept of Rota--Baxter algebra, including a list of instructive
examples as well as Spitzer's identity and a link between the
Magnus recursion known from matrix differential equations and the
new $BC\!H$-recursion. Section~\ref{sect:RBmatrixCK} contains a
presentation of the Rota--Baxter algebra structure underlying the
Birkhoff decomposition of Connes--Kreimer. We establish a matrix
representation of the group of regularized Feynman rules
characters and recover Connes--Kreimer's Birkhoff decomposition in
terms of a classical matrix factorization problem.


\section{Renormalization in \textsf{pQFT}: Bogoliubov's $\bar{\mathrm{R}}$-operation}
\label{sect:renormalization}

We recall some of the essential points from {\it{classical}}
renormalization theory as applied in the context of \textsf{pQFT}.
We follow Kreimer's book \cite{KreimerBuch} and the recent brief
and concise summary given by Collins in \cite{Collins06}.\medskip


\subsection{Feynman rules and renormalizability}

Let us begin with a statement which summarizes perturbative
renormalization theory of \textsf{QFT} in its Lagrangian
formulation. The quintessence of --multiplicative--
renormalization says that under the severe constraint of
maintaining the physical principles of locality, unitarity, and
Lorentz invariance, it is possible to absorb to all orders in
perturbation theory the (ultraviolet) divergencies in a
redefinition of the parameters defining the \textsf{QFT}. Here, in
fact, two very distinct concepts enter, to wit, that of
{\it{renormalizability}}, and the {\it{process of
renormalization}}. The former distinguishes those theories with
only a finite number of parameters, lending them considerably more
predictive power. However, the process of renormalization instead
works order by order, indifferently of the number of
parameters.\smallskip

In the Lagrangian picture we start with a Lagrangian density
defining the \textsf{QFT}, which we assume to be renormalizable.
For instance, the so-called $\phi^4$ and $\phi^3$ theories are
described by the following two Lagrangians
\begin{equation*}
    L = \frac{1}{2} (\partial_\mu \phi)^2 - \frac{1}{2} m_1^2\phi^2 + \frac{\lambda_4}{4!}\phi^4
  \quad {\makebox{ resp. }} \qquad
    L = \frac{1}{2} (\partial_\mu \phi)^2  - \frac{1}{2} m_2^2\phi^2 + \frac{\lambda_3}{3!}\phi^3.
\end{equation*}
The classical calculus of variations applied to $L = \frac{1}{2}
(\partial_\mu \phi)^2 - \frac{1}{2} m_{i}^2 \phi^2$, $i=1,2$
provides the equations of motion
\begin{equation*}
    \partial_\mu \frac{\delta L}{\delta \partial_\mu \phi}
        -\frac{\delta L}{\delta \phi}=0 \Rightarrow \;(\Box + m_{i}^2) \phi(x)=0,\qquad i=1,2.
\end{equation*}
There is only one type of --scalar-- field denoted by
$\phi=\phi(x)$, parametrized by the $D$ dimensional space-time
point $x=(t,{\bf{x}})$. The parameters $m_1,m_2$ and
$\lambda_4,\lambda_3$ are called the mass and coupling constants,
respectively. The equations of motion as well as the Lagrangian
are local in the sense that all fields are evaluated at the same
space-time point. The reader interested in quantum field theories
more relevant to physics, such as \textsf{QED} or quantum
chromodynamics, should consult the standard literature, for
instance \cite{BS59,Collins84,IZ80,Muta,Ryder85,tHV73}.

We may separate the above Lagrangians into a free and an
interaction part, $L=L_f+L_i$, where in the both cases above
$L_f=\frac{1}{2} (\partial_\mu \phi)^2 - \frac{1}{2} m_i^2\phi^2$,
$i=1,2$. The remaining quartic respectively cubic term represents
the interaction part, $L_i$.

The general goal is to calculate the $n$-point Green's functions,
$$
    G(x_1,\ldots,x_n) := \langle 0|{\rm{T}}[\phi(x_1)\cdots\phi(x_n)]|0\rangle
$$
where the time-ordering ${\rm{T}}$-operator is defined by
\allowdisplaybreaks{
\begin{eqnarray*}
  {\rm{T}}[\phi(x_1) \phi(x_2)]  :=\begin{cases}  \phi(x_1)\phi(x_2),\ {\rm{if}}\ t_1>t_2, \\
                                                  \phi(x_2)\phi(x_1),\ {\rm{if}}\ t_2>t_1,
 \end{cases}
\end{eqnarray*}}
since from these vacuum correlators of time-ordered products of
fields one can obtain all $S$-matrix elements with the help of the
LSZ-formalism, see for instance \cite{IZ80} for more details.

So far, perturbation theory is the most successful approach to
Lagrangian \textsf{QFT} with an overwhelming accordance between
experimental and theoretical, that is, perturbative results.

In perturbation theory Green's functions are expanded in power
series of a supposed to be small parameter like the coupling
constant (or Planck's $\hbar$). Its coefficients are given by
complicated iterated integrals of the external parameters defining
the scattering process, respectively the participating particles.
However, these coefficients turn out to show a highly organized
structure. Indeed, Feynman discovered a graphical coding of such
expressions in terms of graphs that later bore his name. A
{\it{Feynman graph}} is a collection of internal and external
lines, or edges, and vertices of several types. Proper subgraphs
of a Feynman graph are determined by proper subsets of the set of
internal edges and vertices.

Eventually, perturbative expansions in \textsf{QFT} are organized
in terms of one-particle irreducible (1PI) Feynman graphs, i.e.,
connected graphs that stay connected upon removal of any of the
internal edges, also called propagators. For example, we have such
simple drawings as \newfish\ or \winecup encoding Feynman
amplitude integrals appearing in the expansion of the $4$-point
Green's function in $\phi^4$-theory. The fact that there is only
one type of --undirected-- lines and only $4$-valent vertices
reflects the fact that the Lagrangian $L$ only contains one type
of --scalar-- field $\phi$ with only a quartic interaction term in
$L_i$. For the $\phi^3$ case we would find drawings with only
$3$-valent vertices and one type of undirected lines.

Feynman rules give a mean to translate these drawings --back--
into the corresponding amplitudes. The goal is to calculate them
order by order in the coupling constants of the theory.
Originally, a graph was somehow `identified' with its amplitude.
Let us remark that it is only in the Hopf algebraic picture that
these objects get properly distinguished into the
algebraic-combinatorial and analytic side of the same underlying
\textsf{pQFT} picture. We will see that in this context graphs are
studied as combinatorial objects organized into (pre-)Lie algebras
and the corresponding Hopf algebras. Whereas Feynman rules provide
particular maps assigning analytic expressions, the so-called
Feynman amplitudes, to these graphs.

As an example for the latter we list here the set of Feynman rules
related to the above Lagrangian of the $\phi^4$-theory in four
space-time dimensions~\cite{KreimerBuch}.
\begin{itemize}
    \item For each internal scalar field line, associate a
          propagator: $i\Delta(p) =\frac{i}{p^2-m^2+i\eta}$.

    \item At each vertex require momentum conservation and place a
          factor of $-i\lambda_4$.

    \item For each closed loop in the graph, integrate over
          $\int \frac{d^4q}{(2\pi)^4},$ where $q$ is the
          momentum associated with the loop.

    \item For any Feynman graph $\Gamma$ built from these propagators and
          vertices divide by the symmetry factor $sym(\Gamma)$ of the graph, which is
          assumed to contain $r$ vertices. The symmetry factor is then given
          by the number of possibilities to connect $r$ vertices to give
          $\Gamma$, divided by $(4!)^r$.
\end{itemize}

A few examples are presented now, see for instance \cite{FG05} and
the standard literature. Let us write down the integral expression
corresponding to the graph called {\em{fish}}, \newfish, made out
of two $4$-valent vertices and two propagators. The unconnected
vertex legs carry the external momenta of the in- and outgoing
particles which are denoted by $k_1,k_2,k_3,k_4$. They are related
by a delta function ensuring the overall momentum conservation.
\begin{equation}
    \label{eq:fish}
    I(\newfish):= \left(-i\lambda_4 \right)^2
    \int \frac{d^4q}{(2\pi)^4} \frac{i}{(q+k_1+k_2)^2-m^2+i\eta}\ \frac{i}{q^2-m^2+i\eta}.
\end{equation}

For the {\em{bikini}} graph,\ \bikini, with three $4$-valent
vertices and four propagators we find after some algebra a product
of two `fishy' integrals
\begin{eqnarray}
    I(\bikini) &:=& \left(-i\lambda_4 \right)^3
                    \int \frac{d^4q}{(2\pi)^4}\frac{i}{(q + k_1 + k_2)^2-m^2 + i\eta}\
                    \frac{i}{q^2-m^2+i\eta} \times \nonumber\\
               &  & \qquad\quad \int \frac{d^4p}{(2\pi)^4}\frac{i}{(p +k_1 +k_2)^2-m^2+i\eta}\ \frac{i}{p^2-m^2+i\eta}.
    \label{eq:bikini}
\end{eqnarray}

Let us go one step further and write down the integral for the
{\em{winecup}} graph,\ \winecup, of order three in the coupling
constant $i\lambda_4$, also made out of four propagators
\begin{eqnarray}
        I\big(\winecup \big) &:=&  \left(-i\lambda_4 \right)^3 \int \frac{d^4p}{(2\pi)^4}\ \int \frac{d^4q}{(2\pi)^4}
                                        \frac{i}{(q +k_1 +k_2)^2-m^2+i\eta}\ \frac{i}{q^2-m^2+i\eta} \times \nonumber\\
                             &  & \qquad\qquad\qquad\qquad \frac{i}{(q^2-m^2+i\eta)\
                                                                            ((p-q -k_3)^2-m^2+i\eta)}.
        \label{eq:winecup}
\end{eqnarray}

\noindent Unfortunately, all of the above integrations over the
loop-momenta are not restricted by any constraint and some
power-counting analysis of the integrands reveals a serious
problem as we observe that the above integrals are not
well-defined for high loop-momenta.\medskip

For the purpose of classifying such problematic integrals within a
\textsf{pQFT}, we must introduce the notion of {\it{superficial
degree of divergence}} associated to a Feynman graph $\Gamma$ and
denoted by $\omega(\Gamma)$. In general, $\omega(\Gamma)$ simply
follows from counting powers of integral loop-momenta in the
corresponding Feynman integral in the limit when they are large.
The amplitude corresponding to the graph $\Gamma$ is superficially
{\it{convergent}} if $\omega(\Gamma)<0$, and superficially
{\it{divergent}} for $\omega(\Gamma) \geq 0$. For
$\omega(\Gamma)=0$ we call the diagram superficially
{\it{logarithmically}} divergent and we say that it is
superficially {\it{linearly}} divergent, if $\omega(\Gamma)=1$. No
doubt, we may continue this nomenclature. Superficially here
reflects the very fact that the integral associated to a graph
might appear to be convergent, but due to ill-defined subintegrals
corresponding to ultraviolet divergent 1PI subgraphs it is
divergent in fact.

Restricting ourselves to scalar theories such as the two examples
at the beginning one can derive a simple expression for
$\omega(\Gamma)$ depending on the structure of the graph, i.e.,
its vertices and edges, as well as the space-time dimension. We
assume a Lagrangian of the general form $L=L_f + \sum_{k=3}
L_{i_k}$ where $L_f$ as usual consists of the free part and the
sum contains the interaction field monomials of the form
$L_{i_k}:=\frac{\lambda_{k}}{k!}\phi^k$. Recall that the canonical
dimension (inverse length) in $D$ space-time dimensions of the
scalar field $\phi$ is given by $[\phi]=(D-2)/2$ in natural units,
following from the free part of the Lagrangian. Therefore we have
$[\phi^k]=(D-2)k/2=:\omega_k$ and the coupling constant must be of
dimension $[\lambda_k]=D-\omega_k$.

For a given connected graph $\Gamma$ consisting of $E=E(\Gamma)$
external lines, $I=I(\Gamma)$ internal lines and $V=V(\Gamma)$
vertices we find $L=I-V+1$ loop integrations in the corresponding
amplitude. By standard considerations it follows that $E - 2I =
\sum_{j=1}^{V} n_j$, where $n_{j}$ denotes the order, or valency,
of the $j$-th vertex in $\Gamma$. Since each $D$ dimensional loop
integration contributes $D$ loop momenta in the numerator and each
propagator contributes two inverse loop momenta it follows that
$$
    \omega(\Gamma) = DL-2I = D - \frac{D-2}{2}E
                               + \sum_{j=1}^{V} \Big(\frac{D-2}{2} n_j - D \Big).
$$
The last term is just $\frac{D-2}{2}
n_j-D=\omega_{n_j}-D=[\lambda_{n_j}]$ and hence
$$
 \omega(\Gamma) = D - \frac{D-2}{2}E
                               + \sum_{j=1}^{V} [\lambda_{n_j}].
$$

\noindent For our $\phi^4$-theory in $D=4$ space-time dimensions
we find a dimensionless coupling constant, $[\lambda_4]=0$, which
implies $\omega(\Gamma)=4-E$. In fact, from this formula it
follows immediately that the above integrals are all superficially
logarithmically divergent, rendering the perturbative series
expansion of --the $4$-point-- Green's functions quite useless
right from the start, to wit, for the $1$-loop fish diagram
\newfish\ corresponding to the first non-trivial term in the
perturbation expansion of the $4$-point Green's functions in $D=4$
space-time dimensions we find $\omega(\newfish)=D-4=4-4=0$, which
indicates the existence of a so-called logarithmic
{\it{ultraviolet}} (UV) {\it{divergence}}. The reader is referred
to the literature for more details on the related dimensional
analysis of \textsf{pQFT} revealing the role played by the number
of space-time dimensions and the related classification into
{\it{non-renormalizable}}, (just) {\it{renormalizable}} and
{\it{super-renormalizable}} theories
\cite{Coleman88,IZ80,PesSch95,Ryder85}. Let us mention that in the
$\phi^4$ example the number of $D=4$ space-time dimensions just
happens to be the critical one, and that we would make a similar
observation for the Lagrangian with cubic interaction in $D=6$
space-time dimensions.

At this level and under the assumption that the perturbative
\textsf{QFT} is {\it{renormalizable}}, the programm of
--multiplicative-- renormalization enters the stage. Its
non-trivial operations will cure the ultraviolet deficiencies in a
self-consistent and physically sound way order by order in the
series expansion.


\subsection{Multiplicative renormalization}

The machinery of multiplicative renormalization consists of the
following parts.\smallskip

{\it{Regularization}}: first, one must regularize the theory,
rendering integrals of the above type formally finite upon the
introduction of new and nonphysical parameters. For instance, one
might truncate, i.e., {\it{cut-off}} the integration limits at an
upper bound $\Lambda$ by introducing a step-function $\Theta$
\begin{equation*}
    \int d^4q \frac{ \Theta[\Lambda - q^2]}{(q+k_1+k_2)^2 - m^2 + i\eta}
                                \frac{1}{q^2 - m^2 + i\eta}.
\end{equation*}
Evaluating such an integral results in $\log$-terms containing
$\Lambda$ so that naively removing the cut-off parameter by taking
the limit $\Lambda \to \infty$ must be avoided since it gives back
the original divergence. The cut-off approach usually interferes
with gauge symmetry.

Another regularization scheme is {\it{dimensional regularization}}
\cite{Collins84,tH73}, which respects --almost-- all symmetries of
the theory under consideration. It introduces a complex parameter
by changing the integral measure, that is, the space-time
dimension $D \in \mathbb{N}$ to $\mathcal{D} \in \mathbb{C}$
$$
    \frac{d^Dq}{(2\pi)^D} \longrightarrow
        \mu^{2\varepsilon} \frac{d^\mathcal{D}q}{(2\pi)^\mathcal{D}},
$$
where $\varepsilon = \frac{(D - \cald)}{2} \in \mathbb{C}$. The
parameter $\mu$ (t'Hooft's mass) is introduced for dimensional
reasons. Hence, forgetting about the other involved variables,
that do not enter our considerations, dimensionally regularized
Feynman rules associate to each graph an element from the field of
Laurent series $\mathbb{C}[\varepsilon^{-1},\varepsilon]]$. In
general, one obtains a proper Laurent series of degree at most $n$
in an $n$-loop calculation, which means that we have at most poles
of order $n$ in the regularized Feynman integral corresponding to
a graph with $n$ independent closed cycles. The pole terms
parametrize the UV-deficiency of the integrals. Obviously, the
limit of $\varepsilon \to 0$ results in the original divergences
of the integral and must be postponed.

Let us emphasize that it is common knowledge that the choice of
the regularization procedure is highly arbitrary and mainly driven
by calculational conveniences and the wish to conserve symmetries
of the original theory.

But, so far we have not achieved much more than a parametrization
of the divergent structure of Feynman integrals. Though important
and useful all this just indicates the need of some more
sophisticated set of rules on how to handle the terms which
diverge with $\Lambda \to \infty$ in the cut-off setting
respectively $\varepsilon \to 0$ for dimensional regularization,
since in the end we must remove these non-physical regularization
parameters.
\medskip

{\it{Renormalization}}: to render the picture complete we must fix
a so-called {\it{renormalization scheme}} (or {\it{condition}}),
which is supposed to isolate the divergent part of a regularized
integral. For instance, in dimensional regularization a natural
choice is the minimal subtraction scheme $R:=R_{ms}$, mapping a
Laurent series to its strict pole part, i.e., $R\big(\sum_{n \geq
-N} a_n \varepsilon^n\big):=\sum_{n = -N}^{-1} a_n
\varepsilon^n$.\smallskip

The next and conceptually most demanding step is to provide a
procedure which renders the integrals finite upon the removal of
the regularization parameters order by order in a --mathematically
and-- physically sound way. This actually goes along with the
proof of perturbative renormalizability of the underlying
\textsf{QFT}. A convenient procedure which accomplishes this and
which is commonly used nowadays was established by Bogoliubov and
Parasiuk, further improved by Hepp and Zimmermann.

Let us go back to the statement we made at the beginning of this
section, characterizing the concept of perturbative
renormalization in a --renormalizable-- \textsf{QFT}. The idea is
to {\it{redefine multiplicatively}} the parameters appearing in
the Lagrangian $L$ so as to absorb all divergent contributions
order by order in the perturbation expansion. We will outline this
idea in the next step where we briefly indicate how to implement
Bogoliubov's {\it{subtraction}} procedure for renormalization in a
{\it{multiplicative}} way on the level of the Lagrangian by
{\it{adding}} the so-called {\it{counterterm}} Lagrangian $L_c$ to
the original Lagrangian $L=L_f + L_i$. Indeed, the result is $L
\xrightarrow{ren.} L_{ren}=L + L_c$ with $L_{ren}$ describing the
perturbatively renormalized quantum field theory. We achieve this
by introducing so-called $\mathrm{Z}$-factors together with the
idea of renormalized respectively bare parameters. In the case of
the $\phi^4$-theory we have
$$
    L_{ren}=L_f + L_i + \underbrace{ \frac{Z_{\phi}-1}{2} (\partial_\mu \phi)^2
                                        - \frac{( m_{1}^2Z_{m_{1}} - m_1^2)}{2}\phi^2
                                        + \frac{(\lambda_{4}Z_{\lambda_4} - \lambda_4)}{4!}\phi^4}_{L_c}.
$$
One can reorganize the right hand side into the {\it{original
form}} upon introducing the notion of bare fields and bare
parameters. First, we replace the field $\phi$ by the bare field
$$
    \phi_0:=Z_{\phi}^{\frac{1}{2}}\phi
$$
defined in terms of the $\mathrm{Z}$-factor
$Z_{\phi}:=1+\delta\phi$. Then, we introduce the
$\mathrm{Z}$-factors $Z_{m_1}$, $Z_{\lambda_4}$ and replace the
parameters $m_1$ and $\lambda_4$ by {\it{bare}} parameters
$$
    m^2_{1,0}:=m_{1}^2Z_{m_{1}}Z^{-1}_{\phi}=m^2_1\frac{(1+\delta m_1)}{(1+\delta\phi)},
 \makebox{ and }\
    \lambda_{4,0}:=\lambda_4Z_{\lambda_4}Z^{-2}_{\phi}=\lambda_4\frac{(1+\delta \lambda_4)}{(1+\delta\phi)^2},
$$
such that we can write the multiplicatively renormalized
Lagrangian as
$$
    L_{ren} = \frac{1}{2}(\partial_\mu \phi_0)^2 - \frac{1}{2} m_{1,0}^2\phi_0^2 + \frac{\lambda_{4,0}}{4!}\phi_0^4.
$$
The property of the original $\phi^4$ Lagrangian, $L=L_f+L_i$, to
be renormalizable is encoded in the fact that the counterterm
Lagrangian, $L_c$, only contains field monomials which also appear
in $L$. This allows us to recover $L_{ren}$ in the original form
of $L$ upon introducing bare parameters respectively the
$\mathrm{Z}$-factors, which implies the notion of
{\it{multiplicative renormalization}}. We observe explicitly that
$Z_\phi$ must be a numerical constant to assure {\it{locality}}
($[Z_\phi,\partial_\mu]=0$). This is also true for the other
$\mathrm{Z}$-factors as well. However, they depend in a highly
singular form on the regularization parameter. In fact,
$Z_X=1+\delta X$ is an order by order determined infinite power
series of {\it{counterterms}}, each of which is divergent when
removing the regulator parameter. They are introduced to
compensate --i.e. renormalize-- the pole parts in the original
perturbation expansion, and --one must prove that-- such a process
assures the cancellation of divergent contributions at each order
in perturbation theory, see e.g.~\cite{Collins84,IZ80}.


\subsection{Preparation of Feynman graphs for simple subtraction}
\label{ssect:BogoRbar}

Recall that a proper connected UV divergent 1PI subgraph $\gamma$
of a 1PI Feynman graph $\Gamma$ is determined by proper subsets of
the set of internal edges and vertices of $\Gamma$.

Bogoliubov's $\bar{\mathrm{R}}$-operation provides a recursive way
to calculate the counter\-terms, which we mentioned above and
denoted by $C(\Gamma)$ for each Feynman graph $\Gamma$ as follows.
The $\bar{\mathrm{R}}$-operation maps a graph $\Gamma$, that is,
the corresponding regularized amplitude to a linear combination of
suitably modified graphs including $\Gamma$ itself
\begin{equation}
    \label{eq:Bogoclassic}
    \bar{\mathrm{R}}(\Gamma) := \Gamma + \sum{}\!^{\prime}_{\{\gamma{}'\} \subsetneq \Gamma}
    \Gamma \big(\{\gamma{}'\} \to C(\gamma{}')\big).
\end{equation}
We follow the notation of~\cite{Collins06}. The primed sum is over
all unions $\{\gamma{}'\}$ of UV divergent 1PI subgraphs sitting
inside $\Gamma$. More concretely, following the terminology used
in \cite{CaswellKennedy82} we call $\mathfrak{s}:=\{\gamma{}'\}$ a
{\it{proper spinney}} of $\Gamma$ if it consists of a nonempty
union of disjoint {\it{proper}} UV divergent 1PI subgraphs,
$\{\gamma{}'\}:=\{\gamma'_1 \dots \gamma'_n\}$, $\gamma'_i
\subsetneq \Gamma$, $\gamma'_i \cap \gamma'_j = \emptyset$ for $i
\neq j$. We call the union of all such spinneys in the graph
$\Gamma$ a {\it{proper wood}} which we denote by
$$
 \mathfrak{W}(\Gamma):=\big\{\{\gamma{}'\} \subsetneq \Gamma\ \big{|}
 \{\gamma{}'\}=\{\mathrm{union\ of\ disjoint\ proper\ 1PI\ subgraphs}\} \big\}.
$$
Let us enlarge this set to $\bar{\mathfrak{W}}(\Gamma)$ which
includes by definition the empty set, $\{\emptyset\}$, and the
graph $\{\Gamma\}$ itself, i.e., $\{\emptyset\},\{\Gamma\} \in
\bar{\mathfrak{W}}(\Gamma)$.\smallskip

For instance, the two loop self-energy
graph $\Gamma\!=\!\!\begin{array}{c}\scalebox{0.6}{\coprod2} \\
\end{array}$, borrowed from $\phi^3$-theory in six dimensions, has two proper 1PI UV
subgraphs $\gamma_l\!=\!\!\begin{array}{c}\scalebox{0.6}{\l2cop} \\
\end{array}$ and
$\gamma_r\!=\!\!\begin{array}{c}\scalebox{0.6}{\r2cop}\\
\end{array}$, but they are not disjoint. Instead, they happen to be
overlapping, i.e. $\gamma_r \cap \gamma_l \neq \emptyset$ in
$\Gamma$. This is why we do not have $\{\gamma_l\ \gamma_r\}
\notin \mathfrak{W}(\Gamma)$, that is, the set of proper spinneys
corresponding to that graph is
\begin{equation}
 \label{graphs:forest1}
    \mathfrak{W}\Big(\!\!\begin{array}{c}\scalebox{0.6}{\coprod2} \\ \end{array}\!\!\Big)
    = \Big\{\big\{\!\!\begin{array}{c}\scalebox{0.6}{\r2cop}\\
    \end{array}\!\!\big\},\big\{\!\!\begin{array}{c}\scalebox{0.6}{\l2cop} \\
    \end{array}\!\!\big\}\Big\}.
 \end{equation}
A commonly used way to denote such a wood is by putting each
subgraph of a spinney into a box
\begin{equation}
 \label{graphs:forest1}
    \mathfrak{W}\Big(\!\!\begin{array}{c}\scalebox{0.6}{\coprod2} \\ \end{array}\!\!\Big)
    = \Big\{\big\{\!\!\begin{array}{c}\scalebox{0.6}{\fra2coprod2}\\
    \end{array}\!\!\big\},\big\{\!\!\begin{array}{c}\scalebox{0.6}{\Fra1coprod2} \\
    \end{array}\!\!\big\}\Big\}.
 \end{equation}
Let us remark that the notion of spinney respectively wood is
particularly well-adapted to Bogoliubov's
$\bar{\mathrm{R}}$-map~\cite{CaswellKennedy82}.
Zimmermann~\cite{Zimmermann69} gave a solution to Bogoliubov's
recursion by using the combinatorial notion of {\it{forests}} of
graphs, which differs from that of spinneys~\cite{Collins84}.

For more examples, this time borrowed from $\phi^4$-theory and
\textsf{QED} in four space-time dimensions, we look at the woods
of the divergent $4$-point graphs
$\begin{array}{c} \\[-0.3cm] \scalebox{0.7}{\roll}
\end{array}$, \scalebox{0.7}{\kite}
\begin{equation}
 \label{graphs:forest2}
    \mathfrak{W}\Big(\!\! \begin{array}{c}
                        \\[-0.3cm] \roll
                      \end{array}\!\! \Big) = \Big\{\big\{ \newfish \big\},\big\{ \newfish \big\},
                                              \big\{ \newfish\ \newfish \big\}\Big\},
\quad
    \mathfrak{W}\Big( \kite \Big) = \Big\{\big\{ \newfish \big\},\big\{ \winecup \big\}\Big\}.
\end{equation}

\noindent and the \textsf{QED} self-energy graphs
\scalebox{0.7}{\tw1}\ , \scalebox{0.7}{\sw1}\ ,
\scalebox{0.7}{\uw1} \ and \scalebox{0.7}{\vw1}
 \allowdisplaybreaks{
\begin{eqnarray}
 \label{graphs:forest3}
    \mathfrak{W}\Big( \tw1 \Big) &\!\!=\!\!& \Big\{\big\{ \emptyset \big\}\Big\}\\
 \label{graphs:forest4}
    \mathfrak{W}\Big( \sw1 \Big) &\!\!=\!\!& \Big\{\big\{ \tw1 \big\}\Big\}\\
 \label{graphs:forest5}
   \mathfrak{W}\Big( \uw1 \Big) &\!\!=\!\!& \Big\{\big\{ \tw1 \big\},\big\{ \sw1 \big\}\Big\}\\
 \label{graphs:forest6}
    \mathfrak{W}\Big( \vw1 \Big) &\!\!=\!\!& \Big\{\big\{ \tw1 \big\},\big\{ \tw1 \big\},
                                         \big\{ \tw1\ \tw1 \big\}\Big\}.
\end{eqnarray}}

\noindent We will later see that the union of 1PI subgraphs, such
as for instance in the spinney $\big\{ \scalebox{0.7}{\tw1}\
\scalebox{0.7}{\tw1} \big\} \in \mathfrak{W}\Big(
\scalebox{0.7}{\vw1} \Big)$, gives rise to the algebra part in
Connes' and Kreimer's Hopf algebra of Feynman graphs in
Section~\ref{ssect:BirkhoffCK}. Also, we should underline that the
subgraphs in a spinney may posses nontrivial spinneys themselves,
see for instance the graph \scalebox{0.7}{\sw1} which appears as a
spinney in the wood (\ref{graphs:forest5}).

We call graphs such as the one-loop diagram \scalebox{0.7}{\tw1}
{\it{primitive}} if their woods contain only the empty set, i.e.,
they have no proper UV divergent 1PI subgraphs and hence the
primed sum in (\ref{eq:Bogoclassic}) disappears.

To each element $\{\gamma'\} \in \mathfrak{W}(\Gamma)$ of a graph
$\Gamma$ corresponds a {\it{cograph}} denoted by $\Gamma /
\{\gamma'\}$, which follows from the contraction of all graphs
$\gamma \in \{\gamma'\}$ in $\Gamma$ to a point at once. For
instance, to the spinneys in the woods (\ref{graphs:forest4}) -
(\ref{graphs:forest6}) correspond the cographs
 \allowdisplaybreaks{
\begin{eqnarray*}
      \sw1 \big{/} \big\{\tw1\big\} \!\!\!&=&\!\!\!\! \tw1, \\
      \uw1 \big{/} \big\{\tw1\big\} \!\!\!&=&\!\!\!\! \sw1\ , \uw1 \big{/} \big\{\sw1\big\} \!=\! \tw1, \\
      \vw1 \big{/} \big\{\tw1\big\} \!\!\!&=&\!\!\!\! \sw1\ , \vw1 \big{/} \big\{ \tw1\ \tw1 \big\} \!=\! \tw1,
\end{eqnarray*}}
respectively. Remember, that we defined
$\bar{\mathfrak{W}}(\Gamma)$ to contain the spinneys $\{\Gamma\}$
and $\{\emptyset\}$. Those spinneys have the following cographs,
$\Gamma/\{\Gamma\}=\emptyset$ and $\Gamma/\{\emptyset\}=\Gamma$,
respectively. Again, as a preliminary remark we mention here that
in the Hopf algebra of Feynman graphs the coproduct of a graph
consists simply of the sum of tensor products of each spinney with
its cograph, see Eqs.~(\ref{def:coprod}) and
(\ref{eq:coproductWOOD}). \smallskip

We call a spinney {\it{maximal}} if its cograph is primitive. Each
of the graphs in the \textsf{QED} example contains only one
maximal spinney. The wood in (\ref{graphs:forest1}) consists of
two maximal spinneys, as the two UV divergent 1PI subgraphs
overlap. In fact, Feynman graphs with overlapping UV divergent 1PI
subgraphs always contain more than one maximal spinney. We will
come back to this in subsection~\ref{sssect:rootedtrees} where we
remark about the correspondence between Feynman graphs and
decorated non-planar rooted trees.

As an example for the sum in Eq.~(\ref{eq:Bogoclassic}) we write
it explicitly for the graph \scalebox{0.7}{\roll}\ using its wood
in (\ref{graphs:forest2}) \allowdisplaybreaks{
\begin{eqnarray*}
 \lefteqn{\bar{\mathrm{R}}\Big(\!\! \begin{array}{c}
                            \\[-0.3cm] \roll
                          \end{array}\!\! \Big) =   \begin{array}{c}
                                                        \\[-0.3cm] \roll
                                                    \end{array} +
                          \sum_{\{\gamma{}'\} \in \big\{ \{ \scalebox{0.7}{\newfish}\},
                                                                      \{ \scalebox{0.7}{\newfish}\},
                                               \{ \scalebox{0.7}{\newfish} \scalebox{0.7}{\newfish}\}\big\}}
                          \begin{array}{c}
                            \\[-0.3cm] \roll
                          \end{array}\!\!\big(\{\gamma{}'\} \to C(\gamma{}')\big)}\\
                 &=&
       \begin{array}{c}
            \\[-0.3cm] \roll
       \end{array}
            + 2\begin{array}{c}
                 \\[-0.3cm] \roll
                \end{array}\!\!\!\Big(\big\{ \newfish \big\} \to C\big( \newfish \big)\Big)
             + \begin{array}{c}
                 \\[-0.3cm] \roll
               \end{array}\!\!\!\Big(\big\{ \newfish \newfish \big\} \to C\big( \newfish \newfish\big)\Big).
\end{eqnarray*}}

\noindent Recall the notational abuse by using a graph just as a
mnemonic for an analytic expression, e.g. a Laurent series. We
must now explain what happens inside the primed sum in
Eq.~(\ref{eq:Bogoclassic})
\begin{equation}
    \label{eq:Bogosum}
    \sum{}\!^{\prime}_{\{\gamma{}'\} \in \mathfrak{W}(\Gamma)}
    \Gamma \big(\{\gamma{}'\} \to C(\gamma{}')\big).
\end{equation}

\noindent More precisely, the expression $\Gamma
\big(\{\gamma{}'\} \to C(\gamma{}')\big)$ denotes the
{\it{replacement}} of each subgraph $\gamma \in \{\gamma{}'\}$ in
$\Gamma$ by its {\it{counterterm}} $C(\gamma)$ defined in terms of
the $\bar{\mathrm{R}}$-operation and the particular
renormalization scheme map $R$
\begin{equation}
    \label{eq:BogoCounterTerm}
    C(\gamma):=-R(\bar{\mathrm{R}}(\gamma)).
\end{equation}

\noindent At this point the recursive nature of Bogoliubov's
procedure becomes evident, making it a suitable tool for applying
inductive arguments in the proof of perturbative renormalization
to all orders. Zimmermann established a solution to Bogoliubov's
recursive formula~\cite{Zimmermann69}. A subtle point arises from
the evaluation of the counterterm map on the disjoint union of
graphs, e.g. $C\big( \newfish \newfish \big)$ in the above
example. In fact, it is natural to demand that $C$ is an algebra
morphism, that is, $C\big( \newfish\ \newfish\big) = C\big(
\newfish \big) C\big(\newfish\big)$. We should remark here that it
is exactly at this point where Connes and Kreimer in the
particular context of the minimal subtraction scheme in
dimensional regularization needed the renormalization scheme map
$R_{ms}$ to satisfy the so-called multiplicativity
constraint~\cite{CK2,KreimerChen}, i.e. $R_{ms}$ must have the
Rota--Baxter property~(\ref{eq:RBrel}), which in turn implies the
property of multiplicativity for the counterterm map $C$. We will
return to this problem later in more detail.

Graphically the replacement operation, $\big(\{\gamma{}'\} \to
C(\gamma{}')\big)$, is represented by shrinking the UV divergent
1PI subgraph(s) from $\{\gamma{}'\}$ to a point --,i.e., becoming
a vertex-- in $\Gamma$ and to mark that point with a cross or
alike to indicate the counterterm. For a graph $\gamma$ having no
proper UV divergent 1PI subgraphs we find $C(\gamma)=-R(\gamma)$.

As we will see, the Hopf algebra approach provides a very
convenient way to describe this recursive graph replacement
operation on Feynman graphs in a combinatorial well-defined manner
in terms of a convolution product.

From Bogoliubov's $\bar{\mathrm{R}}$-operation the renormalized
expression for the graph $\Gamma$ follows simply by a subtraction
\allowdisplaybreaks{
\begin{eqnarray*}
  \Gamma \xrightarrow{ren.} \Gamma_+ := (\id-R)\big(\bar{\mathrm{R}}(\Gamma) \big)
                                     &=&\bar{\mathrm{R}}(\Gamma) + C(\Gamma)\\
                            &=& \Gamma + \sum{}\!^{\prime}_{\{\gamma{}'\} \in
                            \mathfrak{W}(\Gamma)}
                              \Gamma \big(\{\gamma{}'\} \to C(\gamma{}')\big) + C(\Gamma).
\end{eqnarray*}}
This simple subtraction is made possible since the
$\bar{\mathrm{R}}$-operation applied to $\Gamma$, see
Eq.~(\ref{eq:Bogoclassic}), contains the graph $\Gamma$ itself,
such that adding to $\Gamma$ the sum (\ref{eq:Bogosum}) where the
proper subgraphs $\gamma$ in $\Gamma$ are replaced by there
counterterms $C(\gamma)$ prepares the graph in such a particular
way that a simple subtraction finally cancels its superficial
divergence. Indeed, Bogoliubov's $\bar{\mathrm{R}}$-operation
leads to the renormalization of all proper UV divergent 1PI
subgraphs $\gamma$ in $\Gamma$, i.e., they are replaced by their
renormalized value $\gamma_+$ and we may write for Bogoliubov's
$\bar{\mathrm{R}}$-operation
\begin{equation*}
    \bar{\mathrm{R}}(\Gamma) = \sum{}\!^{\prime}_{\{\gamma{}'\} \in \mathfrak{W}(\Gamma)}
    \Gamma \big(\{\gamma{}'\} \to (\gamma{}'_{+})\big).
\end{equation*}

\noindent As before, we encounter a subtle point with respect to
the multiplicativity of the renormalization map, that is, on the
disjoint union of graphs $\{\gamma_1 \cdots \gamma_n\}$ we need,
$(\gamma_1 \cdots \gamma_n)_{+}={\gamma_1}_{+} \cdots
{\gamma_n}_{+}$.

Let us exemplify this intriguing operation by a little toy-model
calculation. We choose the following integral
$f_1(c):=\int_{0}^{\infty}(y+c)^{-1}dy \in \mathbb{R}$, which is
ill-defined at the upper boundary. It is represented pictorially
by the toy-model `self-energy' graph \tw1. Here $c>0$ is an
external dimensional parameter. Its non-zero value avoids a
so-called {\it{infrared divergency}} problem at the lower integral
boundary. We may now iterate those integrals, representing them by
rainbow graphs
 \allowdisplaybreaks{
\begin{eqnarray*}
    \sw1 \sim f_2(c)\!\!&\!\!:=\!\!&\!\!\int_{0}^{\infty} f_1(y) (y+c)^{-1}dy,
\\
    \uw1 \sim f_3(c)\!\!&\!\!:=\!\!&\!\!\int_{0}^{\infty} f_2(y) (y+c)^{-1}dy \cdots
\end{eqnarray*}}
Of course, these graphs just exemplify the simple iterated nature
of the integral expression and we could have used instead rooted
trees without side branchings~\cite{CK1,KreimerChen}. Similar
graphs appear in \textsf{QED}. However, there they represent
different functions corresponding to the \textsf{QED} Feynman
rules~\cite{IZ80,KreimerBuch}. Here and further below, in
Section~\ref{sect:RBmatrixCK}, we will use those graphs without
any reference to \textsf{QED}. In fact, they will appear in the
context of a simple matrix calculus capturing the combinatorics
involved in the Bogoliubov formulae which was inspired by the Hopf
algebra approach to renormalization.

The functions $f_n(c)$ actually appear as coefficients in the
--formal-- series expansion of the perturbative calculation of the
physical quantity represented by the function $F(c)$
$$
    F(c) = 1 + \alpha f_1(c) + \alpha^2 f_2(c) + \alpha^3 f_3(c) + \cdots
$$
where $\alpha$ is our perturbation parameter, which is supposed to
be small and finite. We may write this as a recursive equation due
to the simple iterated structure of the functions $f_n$
$$
    F(c) = 1 + \alpha \int_{0}^{\infty} F(y)\ \frac{ dy}{(y+c)}.
$$
This recursion is a highly simplified version of a
Dyson--Schwinger equation for the quantity $F$ (we refer the
reader to Kreimer's contribution in this volume).

However we readily observe that the integrals $f_n$, $n>0$ are not
well-defined. Indeed, they are logarithmically divergent,
$\omega(f_n)=0$ as a simple power-counting immediately tells.
Hence, we introduce a regularization parameter $\varepsilon \in
\mathbb{C}$ to render them formally finite
$$
    f_n(c;\varepsilon):=\mu^{\varepsilon}
                        \int_{0}^{\infty} f_{n-1}(y;\varepsilon)(y+c)^{-1-\varepsilon}dy
                        \in \mathbb{C}[\varepsilon^{-1},\varepsilon]][[\ln(\mu/c)]],\qquad n>0,
$$
and $f_0=1$. Here, the parameter $\mu$ was introduced for
dimensional reasons so as to keep the $f_n(c;\varepsilon)$
dimensionless. This implies a regularized, and hence formally
finite Dyson--Schwinger equation
$$
    F(c;\varepsilon) = 1 + \alpha \int_{0}^{\infty} F(y;\varepsilon)\
                      \frac{ \mu^{\varepsilon}\ dy}{(y+c)^{1+\varepsilon}}.
$$
The --forbidden-- limit $\varepsilon \to 0$ just brings back the
original divergence. The first coefficient function is
$$
    f_1(c;\varepsilon) = \mu^{\varepsilon} c^{-\varepsilon}
      \varepsilon^{-1} = \varepsilon^{-1} \sum_{n\geq 0} \frac{(\varepsilon)^n \log^n(\mu/c)}{n!}.
$$
For $n>1$ those iterated $\varepsilon$-regularized integrals are
solved in terms of the function
$$
    \int_{0}^{\infty} y^{ - a\varepsilon}(y+c)^{-1-b\varepsilon} dy =
    B\big((a+b)\varepsilon,1-a\varepsilon\big)c^{-(a+b)\varepsilon}\ ,
    \makebox{where}\ B(a,b):=
    \frac{\Gamma(a)\Gamma(b)}{\Gamma(a+b)},
$$
and $\Gamma(a)$ is the usual Gamma-function \cite{KreimerBuch}.
For $n=2$ we find \allowdisplaybreaks{
\begin{eqnarray*}
    f_2(c;\varepsilon) &=& \mu^{2\varepsilon} \int_{0}^{\infty}\int_{0}^{\infty} (y+c)^{-1-\varepsilon} (z+y)^{-1-\varepsilon} dz dy\\
                       &=& \frac{\mu^{2\varepsilon}}{\varepsilon}  \int_{0}^{\infty} y^{-\varepsilon}(y+c)^{-1-\varepsilon} dy
                        = \frac{\mu^{2\varepsilon} c^{-2\varepsilon}}{\varepsilon} B(2\varepsilon,1-\varepsilon)\\
                       &=& \sum_{n\geq 0} \frac{(2\varepsilon)^n
    \log^n(\mu/c)}{n!}\Big(\frac{1}{2\varepsilon^{2}} + h + O(\varepsilon) \Big),
\end{eqnarray*}}
where $h$ is a constant term. Now, first observe that,
$$
    f_1(c;\varepsilon) - f_1(\mu;\varepsilon) =
    f_1(c;\varepsilon) - \Big( \frac{1}{\varepsilon}
    \Big) \in \mathbb{C}[[\varepsilon]][[\ln(\mu/c)]]
                                \xrightarrow{\ \varepsilon \to 0\ } f_{1,+}=\log{\mu/ c}.
$$
So, by subtracting $f_1(\mu;\varepsilon)$ from
$f_1(c;\varepsilon)$ we have achieved a Laurent series without
pole part, such that the limit $\varepsilon \to 0$ is allowed and
gives a finite, that is, renormalized value of the first order
coefficient, $f_1(c) \to f_{1,+}$.

Let us formalize the operation which provides the term to be
subtracted by calling it the {\it{renormalization scheme map
$R$}}, defined to be
$$
    R(g)(t) := g(t)_{|t=\mu},
$$
for an arbitrary function $g$. In fact, we just truncated the
Taylor expansion of $g(t)$, $\mathcal{T}_\mu[g](t) := \sum_{k \geq
0}  g^{(k)}(\mu)\frac{(t-\mu)^k}{k!}$, at zeroth order. The order
at which one truncates the Taylor expansion corresponds to the
superficial degree of divergence, e.g. in our toy model example we
find logarithmic divergencies, $\omega(f_n)=0$, $n>0$. The
counterterm for the primitive divergent regularized one-loop graph
\scalebox{0.7}{\tw1}\ is
$$
    C(\tw1)=-R(f_1)_{|t=\mu} = - f_1(\mu;\epsilon).
$$
Applying the same procedure to the second order term we find
$$
    f_2(c;\varepsilon) -  f_2(\mu;\varepsilon)
    \xrightarrow{} \frac{\log{\mu/ c}}{\varepsilon} + h' + O(\varepsilon),
$$
with a modified constant term $h'$. Hence, our subtraction ansatz
fails already at second order. Indeed, what we find is an artefact
of the --UV divergent 1PI-- one-loop diagram \tw1\ sitting inside
the two loop diagram \sw1\ , reflected in the $1/\varepsilon$-term
with logarithmic coefficient.

Instead, let us apply Bogoliubov's $\bar{\mathrm{R}}$-operation to
the logarithmic divergent two-loop graph \sw1\ with one-loop \tw1\
subdivergence \allowdisplaybreaks{
\begin{eqnarray*}
    \bar{\mathrm{R}}\Big(\sw1\ \Big) &=& \sw1 + \sum_{ \{\gamma'\} \in \big\{\big\{\tw1 \big\}\big\}}
                                                          \sw1\big(\{\gamma'\} \to C(\tw1)\big)\\
        &=&\sw1 + \sw1 \big(\{\tw1\} \to C(\tw1)\big)\\
        &=& \sw1 + C(\tw1)\tw1.
\end{eqnarray*}}
For the sake of notational transparency we omitted to mark the
obvious place in the cograph \tw1\ where the subgraph \tw1\ of
\sw1\ was contracted to a point. In fact, with the function
$f_{1}$ and $f_{2}$ replacing the graphs, we find
\allowdisplaybreaks{
\begin{eqnarray*}
    \sw1 \!+\! C(\tw1)\tw1 \!\!&=&\!\!\! \mu^{\varepsilon}\int_{0}^{\infty} \frac{f_1(y;\varepsilon)dy}{(y+c)^{1+\varepsilon}}
                                    + C(\tw1) \mu^{\varepsilon}\int_{0}^{\infty}\frac{dy}{(y+c)^{1+\varepsilon}}\\
                       &=& \mu^{\varepsilon}\int_{0}^{\infty} \frac{\big(f_1(y)+ C(\tw1) \big)dy}{ (y+c)^{1+\varepsilon}}\\
                       &=& \mu^{\varepsilon}\int_{0}^{\infty} \frac{\big(f_1(y;\varepsilon) - f_{1}(\mu;\varepsilon)\big)dy}
                                                                   {(y+c)^{1+\varepsilon}}\\
                       &=& \mu^{\varepsilon}\int_{0}^{\infty} \frac{f_{1,+}(y;\varepsilon)dy}{(y+c)^{1+\varepsilon}}        .
\end{eqnarray*}}
And hence, we have replaced the subdivergent one-loop graph by its
renormalized expression, see expression (\ref{eq:Bogosum}).
\begin{eqnarray*}
    \bar{\mathrm{R}}\Big(\sw1\ \Big) &=& f_2(c;\varepsilon) +
            C\big(f_1(c;\varepsilon)\big)f_1(c;\varepsilon) = f_2(c;\varepsilon) -
            R[f_1(c;\varepsilon)]f_1(c;\varepsilon)\\
            &=& \sum_{n\geq 0} \frac{(2\varepsilon)^n
                \log^n(\mu/c)}{n!}\Big(\frac{1}{2\varepsilon^{2}} + h +O(\varepsilon) \Big)
             - \frac{1}{\varepsilon^2}\sum_{n\geq 0} \frac{(\varepsilon)^n \log^n(\mu/c)}{n!}.
\end{eqnarray*}
One then shows that $\bar{\mathrm{R}}\big(\sw1\ \big) \in
\mathbb{C}[\varepsilon^{-1},\varepsilon]]$ does not contain any
$\log(\mu/c)$-terms as coefficients in the pole part and that the
simple subtraction
$$
    (\id-R)\bar{\mathrm{R}}\Big(\sw1\ \Big) =
    {\bar{\mathrm{R}}}\big(\sw1\ \big) + C\Big(\sw1\ \Big)
    \xrightarrow{\ \varepsilon \to 0\ }
    f_{2,+}(c) < \infty.
$$

The renormalization scheme $R$, of course, is defined as a map
only on the particular space of regularized amplitudes, e.g., the
field of Laurent series. This provides a recursive way to
calculate the counterterms as well as the renormalized amplitude
corresponding to a graph order by order in the perturbation
expansion. We may remark here that in the classical
--pre-Hopfian-- approach the notion of graph and amplitude are
used in an interchangeable manner.

We have described a way to renormalize each coefficient in the
perturbative expansion of $F$. We may now introduce a
$\mathrm{Z}$-factor, defined as an expansion in $\alpha$ with the
counterterms $C(f_n)=C(f_n(c;\varepsilon))$ as coefficients
$$
    \mathrm{Z}=\mathrm{Z}(\varepsilon):= 1 + \sum_{n>0} \alpha^n C(f_n).
$$
We then multiply the Dyson--Schwinger toy-equation for $F$ with
$\mathrm{Z}$
$$
   F(c;\varepsilon) \xrightarrow{\mathrm{Z}}
   \bar{F}(c;\varepsilon) = \mathrm{Z} + \alpha \int_{0}^{\infty} \bar{F}(y;\varepsilon)\
                                                      \frac{\mu^{\varepsilon}\ dy}{(y+c)^{1+\varepsilon}}.
$$
Upon expanding $\mathrm{Z}$ we find up to second order
$$
    \bar{F}(c;\varepsilon) = 1 + \alpha^1\big(C(f_1) + f_1(c;\varepsilon)\big)
                                    + \alpha^2\big(C(f_2) + C(f_1)f_1(c;\varepsilon) + f_2(c;\varepsilon)\big)
                                    + O(\alpha^3).
$$
Comparing with our calculations of the renormalized amplitudes
$f_{1,+}$ and $f_{2,+}$ above, we readily observe that the
multiplication of the Dyson--Schwinger toy-equation with
$\mathrm{Z}$ has renormalized the perturbative expansion of $F$
order by order in the coupling $\alpha$
$$
    F(c;\varepsilon) \xrightarrow{\mathrm{Z}} \bar{F}(c;\varepsilon) \xrightarrow{\varepsilon \to 0} F_{ren}(c).
$$
This is a simplified picture of the process of multiplicative
renormalization via constant $\mathrm{Z}$-factors,
$\mathrm{Z}F(c;\varepsilon) \xrightarrow{\varepsilon \to 0}
F_{ren}(c)$, which is a power series in $\alpha$ with finite
coefficients, see \cite{KreimerChen,KreimerBuch}. Later we will
establish a more involved but similar result known as
Connes--Kreimer's Birkhoff decomposition of renormalization.
\smallskip

Let us summarize and see what we have learned so far. Feynman
amplitudes appear as the coefficients in the perturbative
expansion of physical quantities in powers of a coupling constant.
In general, they are ill-defined objects. The process of
regularization renders them formally finite upon introducing
nonphysical parameters into the theory which allow for a
decomposition of such amplitudes into a finite part, although
ambiguously defined, and one that contains the divergent
contributions. Renormalization of such regularized amplitudes is
mainly guided by constraints coming from fundamental physical
principles, such as locality, unitarity, and Lorentz invariance
and consists of careful manipulations eliminating, i.e.,
subtracting, though in a highly non-trivial manner, the illness
causing parts, together with a renormalization hypotheses which
fixes the value of the remaining finite parts. Upon introducing
multiplicative $\mathrm{Z}$-factors respectively bare parameters
$m_0,\lambda_0$ and bare fields in the Lagrangian, such that $L
\xrightarrow{\phi_0,m_0,\lambda_{0}}L_{ren}$ we arrive at a local
perturbative renormalization of a --renormalizable-- quantum field
theory in the Lagrangian approach. However, it is crucial for the
$\mathrm{Z}$-factors to be independent of external momenta, as
otherwise such a dependence would obstruct the equations of motion
derived from the Lagrangian. After renormalization one can remove
consistently the regularization parameter in the perturbative
expansion.

Bogoliubov's $\bar{\rm{R}}$-map was exactly invented to extract
the finite parts of --re\-gu\-la\-rized-- Feynman integrals
corresponding to Feynman graphs. The simple strategy used in this
process consists of preparing a superficially divergent Feynman
graph respectively its proper ultraviolet divergent one-particle
irreducible subgraphs in such a way that a final `naive'
subtraction renders the whole amplitude associated with that
particular graph finite. Today this is summarized in the BPHZ
renormalization prescription refereing to Bogoliubov, Parasiuk,
Hepp and Zimmermann \cite{BoPa57,Hepp66,Zimmermann69},
see~\cite{BS59,CaswellKennedy82,Collins84} for more details.


\section{Connes--Kreimer's Hopf algebra of Feynman graphs}
\label{sect:HopfCK}

Before recasting adequately the above classical setting of
perturbative renormalization in Hopf algebraic terms we would like
to recall some basic notions of connected graded Hopf algebras.

In fact, we will provide most of the algebraic notions we need in
the sequel most of which are well-known from the classical
literature, with only two exceptions, namely, we will explore
Rota--Baxter algebras in an extra section. We do this simply
because as we will see much of the algebraic reasoning for
Connes--Kreimer's Birkhoff decomposition follows from key
properties characterizing algebras especially associative ones
with a Rota--Baxter structure. Furthermore, in subsection
\ref{sssect:BCH} we present a general factorization theorem for
complete filtered algebras, which we discovered together with
D.~Kreimer in \cite{E-G-K2,E-G-K3} and which was further explored
jointly with D.~Manchon in \cite{EGMbch06}.

The concept of a Hopf algebra originated from the work of
Hopf~\cite{Hop41} by algebraic topologists~\cite{MilnorMoore65}.
Comprehensive treatments of Hopf algebras can be found
in~\cite{Abe80,Swe69}. Some readers may find Bergman's
paper~\cite{Berg85} a useful reference. Since the 1980s Hopf
algebras became familiar objects in the realm of quantum
groups~\cite{ChariPressley95,FGV01,Kassel95,Majid,ShSt93}, and
more recently in noncommutative geometry~\cite{CMos98}. The class
of Hopf algebras we have in mind appeared originally in the late
1970s in the context of combinatorics, where they were introduced
essentially by Gian-Carlo Rota. The seminal references in this
field are the work by Rota~\cite{Rota78}, and in somewhat
disguised form Joni and Rota~\cite{JoniRota79}. Actually, the
latter article essentially presents a long list of bialgebras,
whereas the Hopf algebraic extra structure, i.e., the antipode map
was largely ignored. Later, one of Rota's students, W.~Schmitt,
further extended the subject considerably into the field of
incidence Hopf algebras \cite{Schmitt87,Schmitt93,Schmitt94}.
Other useful references on Hopf algebras related to combinatorial
structures are \cite{FG05,Majid,NiSw1982,SD97}.\\

In the sequel, $\mathbb{K}$ denotes the ground field of
$char(\mathbb{K})=0$ over which all algebraic structures are
defined. The following results have all appeared in the literature
and we refer the reader to the above cited references, especially
\cite{FG05,Ma01} for more details.


\subsection{Complete filtered algebra}
\label{sec:algebra}

In this section we establish general results to be applied in
later sections. We obtain from the Baker--Campbell--Hausdorff
($\BCH$) series a non-linear map $\chi$ on a complete filtered
algebra $A$ which we called $BC\!H$-recursion. This recursion
gives a decomposition on the exponential level (see
Theorem~\ref{thm:bch}), and a one-sided inverse of the
Baker--Campbell--Hausdorff series with the later regarded as a map
from $A \times A \to A$.\\


\subsubsection{Algebra}

We denote associative $\mathbb{K}$-algebras by the triple
$(A,m_A,\eta_A)$ where $A$ is a $\mathbb{K}$-vector space with a
product $m_A: A \otimes A \to A$, supposed to be associative, $m_A
\circ (m_A \otimes \id_A) = m_A \circ (\id_A \otimes m_A)$, and
$\eta_A: \mathbb{K} \to A$ is the unit map. Often we denote the
unit element in the algebra by $1_A$. A $\mathbb{K}$-subalgebra in
the algebra $A$ is a $\mathbb{K}$-vector subspace $B \subset A$,
such that for all $b,b' \in B$ we have $m_A(b \otimes b') \in B$.
A $\mathbb{K}$-subalgebra $I \subset A$ is called (right-)
left-ideal if for all $i \in I$ and $a \in A$, ($m_A( i \otimes a)
\in I$) $m_A(a \otimes i) \in I$. $I \subset A$ is called a
bilateral ideal, or just an ideal if it is a left- and
right-ideal.

In order to motivate the concept of a $\mathbb{K}$-coalgebra to be
introduced below, we rephrase the definition of a
$\mathbb{K}$-algebra $A$ as a $\mathbb{K}$-vector space $A$
together with a $\mathbb{K}$-vector space morphism $m_A: A \otimes
A \to A$ such that the diagram
\begin{equation}
    \xymatrix{ A \otimes A \otimes A \ar[rr]^{m_A \otimes \id_A}
       \ar[d]_{\id_A \otimes m_A} && A \otimes A \ar[d]^{m_A}\\
        A \otimes A \ar[rr]^{m_A} && A }
    \label{assoc2}
\end{equation}
is commutative. $A$ is a unital $\mathbb{K}$-algebra if there is
furthermore a $\mathbb{K}$-vector space map $\eta_A: \mathbb{K}
\to A$ such that the diagram
\begin{equation}
 \xymatrix{ \mathbb{K} \otimes A \ar[r]^{\eta_A \otimes \id_A}
 \ar[rd]_{\alpha_l} & A \otimes A \ar[d]^{m_A} & A \otimes \mathbb{K}
 \ar[l]_{\id_A \otimes \eta_A}
 \ar[ld]^{\alpha_r} \\
  & A & }
    \label{unit}
\end{equation}
is commutative. Here $\alpha_l$ (resp. $\alpha_r$) is the
isomorphism sending $k \otimes a$ (resp. $a \otimes k$) to $ka$,
for $k \in  \mathbb{K}$, $a \in A$.

Let $\tau:= \tau_{A,A}: A \otimes A \to A \otimes A$ be the flip
map defined by $\tau_{A,A}(x \otimes y):= y \otimes x$. $A$ is a
commutative $\mathbb{K}$-algebra if the next diagram commutes,
\begin{equation}
 \xymatrix{
     A \otimes A \ar[d]_{m_A} \ar[r]^{\tau} &  A \otimes A
                                             \ar[ld]^{m_A} \\
     A & }
 \label{abelian}
\end{equation}

\noindent We denote by $\call(A)$ the Lie algebra associated with
a $\mathbb{K}$-algebra $A$ by anti-sym\-me\-trization of the
product $m_A$. At this point the reader may wish to recall the
definition and construction of the universal enveloping algebra
$\mathfrak{U}(\call)$ of a Lie algebra $\call$
\cite{FG01,Reutenauer}, and the fact that for a Lie algebra
$\call$ with an ordered basis $l_1,l_2,\cdots,l_n,\cdots$ we have
an explicit basis $\{ l_{i_1}^{n_1} \dots l_{i_s}^{n_s} \big{|}
i_1 < \dots < i_s,\: n_i > 0,\ s \geq i \geq 1\}$ for
$\mathfrak{U}(\call)$.\\


\subsubsection{Filtered algebra}

A {\it{filtered $\mathbb{K}$-algebra}} is a
$\mathbb{K}$-algebra $A$ together with a {\it{decreasing}}
filtration, i.e., there are nonunitary $\mathbb{K}$-subalgebras
$A\fil{n+1} \subseteq A\fil{n}$, $n \geq 0$ of $A$ such that
$$
  A = \bigcup_{n\geq 0} A\fil{n}\ , \qquad m_A(A\fil{n} \otimes A\fil{m}) \subseteq A\fil{n+m}.
$$

\noindent It follows that $A\fil{n}$ is an ideal of $A$. In a filtered $\mathbb{K}$-algebra
$A$, we can use the subsets $\{A\fil{n}\}$ to define a {\it{metric}} on
$A$ in the standard way. Define, for $a \in A$,
$$
 l(a)=\left \{
         \begin{array}{ll}
           \max \{k \ \big|\ a \in A\fil{k}\}, & a \notin \cap_n A\fil{n},\\
                                   \infty,& {\rm otherwise}
        \end{array}
        \right.
$$
and, for $a,b \in A$, $d(a,b)=2^{-l(b-a)}$.

A filtered algebra is called {\it{complete}} if $A$ is a complete
metric space with metric $d(a,b)$, that is, every Cauchy sequence
in $A$ converges. Equivalently, we also record that a filtered
$\mathbb{K}$-algebra $A$ with $\{A\fil{n}\}$ is complete if $\cap_n A\fil{n}
= 0$ and if the resulting embedding
$$
  A \to \bar{A}:= \varprojlim A/A\fil{n}
$$
is an isomorphism. When $A$ is a complete filtered algebra, the
functions
\begin{align}
    \exp:& A\fil{1} \to 1_A + A\fil{1},\qquad \exp(a):=\sum_{n=0}^\infty \frac{a^n}{n!},
                   \label{eq:expon} \\
    \log:& 1_A + A\fil{1} \to A\fil{1},\qquad \log(1_A + a):=-\sum_{n=1}^\infty \frac{(-a)^n}{n}
                   \label{eq:logar}
\end{align}
are well-defined and are the inverse of each other.

Let us mention two examples of complete filtered associative
algebra that will cross our ways again further below. First,
consider for $A$ being an arbitrary associative
$\mathbb{K}$-algebra, the power series ring $\mathcal{A}:=A[[t]]$
in one (commuting) variable $t$. This is a complete filtered
algebra with the filtration given by powers of $t$,
$\cala\fil{n}:=t^nA[[t]]$, $n>0$.

\begin{exam}\label{ex:matrix}{\rm{{\it{Triangular matrices:}}
Another example is given by the subalgebra $\mathcal{M}^\ell_n(A)
\subset \mathcal{M}_n(A)$ of (upper) lower triangular matrices in
the algebra of $n \times n$ matrices with entries in the
associative algebra $A$, and with $n$ finite or infinite.
$\mathcal{M}^\ell_n(A)\fil{k}$ is the ideal of strictly lower
triangular matrices with zero on the main diagonal and on the
first $k-1$ subdiagonals, $k>1$. We then have the decreasing
filtration
$$
   \mathcal{M}^\ell_n(A) \supset \mathcal{M}^\ell_n(A)\fil{1} \supset \dots
  \supset \mathcal{M}^\ell_n(A)\fil{k-1} \supset \mathcal{M}^\ell_n(A)\fil{k} \supset \cdots, k<n,
$$
with
$$
  \mathcal{M}^\ell_n(A)\fil{k}\: \mathcal{M}^\ell_n(A)\fil{m} \subset
  \mathcal{M}^\ell_n(A)\fil{k+m}.
$$
For $A$ being commutative we denote by $\mathfrak{M}_n(A)$ the
group of lower triangular matrices with unit diagonal which is
$\mathfrak{M}_n(A) = {\bf{1}} + \mathcal{M}^\ell_n(A)\fil{1}$. Here the
$n \times n$, $n\leq \infty$, unit matrix is given by
\begin{equation}
    \label{def:unitmatrix}
     {\bf 1}
    :=(\delta_{ij}1_A)_{1\leq i,j\leq n }.
\end{equation}
}}
\end{exam}


\subsubsection{Baker--Campbell--Hausdorff recursion}
\label{sssect:BCH}

Formulae (\ref{eq:expon}) and  (\ref{eq:logar}) naturally lead to
the question whether the underlying complete filtered algebra $A$
is commutative or not. In general, we must work with the
Baker--Campbell--Hausdorff ($\BCH$) formula for products of two
exponentials
$$
    \exp(x)\exp(y)=\exp\big(x+y+\BCH(x,y)\big).
$$
$\BCH(x,y)$ is a power series in the non-commutative power series
algebra $Q:=\QQ\langle\langle x,y \rangle\rangle$. Let us recall
the first few terms of~\cite{Reutenauer,Varadarajan}
$$
    \BCH(x,y)=\frac{1}{2}[x,y]+\frac{1}{12}[x,[x,y]]-\frac{1}{12}[y,[x,y]]-
                \frac{1}{24}[x,[y,[x,y]]] +\cdots
$$
where $[x,y]:=xy-yx$ is the commutator of $x$ and $y$ in $Q$.
Also, we denote $C(x,y):=x+y+\BCH(x,y)$. So we have
\begin{equation}
    \label{eq:fullBCH}
    C(x,y) = \log \big(\exp(x)\exp(y)\big).
\end{equation}
Then for any complete filtered algebra $A$ and $u,v \in A\fil{1}$,
$C(u,v) \in A\fil{1}$ is well-defined and we get a map
$$
    C : A\fil{1} \times A\fil{1} \to A\fil{1}.
$$

Now let $P: A \to A$ be any linear map preserving the filtration
of $A$, $P(A\fil{n})\subseteq A\fil{n}$. We define $\tilde{P}$ to be
$\id_A-P$. For $a \in A\fil{1}$, define $\chi(a) = \lim_{n \to \infty}
\chi_{(n)}(a)$ where $\chi_{(n)}(a)$ is given by the
$BC\!H$-recursion \allowdisplaybreaks{
\begin{eqnarray}
    \chi_{(0)}(a) &:=& a, \nonumber\\
    \chi_{(n+1)}(a) &=& a - \BCH \big( P(\chi_{(n)}(a)),(\id_A-P)(\chi_{(n)}(a)) \big),
    \label{eq:chik}
\end{eqnarray}}
and where the limit is taken with respect to the topology given by
the filtration. Then the map $\chi: A\fil{1} \to A\fil{1}$ satisfies
\begin{equation}
    \label{BCHrecursion1}
    \chi(a) = a - \BCH\big(P(\chi(a)),\tilde{P}(\chi(a))\big).
\end{equation}
This map appeared in \cite{E-G-K2,E-G-K3,EG}, see also
\cite{EGMbch06} for more details.

We observe that for any linear map $P: A \to A$ preserving the
filtration of $A$ the (usually non-linear) map $\chi: A\fil{1} \to A\fil{1}$
is  unique and such that $(\chi - \id_A)(A\fil{i}) \subset A\fil{2i}$ for
any $i \ge 1$. Further, with $\tilde P:= \id_A - P$ we have
\begin{equation}
    \label{truc}
    \forall a \in A\fil{1},\ \ a = C \Big(P\big(\chi(a)\big),\,\tilde P\big(\chi(a)\big)\Big).
\end{equation}
This map is bijective, and its inverse is given by
\begin{equation}
    \label{chi-inverse}
    \chi^{-1}(a)=C\big(P(a),\,\tilde{P}(a)\big)
                        =a+\BCH\big(P(a),\,\tilde P(a)\big).
\end{equation}

Now follows the first theorem, which  contains the key result and
which appeared already in \cite{E-G-K2,E-G-K3}. It states a
general decomposition on $A$ implied by the map $\chi$. In fact,
it applies to associative as well as Lie algebras.

\begin{theorem} \label{thm:bch}
Let $A$ be a complete filtered associative (or Lie) algebra with a
linear, filtration preserving map $P: A \to A$. For any $a \in
A\fil{1}$, we have
        \begin{equation}
            \exp(a)=\exp\big(P(\chi(a))\big)\exp\big(\tilde{P}(\chi(a))\big).
            \label{eq:bch}
        \end{equation}
\end{theorem}
In our recent work~\cite{EGMbch06} with D.~Manchon we established
a unified approach of several apparently unrelated factorizations
arisen from quantum field theory~\cite{CK2}, vertex operator
algebras~\cite{B-H-L}, combinatorics~\cite{A-B-S} and numerical
methods in differential equations~\cite{MQZ01}.

In Theorem~\ref{thm:bch} we would like to emphasize the particular
case when the map $P$ is idempotent, $P^2=P$. Hence, let $P: A \to
A$ be such an {\it{idempotent}} linear filtration preserving map.
Let $A = A_- \oplus A_+$ be the corresponding vector space
decomposition, with $A_- := P(A)$ and $A_+ := \tilde{P}(A)$. We
define $A\fil{1,}{}_{-}:=P(A\fil{1})$ and $A\fil{1,}{}_{+}:=\tilde{P}(A\fil{1})$, and
$\chi: A\fil{1} \to A\fil{1}$ is the $BC\!H$-recursion map associated to the
map $P$. Then under these hypotheses we find that for any $\eta
\in 1_A + A\fil{1}$ there are {\it{unique}} $\eta_{-} \in
\exp\big(A_{1,-}\big)$ and $\eta_{+} \in
\exp\big(A_{1,+}\big)$ such that $\eta = \eta_-\, \eta_+$.

The factorization in Theorem \ref{thm:bch} gives rise to a simpler
recursion for the map $\chi$, without the appearance of
$\tilde{P}$. Indeed, for $A$ being a complete filtered algebra
with a filtration preserving map $P: A \to A$ the map $\chi$ in
(\ref{BCHrecursion1}) solves the following recursion for $u \in
A\fil{1}$
 \begin{equation}
   \label{BCHrecursion2}
   \chi(u):=u + \BCH\big( -P(\chi(u)),u \big).
 \end{equation}

As a particularly simple but useful remark we mention the case of
an idempotent algebra morphism on a complete filtered associative
algebra $A$, which deserves some attention especially in the
context of renormalization. In fact, such a map $R$ on $A$
satisfies the weight one Rota--Baxter relation, see
Section~\ref{sect:RBA}. Moreover, in this case the map $\chi$ in
Eq.~(\ref{BCHrecursion2}) simplifies considerably, to wit,
\begin{equation}
   \label{BCHrecur3}
   \chi(u) = u + \BCH\big( -R(u),u \big),
 \end{equation}
for any element $u \in A_1$.

The proof follows from $R(\chi(u))=R(u)$ which results from the
multiplicativity of $R$, i.e., applying $R$ to
Eq.~(\ref{BCHrecursion1}) we obtain
$$
    R\big(\chi(u)\big) = R(u) + \BCH\big(R^2(\chi(u)),(R\circ\tilde{R})(u))\big).
$$
and, since $R$ is idempotent, we have $R \circ \tilde{R}=R-R^2=0$.
Thus $R(\chi(u))=R(u)$.

With the foregoing assumptions on $R$ the factorization in
Theorem~\ref{thm:bch} reduces to
\begin{equation}
    \exp(a) = \exp\big(R(a)\big) \exp\big(\tilde{R}(a) + \BCH\big( -R(a),a \big)\big),\qquad a \in A_1.
    \label{eq:simpleFact}
\end{equation}
An example for such a map is given by the evaluation map $R_a$ on
$\mathcal{C}:=\mathbf{Cont}(\mathbb{R})$ which evaluates a
function $f \in \mathcal{C}$ at the point $a \in \mathbb{R}$,
$R_a(f):=f(a)$.


\subsection{Coalgebra and bialgebra}
\label{sssect:Hopf}

A $\mathbb{K}$-coalgebra is obtained by reversing the arrows in
the relations (\ref{assoc2}) and (\ref{unit}). Thus a
$\mathbb{K}$-{\it{coalgebra}} is a triple $(C,\Delta_{C}
,\epsilon_{C})$, where the coproduct map $\Delta_C: C \to C
\otimes C$ is coassociative, i.e. $(\Delta_C \otimes \id_C) \circ
\Delta_C=(\id_C \otimes \Delta_C) \circ \Delta_C$, and
$\epsilon_C: C \to \mathbb{K}$ is the counit map which satisfies
$(\epsilon_C \otimes \id_C) \circ \Delta_C(x) = x =(\id_C \otimes
\epsilon_C) \circ \Delta_C(x)$. We call its kernel
$\mathrm{ker}(\epsilon_C)$ the {\it{augmentation ideal}}.

For $x \in C$, we use the notation $\Delta_C(x) =
\sum_{(x)} x_{(1)} \otimes x_{(2)}$. Then the coassociativity of
the coproduct map $\Delta_C$ just means
$$
  \sum_{(x)}\bigg(\sum_{(x_{(1)})} x_{(1)(1)} \otimes x_{(1)(2)}\bigg ) \otimes x_{(2)}
 = \sum_{(x)} x_{(1)} \otimes \bigg (\sum_{(x_{(2)})} x_{(2)(1)}\otimes x_{(2)(2)} \bigg)
$$
which allows us to use another short hand notation
$$
    \Delta_C^{(2)}(x):=(\id_C \otimes \Delta_C) \circ \Delta_C(x)
        = (\Delta_C \otimes \id_C) \circ \Delta_C(x) = \sum_{(x)} x_{(1)} \otimes x_{(2)} \otimes x_{(3)}.
$$
We speak of a cocommutative coalgebra if $\tau \circ
\Delta_{C}=\Delta_{C}$, that is, the arrows in (\ref{abelian}) are
reversed.

A simple example of a coalgebra is provided by the field
$\mathbb{K}$ itself, with $ \Delta_\mathbb{K}: \mathbb{K} \to
\mathbb{K} \otimes \mathbb{K}$, $c \to c \otimes 1$, $c \in
\mathbb{K}$ and $\epsilon_\mathbb{K}:=\id_\mathbb{K}: \mathbb{K}
\to \mathbb{K}$. We also denote the multiplication in $\mathbb{K}$
by $m_\mathbb{K}$.

Let $(C,\Delta_C,\epsilon_C)$ be a $\mathbb{K}$-coalgebra. A
subspace $I \subseteq C$ is a {\it{subcoalgebra}} if $\Delta_C(I)
\subseteq I \otimes I$. A subspace $I \subseteq C$ is called a
(left-, right-) {\it{coideal}} if ($\Delta_C(I) \subseteq I
\otimes C$, $\Delta_C(I) \subseteq  C \otimes I$) $\Delta_C(I)
\subseteq I \otimes C + C \otimes I$.

An element $x$ in a coalgebra $(C,\Delta_C,\epsilon_C)$ is called
{\it{primitive}} if $\Delta_C(x) = x \otimes 1_C + 1_C \otimes x$.
We will denote the set of primitive elements in $C$ by $P(C)$.
\smallskip

A $\mathbb{K}$-{\it{bialgebra}} consists of a compatible pair of a
$\mathbb{K}$-algebra structure and a $\mathbb{K}$-coalgebra
structure. More precisely, a $\mathbb{K}$-bialgebra is a quintuple
$(B,m_B,\eta_B,\Delta_B,\epsilon_B)$, where $(B,m_B,\eta_B)$ is a
$\mathbb{K}$-algebra, and $(B,\Delta_B,\epsilon_B)$ is a
$\mathbb{K}$-coalgebra, such that $m_B$ and $\eta_B$ are morphisms
of $\mathbb{K}$-coalgebras, with the natural coalgebra structure
on $B \otimes B$. In other words, we have the commutativity of the
following diagrams.
\begin{equation}
\xymatrix{
          B \otimes B \ar[rr]^{m_B} \ar[d]_{\scriptstyle{(\id_B \otimes \tau \otimes \id_B)
          (\Delta_B \otimes \Delta_B)}}  & & B \ar[d]^{\Delta_B}\\
          B \otimes B \otimes B \otimes B \ar[rr]^{\qquad m_B \otimes m_B} && B \otimes B }
    \label{compat1}
\qquad
  \xymatrix{
            B \otimes B \ar[rr]^{\epsilon_B \:\otimes \epsilon_B} \ar[d]_{m_B} & &
            \mathbb{K} \otimes \mathbb{K} \ar[d]^{m_\mathbb{K}}\\
            B \ar[rr]^{\epsilon_B} && \mathbb{K} }
\end{equation}

\begin{equation}
 \xymatrix{
             \mathbb{K} \ar[rr]^{\eta_B} \ar[d]_{\Delta_\mathbb{K}} && B \ar[d]_{\Delta_B}\\
             \mathbb{K} \otimes \mathbb{K} \ar[rr]^{\eta_B \:\otimes \eta_B} && B \otimes B
            }
\qquad
  \xymatrix{
          \mathbb{K} \ar[rr]^{\eta_B} \ar[rd]_{\id_\mathbb{K}} && B \ar[ld]^{\epsilon_B} \\
           & \mathbb{K} & }.
\end{equation}
Here the flip map $\tau$ is defined as before. We can equivalently
require that $\Delta_B$ and $\epsilon_B$ are morphisms of
$\mathbb{K}$-algebras, with the natural algebra structure on
$B\otimes B$. One often uses a slight abuse of notation and writes
the compatibility condition as
$$
  \Delta_B(bb') = \Delta_B(b)\Delta_B(b'), \qquad b,b' \in B,
$$
saying that the {\it{coproduct of the product is the product of
the coproducts}}. The identity element in $B$ will be denoted by
$1_B$. All algebra morphisms are supposed to be unital. The set
$P(B)$ of primitive elements of a bialgebra $B$ is a Lie
subalgebra of the Lie algebra $\call(B)$.

A bialgebra $B$ is called a {\it{graded bialgebra}} if there are
$\mathbb{K}$-vector subspaces $B^{(n)}$, $n \geq 0$ of $B$ such
that
\begin{enumerate}
\item $B= \bigoplus_{n \geq 0} B\grad{n}\ ,$ \item $m_B(B\grad{n}
\otimes B\grad{m}) \subseteq B\grad{n+m},$ \item
$\Delta_B(B\grad{n}) \subseteq \bigoplus_{p+q=n} B\grad{p} \otimes
B\grad{q}.$
\end{enumerate}
Elements $x \in B\grad{n}$ are given a degree $\mathrm{deg}(x)=n$.
Moreover, $B$ is called {\it{connected}} if $B\grad{0} =
\mathbb{K}$. A graded bialgebra $B=\bigoplus_{n \geq 0} B\grad{n}$
is said to be of {\it{finite type}} if each of its homogeneous
components $B\grad{n}$ is a $\mathbb{K}$-vector space of finite
dimension.

Let $B$ be a connected graded $\mathbb{K}$-bialgebra. The key
observation for such objects is the following result describing
the coproduct for any element $x \in B$
$$
    \Delta_B(x) = x \otimes 1_B + 1_B \otimes x + \sum x'\otimes x'',
$$
where $\sum x'\otimes x''\in \mathrm{ker}(\epsilon_B)
\otimes \mathrm{ker}(\epsilon_B)$. Hence, for any $x \in
B\grad{n}$, the element
$$
    \bar{\Delta}_B(x) := \Delta_B(x) - x \otimes 1_B - 1_B \otimes x
$$
is in $\oplus_{p+q=n,\ p>0,\ q>0} B\grad{p} \otimes B\grad{q}$. The
map $\bar{\Delta}_B$ is coassociative on the augmentation
ideal, which is $\mathrm{ker}(\epsilon_B) = \bigoplus_{n>0}
B\grad{n}$ in the case of a connected graded bialgebra. Elements in
the kernel of $\bar{\Delta}_B$ are just the primitive elements
in $B$.\smallskip

As an example we mention the {\it{divided power bialgebra}}. It is
defined by the quintuple
$(\cald,m_\cald,\eta_\cald,\Delta_\cald,\epsilon_\cald)$ where
$\cald$ is the free $\mathbb{K}$-module $\bigoplus_{n=0}^\infty
\mathbb{K} d_n$ with basis $d_n$, $n \geq 0$. The multiplication
is given by $m_\cald: \cald \otimes \cald \to \cald$, $d_m \otimes
d_n \mapsto {m+n \choose m} d_{m+n}$, and the unital map
$\eta_\cald: \mathbb{K} \to \cald$, $1_{\mathbb{K}} \mapsto
d_{0}:=1_\cald$. For the coproduct we have $\Delta_\cald: \cald
\to \cald \otimes \cald$, $d_{n} \mapsto \sum_{k=0}^{n} d_{k}
\otimes d_{n-k}$, and $\epsilon_\cald: \cald \to \mathbb{K}$,
$d_{n} \mapsto \delta_{0,n}1_{\mathbb{K}}$ where $\delta_{0,n}$ is
the Kronecker delta. Other examples of similar type are the
{\it{binomial bialgebra}}, and the
{\it{shuffle}}~\cite{Swe69,Reutenauer} respectively
{\it{quasi-shuffle bialgebra}}~\cite{Hoffman05}.\smallskip


\subsection{Convolution product and Hopf algebra}

For a $\mathbb{K}$-algebra $A$ and a $\mathbb{K}$-coalgebra
$C$, we define the {\it{convolution product}} of two linear
maps $f,g$ in ${\rm{Hom}}(C,A)$ to be the linear map $f \star
g \in {\rm{Hom}}(C,A)$ given by the composition
$$
  C \xrightarrow{\Delta_C} C \otimes C \xrightarrow{f \otimes g} A \otimes A \xrightarrow{m_A} A.
$$
In other words, for $a \in C$,  we define
$$
    (f \star g)(a) = \sum_{(a)} f(a_{(1)}) \, g(a_{(2)}).
$$
For the maps $f_i \in \mathcal{A}:={\rm{Hom}}(C,A)$, $i=1,\cdots,
n$, $n>1$, we define multiple convolution products by
\begin{equation}
  \label{def:convpowers}
    \ f_1 \star \dots \star f_n :=
        m_{A} \circ (f_1 \otimes f_2 \otimes \dots \otimes f_n) \circ \Delta_C^{(n-1)},
\end{equation}
where we define inductively $\Delta_C^{(0)} := \id_C$ and,
for $n>0$, $\Delta_C^{(n)}:=(\Delta_C^{(n-1)} \otimes
\id_{C}) \circ \Delta_C$.\smallskip

Let $(\calh,m_\calh,\eta_\calh,\Delta_\calh,\epsilon_\calh)$ be a
$\mathbb{K}$-bialgebra. A $\mathbb{K}$-linear endomorphism $S$ of
$\calh$ is called an {\it{antipode}} for $\calh$ if it is the
inverse of $\id_\calh$ under the convolution product
\begin{equation}
     S \star \id_\calh = m \circ (S \otimes \id_\calh)\circ \Delta_\calh
                  = \eta_\calh \circ \epsilon_\calh
                  = m_\calh \circ (\id_\calh \otimes S)\circ \Delta_\calh = \id_\calh \star S .
    \label{eq:antipode}
\end{equation}

A {\it{Hopf algebra}} is a $\mathbb{K}$-bialgebra
$(\calh,m_\calh,\eta_\calh,\Delta_\calh,\epsilon_\calh,S)$ with an
antipode $S$, which is unique. The algebra unit in $\calh$ is
denoted by $1_\calh$.

As an example we mention the universal enveloping algebra
$\mathfrak{U}(\call)$ of a Lie algebra $\call$ which has the
structure of a Hopf algebra.

Let $(\calh,m_\calh,\eta_\calh,\Delta_\calh,\epsilon_\calh,S)$ be
a Hopf algebra. The antipode is an algebra anti-morphism and
coalgebra anti-morphism
$$
  m_\calh(S \otimes S)\circ \tau = S \circ m_\calh
  \qquad
  \Delta_\calh \circ S = \tau \circ S \otimes S \circ \Delta_\calh.
$$
If the Hopf algebra $\calh$ is commutative or cocommutative, then
$S \circ S=\id_\calh$.\\

Let $A$ be an $\mathbb{K}$-algebra, $\calh$ a Hopf algebra. By
abuse of language we call an element $\phi \in
{\rm{Hom}}(\calh,A)$ a {\it{character}} if $\phi$ is an algebra
morphism, that is, if it respects multiplication, $\phi \big(
m_\calh(x \otimes y) \big) = m_A\big(\phi(x) \otimes
\phi(y)\big)$. An element $Z \in {\rm{Hom}}(\calh,A)$ is called a
{\it{derivation}} (or {\it{infinitesimal character}}) if
\begin{equation}
 \label{def:inifiniChar}
    Z\big(m_\calh(x \otimes y)\big) = m_A \big(e_A(x)\otimes Z(y)\big)
                                     + m_A \big(Z(x)\otimes e_A(y)\big),
\end{equation}

\noindent for all $x,y \in \calh$ and with $e_A:=\eta_A \circ
\epsilon_\calh$. The set of characters (respectively derivations)
is denoted by $G_A := \mathrm{char}(\calh,A) \subset
{\rm{Hom}}(\calh,A)$ (respectively $g_A:=\partial
\mathrm{char}(\calh,A) \subset {\rm{Hom}}(\calh,A)$). We remark
that `proper' (infinitesimal) characters live in
${\rm{Hom}}(\calh,\mathbb{K})$. We note that $\phi(1_\calh)=1_A$
if $\phi$ is a character and $Z(1_\calh)=0$
for $Z$ being a derivation.\\

In \cite{Ma01} one may find the proof of the following statements
which will be important later.

\begin{prop}
Let $(A,m_A,\eta_A)$ be a unital $\mathbb{K}$-algebra.
 \begin{enumerate}
    \item \label{it:dual}
          Let $(C,\Delta_C,\epsilon_C)$ be a $\mathbb{K}$-coalgebra.
          Then the triple $({\rm{Hom}}(C,A),\star,e_A)$
          is a unital $\mathbb{K}$-algebra, with $e_A:=\eta_A \circ \epsilon_C$ as the unit.

    \item \label{it:comdual}
          Let $B = \oplus_{n\geq 0} B\grad{n}$ be a connected graded
          bialgebra. Let $\mathcal{A}:={\rm{Hom}}(B,A)$, and define
            $$
                \mathcal{A}\fil{n}=\big\{f \in {\rm{Hom}}(B,A) \big | \ f(B\grad{k})=0, k \leq n-1 \big\}
            $$
          for $n \geq 0$ with the convention that $B\grad{-1}=\emptyset$. Then
          $\mathcal{A}$ is a complete filtered unital $\mathbb{K}$-algebra.
\end{enumerate}
\label{pp:filter}
\end{prop}

\noindent Of course, we can replace the target space algebra $A$
in item (\ref{it:dual}) or (\ref{it:comdual}) by the base field
$\mathbb{K}$. And in case that $B$ is a bialgebra we have the
convolution algebra structure on
${\rm{Hom}}(B,B)$ with unit $e_B:=\eta_B \circ \epsilon_{B}$.\\

For a connected graded bialgebra, we have the well-known result
that any such $\mathbb{K}$-bialgebra $\calh$ is a Hopf algebra.
The antipode is defined by the geometric series $\id_\calh^{\star
(-1)} = \big(\eta_\calh \circ \epsilon_\calh - (\eta_\calh \circ
\epsilon_\calh - \id_\calh)\big)^{\star (-1)}$
\begin{equation}
  \label{def:antipode1}
  S(x) = \sum_{k\ge 0}(\eta_\calh \circ \epsilon_\calh - \id_\calh)^{\star k}(x),
\end{equation}
well-defined because of Proposition~\ref{pp:filter}. The proof of
this result is straightforward, see \cite{FG01} for more details.
The antipode preserves the grading, $S(\calh\grad{n}) \subseteq
\calh\grad{n}$. The projector $P := \id_\calh - \eta_\calh \circ
\epsilon_\calh$ maps $\calh$ to its augmentation ideal,
$\mathrm{ker}(\epsilon_\calh)$.

The antipode $S$ for connected graded Hopf algebras may also be
defined recursively in terms of either of the following two
formulae \allowdisplaybreaks{
\begin{eqnarray}
        \label{antipode2}
        S(x) = - S \star \id_\calh \circ P(x) &=& -x - \sum_{(x)} S(x_{(1)})x_{(2)},
        \\
        S(x) = - \id_\calh \circ P \star S(x) &=& -x - \sum_{(x)} x_{(1)}S(x_{(2)}), \nonumber
\end{eqnarray}}
for $x \in \mathrm{ker}(\epsilon_\calh)$, following readily from
(\ref{eq:antipode}) by recalling that
$\mathrm{ker}(\epsilon_\calh) = \bigoplus_{n>0} \calh\grad{n}$, and
$S(1_\calh):=1_\calh$.
The first formula will cross our way in disguised
form in later sections.

An important fact due to {{Milnor and Moore}} \cite{MilnorMoore65}
concerning the structure of cocommutative connected graded Hopf
algebras of finite type states that any such Hopf algebra $\calh$
is isomorphic to the universal enveloping algebra of its primitive
elements, $\calh \cong \mathfrak{U}(P(\calh))$, see also
\cite{FG01}.\medskip

Let $A$ be a commutative $\mathbb{K}$-algebra and $\calh$ a
connected graded Hopf algebra. One can show that the subalgebra
$\bar{g}:=\big\{f \in {\rm{Hom}}(\calh,A)\ \big|\ f(1_\calh)=0
\big\}$ in the filtered algebra
$\mathcal{A}=({\rm{Hom}}(\calh,A),\star,e_A)$, endowed with Lie
brackets defined by anti-symmetrization of the convolution product
is a Lie algebra.

Moreover, $\calg:=\big\{f \in {\rm{Hom}}(\calh,A)\ \big|\
f(1_\calh)=1_A \big\} \subset \cala$ endowed with the convolution
product forms a group in $\cala$. The inverse of $f \in \calg$ is
given by composition with the antipode of $\calh$, $f^{-1} =
f\circ S$, and we have that $\calg = e_A + \bar{g}$. As a matter
of fact we find that
\begin{enumerate}
    \item $A$-valued characters, $G_A =\mathrm{\mchar}(\calh,A)$, form a subgroup of
          $\calg = e_A + \bar{g}$ under convolution;

    \item $A$-valued derivations, $g_A = \partial \mathrm{\mchar}(\calh,A)$
          form a Lie subalgebra of $\bar{g}$;

    \item The bijection $\exp: \bar{g} \to e_A + \bar{g}$
          defined by its power series with respect to convolution restricts
          to a bijection $\exp: g_A \to G_A.$
\end{enumerate}

As $\mathcal{A}=({\rm{Hom}}(\calh,A),\star,e_A)$ is a complete
filtered associative algebra by Proposition~\ref{pp:filter}, we
may immediately apply Theorem~\ref{thm:bch} giving rise to a
factorization in the group $G_A =\mathrm{\mchar}(\calh,A)$ of
$A$-valued characters.

\begin{theorem} \label{thm:factHopf}
Let $A$ be a commutative $\mathbb{K}$-algebra and $\calh$ be a
connected graded commutative Hopf algebra. Let
$\mathcal{A}=({\rm{Hom}}(\calh,A),\star,e_A)$. Let $P: \cala
\to \cala$ be any filtration preserving linear map. Then we have
for all $\phi=\exp(Z) \in G_A$, $Z \in g_A$ the characters
$\phi^{-1}_{-}:=\exp\big(P(\chi(Z))\big)$ and
$\phi_{+}:=\exp\big(\tilde{P}(\chi(Z))\big)$ such that
\begin{equation}
    \label{eq:fact}
    \phi = \phi^{-1}_{-} \star \phi_{+}.
\end{equation}
If $P$ is idempotent this decomposition is unique.
\end{theorem}

As a simple example take the even-odd decomposition described in
\cite{A-B-S}, see also \cite{EGMbch06}. Take an arbitrary
connected graded Hopf algebra $\calh$ and let $Y$ denote the
grading operator, $Y(h)=deg(h)h=nh$ for a homogeneous element $h
\in \calh\grad{n}$. We may define an involutive automorphism on
$\mathcal{H}$, denoted by $\phantom{i}\overline{\phantom{a}}:
\mathcal{H} \to \mathcal{H}$, $\overline{h} := (-1)^{Y}h =
(-1)^{deg(h)}h$, for $h \in \mathcal{H}\grad{n}$. It induces by
duality an involution on $\Hom(\mathcal{H},\mathbb{K})$,
$\overline{\phi}(h) := \phi(\bar{h})$ for $\phi \in
\Hom(\mathcal{H},\mathbb{K})$, $h \in \mathcal{H}$. $\mathcal{H}$
naturally decomposes on the level of vector spaces into elements
of odd respectively even degree
$$
  \mathcal{H}=\mathcal{H}_{-} \oplus \mathcal{H}_{+},
$$
with projectors $\pi_{\pm}: \mathcal{H} \to \mathcal{H}_{\pm}$.
Such that for $\overline{\pi_{+}(h)}=\pi_{+}(h)=:h_+$ and
$\overline{\pi_{-}(h)} = \overline{h_-}= -h_-$, and $h = h_- + h_+
\in \mathcal{H}$. Let $\phi$ be a character in the group $G$. It
is called even if it is a fixed point of the involution,
$\overline{\phi} =\phi$, and is called odd if it is an anti-fixed
point, $\overline{\phi}=\phi^{-1}=\phi \circ S$. The set of odd
and even characters is denoted by $G_-$, $G_+$, respectively. Even
characters form a subgroup in $G$. Whereas the set of odd
characters forms a symmetric space. Theorem~\ref{thm:factHopf}
says that any $\phi \in G$ has a unique decomposition
\begin{equation}
\label{eq:bchdecomp}
    \phi = \exp(Z) = \exp\big(\pi_{-}(Z) + \pi_{+}(Z)\big)
         = \exp \big(\pi_{-}(\chi(Z))\big)
            \star
           \exp\big(\pi_{+}(\chi(Z))\big),
\end{equation}
with
$$
    \phi_-^{-1}:= \exp \big(\pi_{-}(\chi(Z))\big) \in G_-
$$
being an odd character, and
$$
    \phi_+ := \exp\big(\pi_{+}(\chi(Z))\big)\in G_+
$$
being an even character. From the factorization we derive a closed
form for the $BC\!H$-recursion~ \cite{EGMbch06}
\begin{equation*}
    \chi(Z) = Z + \BCH\Big( -\pi_{-}(Z) - \frac{1}{2} \BCH\big(Z, Z-2\pi_{-}(Z)\big) , Z
    \Big).\\
\end{equation*}

Finally, by abuse of language we shall call a connected graded
commutative bialgebra $(\calh=\bigoplus_{n \geq 0} \calh\grad{n}
,m_\calh,\eta_\calh,\Delta_\calh,\epsilon_\calh)$ of finite type a
{\it{renormalization Hopf algebra}} if it is polynomially
generated by a graded vector space $\calh_{\ell in}$ which is also
a right-coideal. This implies that the right hand side of
$\bar{\Delta}_\calh(x) \in \mathrm{ker}(\epsilon_\calh) \otimes
\mathrm{ker}(\epsilon_\calh)$ is linear for $x \in \calh_{\ell
in}$
\begin{equation}
  \label{DeltacombHopf}
  \calh_{\ell in} \xrightarrow{\Delta_\calh} \mathcal{H} \otimes \calh_{\ell in}.
\end{equation}
In the sequel we will also use the name {\it{combinatorial Hopf
algebra}} if we want to underline its combinatorial, i.e.
graphical structure.


\subsection{Hopf algebra of non-decorated non-planar rooted trees}
\label{ssect:rootedtrees}

As a key-example for such a type of Hopf algebra we mention
briefly the Hopf algebra of non-decorated non-planar rooted trees.
Connes and Kreimer introduced it in \cite{CK1}. For more details
on its relation to renormalization we refer the reader so
Section~\ref{sssect:rootedtrees}. The reader may also consult the
references~\cite{BergbauerKreimer05a,FGV01,Foissy02,Hoffman03,Holtkamp03}.\\

A rooted tree $t$ is a connected and simply-connected set of
vertices $V(t)$ and oriented edges $E(t)$ such that there is
precisely one distinguished vertex, called the root, with no
incoming edge. We draw the root on top of the tree with its
outgoing edges oriented towards the root. The empty tree is
denoted by $1_{\mathcal{T}}$
$$
 \ta1           \;\;\;\;
 \tb2           \;\;\;\;
 \tc3           \;\;\;\;
 \td31          \;\;\;\;
 \te4           \;\;\;\;
 \tf41          \;\;\;\;
 \th43          \;\;\;\;
 \thj44          \;\;\;\;
 \ti5           \;\;\;\;
 \tj51         \;\;\;\;
 \tm54         \;\;\; \cdots \;\;\; \tp56  \;\;\; \tr58 \;\;\; \cdots
$$
Let $T$ be the set of isomorphic classes of rooted trees. Let
$\calt$ be the $\mathbb{K}$-vector space generated by $T$, which
is graded by the number of vertices, denoted by $\deg(t):=|V(t)|$
with the convention that $\deg(1_\calt)=0$. Let $\calh_\calt$ be
the graded commutative polynomial algebra of finite type over
$\mathbb{K}$ generated by $\calt$,
$\calh_\calt:=\mathbb{K}[\calt]=\bigoplus_{n\geq 0}
\mathbb{K}\calh\grad{n}$. Monomials of trees are called forests.
We will define a coalgebra structure on $\calt$. The counit is
defined by
\begin{equation} \label{counit}
  \epsilon(t_1 \cdots t_n):= \begin{cases}
                             0, & \;t_1\cdots t_n \ne 1_\calt\\
                             1, & \;t_1\cdots t_n = 1_\calt.
                             \end{cases}
\end{equation}
The coproduct is defined in terms of {\it{cuts}} $c(t) \subset
E(t)$ on a tree $t \in \calt$. A {\it{primitive cut}} is the
removal of a single edge, $|c(t)|=1$, from the tree $t$. The tree
$t$ decomposes into two parts, denoted by the pruned part $P_c(t)$
and the rooted part $R_c(t)$, where the latter contains the
original root vertex. An {\it{admissible cut}} of a rooted tree
$t$ is a set of primitive cuts, $|c(t)|>1$, such that any path
from any vertex of $t$ to its root has at most one cut.

The {\it{coproduct}} is then defined as follows. Let $C_t$ be the
set of all admissible cuts of the rooted tree $t \in \calt$. We
exclude the empty cut $c^{(0)}(t)$, $P_{c^{(0)}}(t)= \emptyset$,
$R_{c^{(0)}}(t)=t$ and the full cut $c^{(1)}(t)$, $P_{c^{(1)}}(t)=
t$, $R_{c^{(1)}}(T)=\emptyset$. Also let $C_t'$ be $C_t \cup
\{c^{(0)}(t),c^{(1)}(t)\}.$ Define
 \begin{equation}
    \Delta(t) := t \otimes 1_\calt + 1_\calt \otimes t + \sum_{c_t \in C_t} P_{c}(t) \otimes R_{c}(t)
               = \sum_{c_t \in C'_t} P_{c}(t) \otimes R_{c}(t).
    \label{coprod1}
 \end{equation}
We see easily, that $\deg(t) = \deg(P_{c}(t)) + \deg(R_{c}(t))$, for
all admissible cuts $c_t \in C_t$, and therefore
$$
  \sum_{c_t \in C_t} P_{c}(t) \otimes R_{c}(t) \in
                    \sum_{\substack{p+q= \deg(t),\ p,\,q>0}} \calh\grad{p} \otimes \calh\grad{q}.
$$
Furthermore, this map is extended by definition to an algebra
morphism on $\calh_\calt$
$$
    \Delta\big(\prod_{i=1}^{n}t_i\big):=\prod_{i=1}^{n}\Delta(t_i).
$$
We see here a very concrete instance of the coproduct map of a
Hopf algebra. The best way to get use to this particular coproduct
is to present some examples,
 \allowdisplaybreaks{
\begin{eqnarray}
 \Delta(\ta1) &=&\;\: \ta1  \otimes 1_\mathcal{T} + 1_\mathcal{T} \otimes \ta1 \notag \\
 \Delta\big(
 \!\!\!\begin{array}{c}
                 \\[-0.5cm]\tb2 \\
               \end{array}\!\! \big) &=&
                         \begin{array}{c}
                           \\[-0.5cm]\tb2 \\
                         \end{array} \!\! \otimes 1_\mathcal{T}
                            + 1_\mathcal{T} \otimes \!\!\!
                         \begin{array}{c}
                           \\[-0.5cm]\tb2 \\
                         \end{array}
                               + \ta1 \otimes \ta1 \label{lad2}\\
 \Delta(\ta1\ta1) &=& \Delta(\ta1)\Delta(\ta1) = \;\: \big(\ta1  \otimes 1_\mathcal{T}
                                                                + 1_\mathcal{T} \otimes \ta1\big)
                          \big(\ta1  \otimes 1_\mathcal{T} + 1_\mathcal{T} \otimes \ta1\big) \notag \\
                  &=& \;\: \ta1 \ta1 \otimes 1_\mathcal{T} + 1_\mathcal{T} \otimes \ta1\ta1
                                                                + 2\ta1 \otimes \ta1 \\
 \Delta\Big(\!\!\begin{array}{c}
                   \\[-0.5cm]\tc3 \\
                 \end{array}\!\!\Big)    &=&
                            \begin{array}{c}
                             \\[-0.5cm]\tc3 \\
                            \end{array}\!\! \otimes 1_\mathcal{T}
                         + 1_\mathcal{T} \otimes \!\!\!
                         \begin{array}{c}
                           \\[-0.5cm]\tc3 \\
                         \end{array}\!\!
                          + \ta1 \otimes\!\!\!
                         \begin{array}{c}
                           \\[-0.5cm]\tb2 \\
                         \end{array} +
                         \begin{array}{c}
                           \\[-0.5cm]\tb2 \\
                         \end{array}\!\! \otimes\! \ta1 \label{lad3} \\
 \Delta\big(\td31\big) &=& \!\! \begin{array}{c}
                                 \\[-0.5cm]\td31 \\
                              \end{array}\!\! \otimes 1_\mathcal{T}
                               + 1_\mathcal{T} \otimes\!\!\!
                              \begin{array}{c}
                                 \\[-0.5cm]\td31 \\
                              \end{array}\!\! +
                                  2\ta1 \otimes \!\!\!
                                       \begin{array}{c}
                                         \\[-0.5cm]\tb2 \\
                                       \end{array} + \ta1\ta1\otimes \ta1 \label{ex:copTree} \\
 \Delta\big(\ta1\tb2\big) &=& \Delta\big(\tb2\ta1\big) = \Delta(\ta1)\Delta\big(\tb2\big) \\
                  &=& \;\: \big(\ta1 \otimes 1_\mathcal{T} + 1_\mathcal{T} \otimes\ta1\big)
                    \Big(\begin{array}{c}
                           \\[-0.5cm]\tb2 \\
                         \end{array} \!\! \otimes 1_\mathcal{T}
                            + 1_\mathcal{T} \otimes \!\!\!
                         \begin{array}{c}
                           \\[-0.5cm]\tb2 \\
                         \end{array}
                               + \ta1 \otimes \ta1\Big) \nonumber\\
                  &=& \;\: \ta1 \tb2 \otimes 1_\mathcal{T} + 1_\mathcal{T} \otimes \ta1\tb2
                                                            + \ta1 \ta1 \otimes \ta1
                            + \ta1 \otimes \ta1\ta1 +  \tb2 \otimes \ta1 +  \ta1 \otimes \tb2\\
 \Delta\bigg(\!\! \begin{array}{c}
                   \\[-0.5cm]\te4 \\
                  \end{array}\!\!\bigg)
                     &=& \begin{array}{c}
                           \\[-0.5cm]\te4 \\
                         \end{array} \!\! \otimes 1_\mathcal{T}
                         + 1_\mathcal{T} \otimes \!\!\!
                         \begin{array}{c}
                           \\[-0.5cm]\te4 \\
                         \end{array}
                         + \ta1 \otimes \!\!\!
                         \begin{array}{c}
                           \\[-0.5cm]\tc3 \\
                         \end{array}
                            + \begin{array}{c}
                        \\[-0.5cm]\tb2 \\
                         \end{array} \!\! \otimes \!\!\!
                         \begin{array}{c}
                        \\[-0.5cm]\tb2 \\
                         \end{array} +
                         \begin{array}{c}
                        \\[-0.5cm]\tc3 \\
                         \end{array}\!\! \otimes \! \ta1 \label{lad4}
\end{eqnarray}}
Connes and Kreimer showed that $(\calh_{\calt},
m,\eta,\Delta,\epsilon)$ with coproduct $\Delta: \calh_{\calt} \to
\calh_{\calt} \otimes \calh_{\calt}$ defined by (\ref{coprod1}),
and counit $\epsilon: \calh_{\calt} \to \mathbb{K}$ (\ref{counit})
is a $\mathbb{Z}_{\geq 0}$ connected graded commutative, but
non-cocommutative bialgebra of finite type. $\calh_\calt$ is
connected since $\calh\grad{0}\simeq \mathbb{K}$, and hence a Hopf
algebra with antipode $S$. See Eqs.~(\ref{def:antipode1}),
(\ref{antipode2}), defined recursively by $S(1_\calt)=1_\calt$ and
$$
    S(t):=- t  - \sum_{c_t \in C_t} S(P_{c}(t))R_{c}(t).
$$
Again, a couple of examples might be helpful here.
{\allowdisplaybreaks{
\begin{eqnarray*}
           S(\ta1) &=& - \ta1 \nonumber\\
           S(\tb2) &=& - \tb2 - S(\ta1)\ta1
                    =  - \tb2 + \ta1 \ta1 \nonumber\\
  S\big(\td31\big) &=& - \;\td31 - 2S(\ta1) \tb2 -S(\ta1\ta1)\ta1
                    =  - \td31 + 2 \ta1 \tb2  - \ta1 \ta1 \ta1 \\
  S\Big(\!\!\! \begin{array}{c}
                   \\[-0.5cm]\th43  \\
                  \end{array}\!\!\Big)  &=& - \th43 - 3 S(\ta1)\td31 - 3 S(\ta1\ta1)\tb2 - S(\ta1\ta1\ta1)\ta1
                                         =  - \th43 + 3 \ta1 \td31   - 3 \ta1 \ta1 \tb2  + \ta1 \ta1 \ta1 \ta1 \\
  S\bigg(\!\!\! \begin{array}{c}
                   \\[-0.5cm]\thj44  \\
                  \end{array}\!\!\bigg)
                     &=&
                       - \thj44
                       - 2\: S(\ta1)\tc3
                       - S(\ta1\ta1)\tb2
                       - S\big(\td31 \big) \ta1
                      = - \thj44
                        + 2\ta1 \tc3
                        - 3\ta1 \ta1 \tb2
                        +  \ta1 \td31
                        +  \ta1 \ta1 \ta1 \ta1
\end{eqnarray*}}}

\begin{remark}{\rm{
\begin{enumerate}
    \item The coproduct (\ref{coprod1}) can be written in a recursive
way, using the $B_+$ operator, which is a closed but not exact
Hochschild 1-cocycle \cite{KreimerChen,FGV01,BergbauerKreimer05a},
hence
\begin{equation}
    \Delta(t)=\Delta(B_+(t_{i_1} \cdots t_{i_n}))= t \otimes 1_{\calt}
                                            + \{\id_\calt \otimes B_+ \}\Delta(t_{i_1} \cdots t_{i_n}).
    \label{jcoprod2}
\end{equation}
$B_+: \calh_{\calt} \to \calh_{\calt}$, in fact $B_+:
\calh_{\calt} \to \calt$, is a linear operator, mapping a (forest
of) rooted tree(s) to a rooted tree, by putting a new root on top
of the (forrest) tree and connecting the old root(s) to this new
adjoined root. A couple of examples tell everything
$$
  B_+(1_\calt)=\ta1\ ,\; B_+(\ta1)=\tb2\ ,\; B_+(\ta1 \ta1)= \td31\ ,
                \; B_+(\ta1 \tb2 \ta1)=\!\!\!\!\tr58 \; \cdots
$$
It therefore raises the degree by $1$, $\deg\big(B_+(t_{i_1} \cdots
t_{i_n})\big) = \deg\big(t_{i_1} \cdots t_{i_n}\big) + 1$. The
inductive proof of coassociativity of the coproduct
(\ref{coprod1}) formulates easily in terms this map. Every rooted
tree lies in the image of the $B_+$ operator. The notion of
subtrees becomes evident from this fact. The conceptual importance
of the $B_+$ map with respect to fundamental notions of physics
was further elaborated in recent work
~\cite{BergbauerKreimer05b,KreimerLesHouches,KreimerDS}.

    \item It is important to notice that the right hand side of
$\Delta(t) \in \mathcal{H}_{\mathcal{T}} \otimes
\mathcal{H}_{\mathcal{T}}$ is linear for $t \in \calt$. Therefore
we may write
\begin{equation}
  \label{linMap}
  \calt \xrightarrow{\Delta} \mathcal{H}_{\mathcal{T}} \otimes \mathcal{T}.
\end{equation}
This is of course not true for the coproduct of proper forests of
rooted trees, $t = t_1\cdots t_n$, $n>1$.
\end{enumerate}}}
\end{remark}

\vspace{0.5cm}


\subsection{Hopf and Lie algebra of Feynman graphs}
\label{ssect:BirkhoffCK}

In the work of Kreimer \cite{KreimerHopf}, and Connes and Kreimer
\cite{CK2,CK3} Feynman graphs as the main building blocks of
perturbative \textsf{QFT} are organized into a Hopf algebra. In
fact it is this Hopf algebra of Feynman graphs that one calls the
renormalization Hopf algebra corresponding to the \textsf{pQFT}.
Theorem~\ref{thm:factHopf} establishes the Birkhoff decomposition
upon replacing the linear filtration preserving map $P$ by the
regularization prescription and the corresponding renormalization
scheme map $R$.\\


\subsubsection{Hopf algebra of Feynman graphs}

We first give a more detailed description of a {\it{connected Feynman graph}} $\Gamma$.
It is a collection of several types of internal and external
lines, and vertices. We denote by $\Gamma^{[0]}$ its set of
vertices and by $\Gamma^{[1]}:=\Gamma^{[1]}_{\rm int} \cup
\Gamma^{[1]}_{\rm ext}$ its set of possibly oriented internal and
external edges. Internal lines are also called propagators,
whereas external ones are called legs. For instance, for
\textsf{QED} we have just two types of edges,
$\begin{array}{c}\\[-.6cm]\!\! \scalebox{0.65}{\FERMprop} \end{array}\!$
and \scalebox{0.65}{\BOSONprop}${}$, together with one type of
vertex, $\begin{array}{c}\\[-.4cm]
\!\scalebox{0.4}{\QEDvertex}\end{array}\!\!$.
A proper subgraph of a Feynman graph is determined by proper
subsets of the set of internal edges and vertices.

Of vital importance is the class of so-called {\it{one-particle
irreducible}} (1PI) Feynman graphs, which consists of connected
graphs that cannot be made disconnected by removing any of its
internal edges. In general, a 1PI Feynman graph $\Gamma$ is
parameterized by attaching to the set of external edges
$\Gamma^{[1]}_{\rm ext}$ the quantum numbers, such as masses,
momenta, and spin, corresponding to the particles that enter
respectively exit the scattering process described by the graph.
In fact, the set of such quantum numbers specifies in a precise
manner the external leg structure, denoted by $\underline{r}$, of
a 1PI graph $\Gamma=\Gamma^{\underline{r}}$, that is, the physical
process to which that graph respectively its amplitude
contributes. We denote the set of all external leg structures
$\underline{r}$ by $\mathfrak{R}$ and observe that for a
renormalizable \textsf{QFT} it consists of those edges and
vertices corresponding to the monomials in the defining
Lagrangian.

\begin{figure}\centering
\includegraphics[scale=0.8]{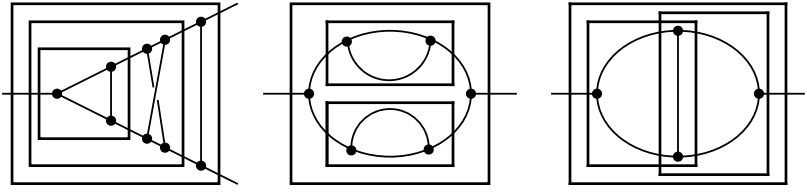}
\caption{{\small{{\cite{KreimerBuch} There are various ways
subdivergences can appear. They can be nested as in (a), disjoint
as in (b) or overlapping as in (c). We put every subdivergence in
a box. These boxes are called forests.}}}}{\label{fig:graphs}}
\end{figure}

Beyond one-loop order, the process of the renormalization of a 1PI
graph is characterized essentially by the appearance of its
ultraviolet (UV) divergent 1PI Feynman subgraphs $\gamma_i \subset
\Gamma$. For instance, two proper Feynman subgraphs $\gamma_1,\
\gamma_2 \subset \Gamma$ might be strictly {\it{nested}},
$\gamma_1 \subset \gamma_2 \subset \Gamma$, or {\it{disjoint}},
$\gamma_1 \cap \gamma_2 =\emptyset$. This hierarchy in which
subgraphs are either located inside another subgraph or appear to
be disjoint is best represented by a decorated rooted tree, see
Section~\ref{sssect:rootedtrees}. However, there is a third
possibility, consisting of subgraphs which might be
{\it{overlapping}}, see Figure~\ref{fig:graphs}. In fact, such
Feynman graphs are represented by linear combinations of decorated
rooted trees. For a detailed treatment of this important case of
overlapping graph structures in \textsf{pQFT} we refer
to~\cite{KreimerOver}.\smallskip

We denote by $F$ the set of all equivalence classes of
one-particle irreducible Feynman graphs. Let $\calf:=\mathbb{K}F$
be the $\mathbb{K}$-vector space with basis $F$. Let $\calh_\calf$
be the commutative polynomial algebra $\mathbb{K}[\calf]$ with the
product denoted by the disjoint union and with the empty graph
$1_\calf$ as the unit. Those 1PI Feynman graphs come with a
grading by the number of loops. This gives rise to a grading on
$\calh_\calf$ by $\deg(\prod_{i=1}^k\Gamma_i)=\sum_{i=1}^k
\deg(\Gamma_i)$, making it into a connected graded commutative
algebra ${\calh}_{\calf} = \oplus_{n=0}^\infty \calh\grad{n}$ with
$\calh\grad{n}$ the subspace spanned by degree $n$ graphs. We have
$\calh\grad{0} \simeq \mathbb{K}$.

Define a counit $\epsilon$ on ${\calh}_{\calf}$ by
$\epsilon(1_\calf)=1_\mathbb{K}$ and zero else. The key map is the
coproduct $\Delta:{\calh}_{\calf} \to {\calh}_{\calf}  \otimes
{\calh}_{\calf}$ defined on 1PI Feynman graphs $\Gamma \in F$ by
\begin{equation}
    \label{def:coprod}
    \Delta(\Gamma) = \Gamma \otimes 1_\calf  + 1_\calf \otimes \Gamma +
                            \sum_{\gamma{\subset}\Gamma} \gamma \otimes \Gamma/\gamma,
\end{equation}
where by abuse of notation the sum is over all unions of disjoint
divergent proper 1PI subgraphs $\gamma:=\{\gamma_1 \cdots
\gamma_n\}$, $\gamma_i \cap \gamma_j=\emptyset$, $i \neq j$, of
$\Gamma$, and $\Gamma/\gamma$ denotes the cograph which follows
from the contraction of $\gamma$ in $\Gamma$.

\begin{remark}\label{rmk:forestcoprod}{\rm{ In fact, one may define the coproduct of a graph
$\Gamma$ directly by the use of its spinneys in the wood
$\bar{\mathfrak{W}}(\Gamma)$
\begin{equation}
    \label{eq:coproductWOOD}
    \Delta(\Gamma) = \sum_{\{\gamma'\} \in
    \bar{\mathfrak{W}}(\Gamma)} \gamma' \otimes \Gamma/\{\gamma'\}.
\end{equation}
Where the first two terms $\Gamma \otimes 1_\calf$ and $1_\calf
\otimes \Gamma$ follow from the spinneys $\{\Gamma\}$ and
$\{\emptyset\}=:1_\calf$, respectively. Hence, the coproduct can
be seen as a tensor list, systematically storing term by term the
spinneys and associated cographs for each Feynman graph.}}
\end{remark}

We extend this definition to products of graphs (forests) in
$\calh_\calf$, $\Delta\big(\prod_{i=1}^k \Gamma_i
\big)=\prod_{i=1}^{k}\Delta(\Gamma_i)$, so that we get a connected
graded commutative non-cocommutative bialgebra, hence a Hopf
algebra of that type with antipode map $S$, as defined in
Eq.~(\ref{def:antipode1}) . This result was first established by
Connes and Kreimer in \cite{CK2}. By the definition of the
coproduct in (\ref{def:coprod}) we may call $\calh_{\calf}$ a
{\it{renormalization Hopf algebra of Feynman graphs}}, since for
$\Gamma \in F$ the cograph $\Gamma/\gamma$ in (\ref{def:coprod})
always lies in $F$, such that
\begin{equation}
    \label{def:coprod2}
    \calf \xrightarrow{\Delta} \calh_{\calf} \otimes \calf.
\end{equation}
As an example we calculate the coproduct of the graph
$\begin{array}{c}\\[-0.4cm]\!\scalebox{0.6}{\coprod2}\!\!\end{array}$, borrowed
from $\phi^3_6$ theory, which is the same as the graph in case (c)
in Figure~\ref{fig:graphs}. We hope that the way we draw this
two-loop Feynman self-energy graph now helps to identify its two
one-loop vertex subgraphs,
$\begin{array}{c} \\[-0.4cm] \!\scalebox{0.6}{\l2cop}\!\!
\end{array}$ and
$\begin{array}{c} \\[-0.4cm] \!\scalebox{0.6}{\r2cop}\!\! \end{array}$.
\allowdisplaybreaks{
\begin{eqnarray*}
    \Delta\left(\!\!\!\begin{array}{c}\scalebox{0.6}{\coprod2} \\ \end{array}\!\!\! \right)
    \!\!\!&=&\!\!\!\!
    \begin{array}{c}\scalebox{0.6}{\coprod2} \\ \end{array}\!\! \otimes 1_{\mathcal{F}}
    +
    1_{\mathcal{F}} \otimes \!\!\begin{array}{c}\scalebox{0.6}{\coprod2} \\ \end{array}
    \!+\!
    \begin{array}{c}\scalebox{0.6}{\l2cop} \\ \end{array}\!\! \otimes
    \!\!\begin{array}{c}\scalebox{0.6}{\c2cop} \\ \end{array}
    \!+\!
    \begin{array}{c}\scalebox{0.6}{\r2cop}\\ \end{array}\!\! \otimes
    \!\!\begin{array}{c}\scalebox{0.6}{\c2cop} \\ \end{array}.
\end{eqnarray*}}
Compare this with the wood containing the spinneys,
$\bar{\mathfrak{W}}\big(\!\! \begin{array}{c} \\[-0.35cm] \scalebox{0.6}{\coprod2}
\end{array}\!\!\big)$, which is just given by Eq.(\ref{graphs:forest1}) enlarged by the
empty graph $\{\emptyset\}$ and the graph itself
$\{\!\!\begin{array}{c}\\[-0.35cm] \scalebox{0.6}{\coprod2}
\end{array}\!\}$.

Using Eq.~(\ref{antipode2}) the antipode of this graph is given by
 \allowdisplaybreaks{
\begin{eqnarray*}
    S\left(\!\!\!\begin{array}{c}\scalebox{0.6}{\coprod2} \\\end{array}\!\!\! \right)
    \!\!\!&=&\!\!\!\!
    -\begin{array}{c}\scalebox{0.6}{\coprod2} \\ \end{array}
    -S\left(\!\!\!\begin{array}{c}\scalebox{0.6}{\l2cop} \\ \end{array}\!\!\right)\!\!
    \begin{array}{c}\scalebox{0.6}{\c2cop} \\ \end{array}
    -S\left(\!\!\begin{array}{c}\scalebox{0.6}{\r2cop}\\ \end{array}\!\!\right)\!\!
    \begin{array}{c}\scalebox{0.6}{\c2cop} \\ \end{array}\\[-0.3cm]
\end{eqnarray*}}
\noindent We must compare this with Bogoliubov's classical formula
for the counterterm $C(\Gamma)$ of a Feynman graph $\Gamma$, see
Eq.~(\ref{eq:BogoCounterTerm}) of
Section~\ref{sect:renormalization}.
 \allowdisplaybreaks{
\begin{eqnarray*}
    C\left(\!\!\!\begin{array}{c}\scalebox{0.6}{\coprod2} \\ \end{array}\!\!\! \right)
    \!\!\!&=&\!\!\!\!
    -R\left(\!\!\!\begin{array}{c}\scalebox{0.6}{\coprod2} \\ \end{array}
    +C\left(\!\!\!\begin{array}{c}\scalebox{0.6}{\l2cop} \\ \end{array}\!\!\right)\!\!
    \begin{array}{c}\scalebox{0.8}{\c2cop} \\ \end{array}
    +C\left(\!\!\begin{array}{c}\scalebox{0.6}{\r2cop}\\ \end{array}\!\! \right) \!\!
    \begin{array}{c}\scalebox{0.6}{\c2cop} \\ \end{array}\!\!\right)\\[-0.3cm]
\end{eqnarray*}}
Here, of course, each graph represents its corresponding
--regularized-- Feynman amplitude, and $R$ denotes the
renormalization scheme map.\\


\subsubsection{Lie and pre-Lie algebra of Feynman graphs}
\label{sssect:preLieFeynman}

The combinatorial Hopf algebra of Feynman graphs has a dual
counter part. To see this let us go back to the coproduct and
antipode respectively the counterterm expression in the above
example. In fact, we readily observe that the Connes-Kreimer
coproduct just {\it{disentangles}} a graph analogously to
Bogoliubov's $\bar{\mathrm{R}}$-map which in turn gives the
counterterm map $C$.

One might therefore think about an opposite operation in terms of
gluing graphs into another graph. Let us define such a bilinear
{\it{gluing operation}} on the vector space $\calf$ of 1PI Feynman
graphs
 \begin{equation}
    \Gamma_1 * \Gamma_2 = \sum_{\Gamma \in F} n(\Gamma_1,\Gamma_2;\Gamma)\Gamma.
 \end{equation}
The sum on the right hand side runs over all 1PI graphs $\Gamma$
in $F$. The number $n(\Gamma_1,\Gamma_2;\Gamma) \in \mathbb{N}$
for $\Gamma,\Gamma_1,\Gamma_2 \in F$ is a section coefficient
which counts the number of ways a subgraph $\Gamma_2$ in $\Gamma$
can be reduced to a point such that $\Gamma_1$ is obtained. The
above sum is finite as long as $\Gamma_1$ and $\Gamma_2$ are
finite graphs. The graphs which contribute to this sum necessarily
fulfill $\mathrm{deg}(\Gamma) = \mathrm{deg}(\Gamma_1) +
\mathrm{deg}(\Gamma_2)$ and ${\bf res}(\Gamma) = {\bf
res}(\Gamma_1)$, where we let ${\bf res}(\Gamma)$, called the
{\it{residue of $\Gamma$}}, be the graph obtained when all
internal edges are shrunk to a point. The graph we obtain in this
manner consists of a single vertex with ${\bf
res}(\Gamma)^{[0]}_{\mathrm{ext}} = \Gamma_{\mathrm{ext}}^{[0]}$.
In case the initial graph was a self-energy graph, we regard its
residue as a single edge.

As a matter of fact it was shown in several references
\cite{ChapotonLivernet01,CK1,CK4} that the operation $\ast$ on
$\calf$ defines a (right) {\it{pre-Lie}} product, satisfying the
{\it{pre-Lie relation}}
    \begin{equation}
        (\Gamma_1\ast\Gamma_2)\ast \Gamma_3  -  \Gamma_1\ast(\Gamma_2\ast \Gamma_3) =
        (\Gamma_1\ast \Gamma_3)\ast \Gamma_2  - \Gamma_1\ast(\Gamma_3\ast \Gamma_2)
        \label{pre-Lie}
    \end{equation}
and hence a pre-Lie algebra $\calp:=\calp(\calf)$. This pre-Lie
algebra structure on Feynman graphs was further explored by
Mencattini and Kreimer~\cite{KM1,KM2}. See also \cite{EMK03} for
more details. The pre-Lie relation is sufficient for the
anti-symmetrization of this product to fulfil the Jacobi identity.
Hence, we get a graded Lie algebra ${\call}(\calp):=\bigoplus_{n
\geq 0} \call\grad{n}$ of finite type, with commutator bracket
 \begin{equation}
    [\Gamma_1,\ \Gamma_2] := \Gamma_1 \ast \Gamma_2 - \Gamma_2 \ast \Gamma_1,
    \label{Lie}
 \end{equation}
for $\Gamma_{1}, \Gamma_2 \in \calf$. The above pre-Lie respectively Lie
algebra structures are available once one has decided on the set
of 1PI graphs $F$ of interest. The Milnor--Moore theorem
\cite{MilnorMoore65} then implies the structure of a corresponding
Hopf algebra of Feynman graphs and vice versa.

So far we have two complementary operations on Feynman graphs
forming either a (pre-)Lie or a Hopf algebra. In the former case
we have a composition of Feynman graphs in terms of a pre-Lie
gluing product, where we replace vertices by Feynman graphs with
compatible external leg structure. Dually, we have the
decomposition of graphs in terms of a Hopf algebra coproduct,
i.e., replacing non-trivial UV divergent 1PI subgraphs by their residues.\\


\subsubsection{Correspondence between non-planar rooted trees and Feynman graphs}
\label{sssect:rootedtrees}

Connes and Kreimer developed in \cite{CK1} a detailed picture of a
connected graded commutative Hopf algebra structure of finite type
on non-planar decorated rooted trees giving rise to an interesting
interplay between perturbative renormalization and non-commutative
geometry. This Hopf algebra of rooted trees serves as the role
model for combinatorial Hopf algebras of the above type due to its
universal property. In Section~\ref{ssect:rootedtrees} we gave a
brief account of its main properties. Here we would like to
indicate its link to Feynman graphs and renormalization.

In the renormalization problem we associated to each graph
$\Gamma$ its wood, i.e. the set of its spinneys, denoted by
$\mathfrak{W}(\Gamma)$. We have seen that the coproduct is a
storing list of those spinneys and the corresponding cographs.
Indeed, the only information needed in this disentanglement
problem of a graph is the hierarchical structure in which the
spinneys appear to sit inside the graph. Each spinney of a graph
consists of a union of disjoint UV divergent 1PI subgraphs $\gamma
\subsetneq \Gamma$ which themselves may carry a nontrivial spinney
structure, e.g. \scalebox{0.8}{\kite}. Its set of proper spinneys
was given in (\ref{graphs:forest2})
$$
    \mathfrak{W}\Big(\kite\Big) = \big\{ \{\newfish\}, \{\winecup\}\big\},
$$
where \winecup is a maximal spinney. We observe that this UV
divergent 1PI subgraph \winecup\ contains itself one --maximal--
(sub)spinney
$$
    \mathfrak{W}(\winecup) = \big\{ \{\newfish\} \big\}.
$$
For the graph $\begin{array}{c} \\[-0.4cm] \scalebox{0.8}{\roll}
\end{array}$ we found
$$
  \mathfrak{W}\Big(\!\!\begin{array}{c}
                     \\[-0.5cm] \scalebox{1}{\roll}
                   \end{array}\!\!\Big) = \big\{\{\newfish\},\{\newfish\}, \{\newfish\ \newfish\}\big\}.
$$
Each of its 1PI subgraphs is primitive, and we have one maximal
spinney which is the union $\{\newfish\ \newfish\}$. These
hierarchies are best represented by rooted trees with vertices
decorated by the primitive UV divergent 1PI $1$-loop graph
\newfish\
$$
    \begin{array}{cc}  \scalebox{0.6}{\ladtr1} & \\[-1.8cm]
                               & \sim \scalebox{2}{\kite}\\
    \end{array} \;\; \qquad \;\;
    \begin{array}{cc} \scalebox{0.6}{\chertr1} &\\[-1.4cm]
                               & \hspace{-0.3cm} \sim \scalebox{1.5}{\roll}\\
    \end{array}\\[1cm]
$$
The orientation of the graphs is of no importance at this level.
The branching of the tree on the right hand side reflects the
disjoint location of the subgraphs in $\begin{array}{c} \\[-0.4cm] \scalebox{0.8}{\roll}
\end{array}$, i.e. the spinney $\{\newfish\ \newfish\}$. Of course, different
decorations may appear naturally in more complicated situations,
like for instance in \textsf{QED}.

The property that both graphs have exactly one maximal spinney,
that is, they contain no overlapping subgraphs, is reflected by
the fact that their subdivergences hierarchy can be properly
represented by a single decorated rooted tree. For graphs with
overlapping subgraphs we have in general several maximal spinneys.
For instance, recall the wood, i.e. the set of proper spinneys
corresponding to the $3$-loop \textsf{QED} graph
$\begin{array}{c} \\[-0.4cm] \scalebox{0.5}{\overQED} \end{array}$
$$
    \mathfrak{W}\left(\!\!\!\!\begin{array}{c} \\[-0.4cm] \scalebox{0.55}{\overQED} \end{array} \!\!\!\!\right) =
     \left\{ \Big\{\!\!\begin{array}{c} \\[-0.4cm] \scalebox{0.55}{\LoverQED} \end{array} \!\! \Big\},
             \Big\{\!\!\begin{array}{c} \\[-0.4cm] \scalebox{0.55}{\RoverQED} \end{array} \!\! \Big\},
             \Big\{\!\!\begin{array}{c} \\[-0.4cm] \scalebox{0.55}{\LLoverQED} \end{array}\!\! \Big\},
             \Big\{\!\!\begin{array}{c} \\[-0.4cm] \scalebox{0.55}{\RRoverQED} \end{array} \!\!\Big\},
             \Big\{\!\!\begin{array}{c} \\[-0.4cm] \scalebox{0.55}{\RoverQED} \end{array}\!
                   \begin{array}{c} \\[-0.4cm] \scalebox{0.55}{\LoverQED} \end{array}\!\!\Big\}
             \right\}
$$
with the following cographs
$$
    \begin{array}{c} \\[-0.3cm] \scalebox{0.55}{\overQED} \end{array}
    \big{/}
    \Big\{\!\!\begin{array}{c} \\[-0.4cm] \scalebox{0.55}{\LoverQED} \end{array}\!\!\Big\}
    =\!\!
    \begin{array}{c} \\[-0.4cm] \scalebox{0.55}{\over1QED} \end{array}
    =\!\!
    \begin{array}{c} \\[-0.4cm] \scalebox{0.55}{\overQED} \end{array}
    \big{/}\
    \Big\{\!\!\begin{array}{c} \\[-0.4cm] \scalebox{0.55}{\RoverQED} \end{array}\!\!\Big\}
$$
and \allowdisplaybreaks{
\begin{eqnarray*}
    \begin{array}{c} \\[-0.4cm] \scalebox{0.55}{\overQED} \end{array}
    \big{/}\
    \Big\{\!\!\begin{array}{c} \\[-0.4cm] \scalebox{0.55}{\LLoverQED}\end{array}\!\!\Big\}
   &=&
    \begin{array}{c} \\[-0.4cm] \scalebox{0.55}{\overQED} \end{array}
    \big{/}\
    \Big\{\!\!\begin{array}{c} \\[-0.4cm] \scalebox{0.55}{\RRoverQED}\end{array}\!\!\Big\}\\
   &=&
    \begin{array}{c} \\[-0.4cm] \scalebox{0.55}{\overQED} \end{array}
    \big{/}\
    \Big\{\!\!\!\!\begin{array}{c} \\[-0.4cm] \begin{array}{c} \\[-0.4cm] \scalebox{0.55}{\RoverQED} \end{array}\!
                   \begin{array}{c} \\[-0.4cm] \scalebox{0.55}{\LoverQED} \end{array}
                   \end{array}\!\!\!\!\Big\}
    =
    \begin{array}{c} \\[-0.4cm] \scalebox{0.55}{\loopQED} \end{array}.
\end{eqnarray*}
From the last line we see immediately that we have three maximal
spinneys in  $\begin{array}{c} \\[-0.4cm] \scalebox{0.5}{\overQED}
\end{array}$, $\mathfrak{W}_{max}\big(\!\!\begin{array}{c} \\[-0.4cm] \scalebox{0.5}{\overQED}
\end{array}\!\!\big):=\left\{
             \Big\{\!\!\begin{array}{c} \\[-0.4cm] \scalebox{0.5}{\LLoverQED} \end{array}\!\! \Big\},
             \Big\{\!\!\begin{array}{c} \\[-0.4cm] \scalebox{0.5}{\RRoverQED} \end{array}\!\!\Big\},
             \Big\{\!\!\begin{array}{c} \\[-0.4cm] \scalebox{0.5}{\RoverQED} \end{array}\!
                   \begin{array}{c} \\[-0.4cm] \scalebox{0.5}{\LoverQED} \end{array}\!\!\Big\}
             \right\}$, all resulting in the primitive $1$-loop self energy
\textsf{QED} graph after contraction. The corresponding tree
representation consists of a linear combination of three rooted
trees each of degree three
$$
  \begin{array}{c} \\[-0.4cm] \scalebox{0.55}{\overQED} \end{array}
  \sim
  2\begin{array}{c} \\[-0.4cm] \cb1 \end{array}+  \begin{array}{c} \\[-0.4cm] \ab1\end{array}
$$
For the sake of completeness we should mention that
the \textsf{QED} $2$-loop self-energy graph $\begin{array}{c} \\[-0.4cm] \scalebox{0.5}{\over1QED}
\end{array}$ has two overlapping divergent $1$-loop vertex subgraphs. Its union
of maximal spinneys is given by
$\mathfrak{W}_{max}\big(\!\!\begin{array}{c} \\[-0.4cm] \scalebox{0.5}{\over1QED} \end{array}\!\!\big):=
             \left\{
             \Big\{\!\!\begin{array}{c} \\[-0.4cm] \scalebox{0.5}{\LoverQED} \end{array}\!\! \Big\},
             \Big\{\!\!\begin{array}{c} \\[-0.4cm] \scalebox{0.5}{\RoverQED} \end{array}\!\!\Big\}
             \right\}$, such that all contractions result in the primitive $1$-loop self energy
\textsf{QED} graph and
$$
  \begin{array}{c} \\[-0.4cm] \scalebox{0.55}{\over1QED} \end{array}
  \sim
  2\begin{array}{c} \\[-0.4cm] \bb1 \end{array}.
$$

The rule may be summarized as follows. To each maximal spinney
$\{\gamma'\}\in\mathfrak{W}_{max}(\Gamma)$ of a graph $\Gamma$
corresponds a decorated rooted tree $t(\{\gamma'\})$, with as many
vertices as the graph has loops and the root decorated by the
primitive graph $\Gamma/\{\gamma'\}$ resulting from the
contraction of the maximal spinney $\{\gamma'\}$ in $\Gamma$. The
other vertices of this tree, representing the hierarchy of
subgraphs in that spinney, are decorated by those spinney
elements. Eventually, the graph $\Gamma$ is represented by the
linear combination of those trees,
$$
    \Gamma \sim \sum_{\{\gamma'\}\in\mathfrak{W}_{max}(\Gamma)} t(\{\gamma'\}).
$$
One verifies readily that the coproduct on both objects, defined
in $\calh_\calf$ respectively $\calh_\calt$, agree modulo the
identification between graphs and decorated rooted trees. As an
example compare the coproduct of the Feynman graph
\scalebox{0.8}{\roll}
$$
 \Delta\Big(\scalebox{0.8}{\roll} \Big) = \scalebox{0.8}{\roll} \otimes 1_{\mathcal{F}}
                           + 1_{\mathcal{F}}\otimes \scalebox{0.8}{\roll} +
                           2\newfish \otimes \winecup +
                            \newfish \newfish \otimes \newfish.
$$
with that of the cherry tree \td31\ in (\ref{ex:copTree}).
$$
\Delta\big(\scalebox{1.2}{\td31}\big) = \!\! \begin{array}{c}
                                 \\[-0.5cm]\scalebox{1.2}{\td31} \\
                              \end{array}\!\! \otimes 1_\mathcal{T}
                               + 1_\mathcal{T} \otimes\!\!\!
                              \begin{array}{c}
                                 \\[-0.5cm]\scalebox{1.2}{\td31} \\
                              \end{array}\!\! +
                                  2\scalebox{1.2}{\ta1} \otimes \!\!\!
                                       \begin{array}{c}
                                         \\[-0.5cm]\scalebox{1.2}{\tb2} \\
                                       \end{array} + \scalebox{1.2}{\ta1}\scalebox{1.2}{\ta1}\otimes \scalebox{1.2}{\ta1},
$$
where one needs to put the decoration \newfish\ on each tree
vertex. Observe that the winecup graph \winecup\ contains
\newfish\ as the only UV divergent 1PI graph and hence is represented by
the ladder tree \tb2\ decorated by \newfish\ . To phrase it
differently, the linear combinations of decorated rooted trees
representing graphs form a Hopf subalgebra in
$\calh_\calt$~\cite{KreimerOver}.


\subsection{Feynman characters and Birkhoff decomposition}
\label{ssect:FeynmanChar}

Recall from the mathematical preliminaries in an earlier section
that the space $\mathrm{Hom}(\calh_\calf,\mathbb{K})$ together
with the convolution product $\star$ and the counit map $\epsilon:
\mathcal{H}_{\calf} \to \mathbb{K}$ as unit forms a unital,
associative and non-commutative $\mathbb{K}$-algebra, which
contains the group of characters,
$G:=\mathrm{char}(\mathcal{H}_{\calf},\mathbb{K})$, i.e., linear
functionals $\phi$ from $\mathcal{H}_{\calf}$ to $\mathbb{K}$
respecting multiplication, $\phi(\Gamma_1
\Gamma_2)=\phi(\Gamma_1)\phi(\Gamma_2)$, $\Gamma_1$, $\Gamma_2 \in
\mathcal{H}_{\calf}$. This group of multiplicative maps possesses
a corresponding Lie algebra, $\mathcal{L}=\partial
\mathrm{char}(\mathcal{H}_{\calf},\mathbb{K})$, of derivations, or
infinitesimal characters, i.e., linear maps $Z$ satisfying the
Leibniz rule
$$
  Z(\Gamma_1 \Gamma_2) = Z(\Gamma_1)\epsilon(\Gamma_2)
                                  +\epsilon(\Gamma_1) Z(\Gamma_2)
$$
for all $\Gamma_1$, $\Gamma_2 \in \mathcal{H}_{\calf}$. The
grading of $\mathcal{H}_{\calf}$ implies a decreasing filtration
on the algebra $\Hom(\calh_\calf,\mathbb{K})$, making it a unital
complete filtered algebra. The exponential map
$\mathrm{exp}^{\star}$ gives a bijection between the Lie algebra
$\mathcal{L}$ and its corresponding group $G$.\\


\subsubsection{Feynman rules as characters}

As a matter of fact, general Feynman rules for any interesting
perturbative \textsf{QFT} give a linear map $\phi \in
\mathrm{Hom}(\calh_\calf,\mathbb{K})$, which can be extended
multiplicativity to form a subclass of characters in $G$.

The reader may wish to recall Theorem~\ref{thm:factHopf} which
already implies at this level a --unique-- factorization of the
group $G:=\mathrm{char}(\mathcal{H}_{\calf},\mathbb{K})$ upon the
choice of an arbitrary --idempotent-- linear map $P$ on
$\mathrm{Hom}(\calh_\calf,\mathbb{K})$. Eventually, the map
 $P$ turns out to be naturally given in renormalization.\smallskip

Amplitudes, viz those Feynman integrals associated with a Feynman
graph via Feynman rules, are given by in general ill-defined
integrals plagued with ultraviolet divergences coming from high
momenta integrations. These divergences demand for a
regularization. In general, a {\it{regularization prescription}}
introduces extra nonphysical parameters into the theory rendering
such Feynman integrals finite but changing the nature of the
target space of Feynman rules.

The particular choice of such a prescription is largely arbitrary,
and mainly guided by two desires. One is the more practical
interest, especially to practitioners in \textsf{pQFT}, for
calculational convenience driven by the fact that multiloop
calculations are in general very complicated problems. The other
desire, though of more fundamental nature, is to maintain as many
--if not all-- of the physical properties of the original
--non-regularized-- theory as possible. However, what one must
assure is that the final physical result is completely independent
of such nonphysical intermediate steps.

Motivated by the need for regularization of \textsf{pQFT}, due to
ultraviolet divergencies we take here the more general point
of view, replacing the base field $\mathbb{K}$ as target space by
a suitable commutative and unital algebra $A$. As an example we
mention $A = \mathbb{C}[\varepsilon^{-1},\varepsilon]]$, the field
of Laurent series that enters in dimensional regularization.

Let us denote by $G_A := \mathrm{char}(\mathcal{H}_\calf,A)
\subset {\mathrm{Hom}}(\mathcal{H}_\calf,A)$ the group of
$A$-valued, or regularized, algebra morphisms. The group law is
given by the convolution product
\begin{equation}
  \phi_1 \star \phi_2 := m_A \circ (\phi_1 \otimes \phi_2) \circ \Delta:\
   \mathcal{H}_{\calf} \xrightarrow{\Delta}
   \mathcal{H}_{\calf} \otimes \mathcal{H}_{\calf}
   \xrightarrow{\phi_1 \otimes \phi_2} A \otimes A
   \xrightarrow{m_A} A
   \label{def:convol}
\end{equation}
With the now regularized Feynman rules understood as canonical
$A$-valued characters $\phi$ we will have to make one further
choice: a {\it{renormalization scheme}}. We do this by demanding
the existence of a linear idempotent map $R$ satisfying the
{\it{Rota--Baxter relation}} of weight one
\begin{equation}
    \label{eq:RBrel}
    R(x)R(y)+R(xy) = R\big( R(x)y \big) + R \big( xR(y) \big)
\end{equation}
for all $x,y \in A$. For $R$ being such a Rota--Baxter map,
$\tilde{R}:=\id_A - R$ also satisfies relation~(\ref{eq:RBrel}).
This equation turns out to lie at the heart of Connes--Kreimer's
Birkhoff decomposition to be explored below. It appeared in
\cite{CK1,KreimerChen} under the name {\it{multiplicativity
constraint}} and tells us that the algebra $A$ splits into two
parallel subalgebras given by the image and kernel of $R$.

As a paradigm we mention again dimensional regularization together
with the minimal subtraction scheme, that is, the pole part
projection $R:=R_{ms}$, which is a Rota--Baxter map. At this point
we must postpone further comments on a complete census of
renormalization schemes used in physics in the light of
Rota--Baxter algebras.\\


\subsubsection{Connes--Kreimer's Birkhoff decomposition of Feynman
rules} \label{sssect:BirkhoffCK}

In the sequel it is our goal to shed more light on the meaning of
this operator relation in the context of Connes--Kreimer's work.

But before this, let us take a shortcut for the moment and see
briefly how all the above structure comes together. Starting with
regularized Feynman rules characters $\phi$ dictated by the
\textsf{pQFT} and taking values in the field of Laurent series
$A:=\mathbb{C}[\varepsilon^{-1},\varepsilon]]$ with
$$
    R:=R_{ms}:A \to A,\qquad \sum_{i=-n}^\infty a_i \varepsilon^i \mapsto \sum_{i=-n}^{-1}a_i \varepsilon^i
$$
defining the renormalization scheme map on $A$, Connes and Kreimer
observed that Bogoliubov's recursive formula for the counterterm
in renormalization has a Hopf algebraic expression given
inductively by the map
\begin{equation}
    \label{eq:CK-ct}
    \phi_-(\Gamma) = -R\big(\phi(\Gamma)
                          + \sum_{\gamma\subset \Gamma} \phi_-(\gamma)\phi(\Gamma /
                          \gamma)\big), \qquad \Gamma \in
                          \ker(\epsilon),
\end{equation}
with $\phi_-(1_{\calf}):=1_A$ and $\phi_-(\Gamma) = -R(\Gamma)$ if
$\Gamma$ is a primitive element in $\mathcal{H}_\mathcal{F}$,
i.e., contains no subdivergences. We denoted this map earlier by
$C$, the counterterm. The argument of $R$
\begin{equation}
    \label{eq:BogoClassic}
    \bar{{\rm{R}}}[\phi](\Gamma):=\phi(\Gamma) +
                                    \sum_{\gamma\subset \Gamma} \phi_-(\gamma)\phi(\Gamma / \gamma)
\end{equation}
for $\Gamma \in \ker(\epsilon)$, is Bogoliubov's preparation map or the
$\mathrm{\bar{R}}$-map. This leads to the Birkhoff decomposition
of Feynman rules found by Connes and Kreimer
\cite{CK1,CK2,CK3,KreimerChen}, described in the following
theorem.

\begin{theorem} \label{thm:ck}
The renormalization of the dimensionally regularized Feynman rules
character $\phi \xrightarrow{ren.} \phi_+$ follows from the
convolution product of the counterterm $\phi_-:=S_R^\phi$ with
$\phi$, $\phi_+:= S_R^\phi \star \phi$, implying the inductive
formula for $\phi_+$
$$
    \phi_{+}(\Gamma) = \phi(\Gamma) + \phi_-(\Gamma)+
           \sum_{\gamma\subset \Gamma} \phi_-(\gamma)\phi(\Gamma / \gamma),\qquad \Gamma \in \ker(\epsilon).
$$
Furthermore, the maps $\phi_-$ and $\phi_+$ are the unique
characters such that $\phi = \phi_-^{-1} \star \phi_+$ gives the
algebraic Birkhoff decomposition of the regularized Feynman rules
character $\phi \in G_{A}$.
\end{theorem}
In the proof of this theorem relation (\ref{eq:RBrel}) enters in
the last statement, that is, in showing that $S_R^\phi$ is a
character. We will see that its appearance has deeper reasons. Let
us postpone to give more details. In Section~ \ref{ssect:Matrep},
we will capture this theorem in the context of a Rota--Baxter
matrix calculus.


\section{Rota--Baxter algebras}
\label{sect:RBA}

Relation (\ref{eq:RBrel}) plays a particular role in the
decomposition result of Connes and Kreimer. We know already from
Theorem~\ref{thm:factHopf} that we have a natural factorization of
$A$-valued (or regularized) characters in the complete filtered
algebra $\cala:=\mathrm{Hom}(\calh_\calf,A)$ upon the arbitrary
choice of a filtration preserving idempotent map $P$ on $\cala$.
Now, we explore the implications due to the Rota--Baxter relation.


\subsection{Definition, history and examples}
\label{ssect:basics}

As a matter of fact, relation (\ref{eq:RBrel}) is a well-known
object in mathematics. In the 1950s and early 1960s, several
interesting results were obtained in the fluctuation theory of
probability. One of the most well-known is Spitzer's classical
identity~\cite{Spitzer56} in the theory of sums of independent
random variables, see also~\cite{Spitzer76}.

In an important 1960 work \cite{Baxter60} the American
mathematician Glen Baxter (1930-1983) deduced Spitzer's identity
from the above operator identity
\begin{equation}
    \label{def:RBR}
    R(x)R(y) = R\big( R(x) y + x R(y) - \theta xy \big),
\end{equation}
for all $x,y \in A$, where $R$ is a $\mathbb{K}$-linear
endomorphism on a $\mathbb{K}$-algebra $A$, which he assumed to be
associative, unital and commutative. Here $\theta$ is a fixed
element in the base field $\mathbb{K}$ called the weight of the
algebra $A$.

Gian-Carlo Rota (1932-1999) started a careful in depth elaboration
of Baxter's article in his 1969 papers \cite{Rota1}, where he
solved the crucial `word problem'. Together with one of his
students, David Smith, he remarked --in the context of the Hilbert
transform of a function-- in a 1972 paper \cite{RotaSmith72} that
Baxter's identity is equivalent, in characteristic zero, to the
modified Rota--Baxter relation
\begin{equation}
    \label{def:modRBR}
    B(x)B(y) = B\big( B(x) y + x B(y)\big) - \theta^2 xy, \qquad \forall x,y \in A,
\end{equation}
where $B:=\theta \id_A - 2R$. During the early 1960s and 1970s,
further algebraic, combinatorial and analytic aspects of Baxter's
identity were studied by several
people~\cite{Cartier72,Kingman62,Miller66,Miller69,Thomas77,Vogel63}.
\smallskip

Acknowledging Rota's seminal contributions to the subject as well
as his efforts to promote its further development and application
in several fields of mathematics \cite{Rota95,Rota98} we call a
pair $(A,R)$ where $A$ is a not necessarily associative
$\mathbb{K}$-algebra and $R$ is a $\mathbb{K}$-linear operator $R:
A \to A$ satisfying the {\it{Rota--Baxter relation}}
(\ref{def:RBR}) a {\it{Rota--Baxter}} $\mathbb{K}$-{\it{algebra}}
of weight $\theta$.\\


\subsubsection{Riemann and Jackson integration}

Let $A={\bf Cont}(\RR)$ be the ring of continuous functions on
$\mathbb{R}$. The indefinite Riemann integral is seen as a linear
map
\begin{equation}
    \label{def:Riemann}
    I: A\to A,\quad  I[f](x):=\int_0^x f(t)dt.
\end{equation}
Then $I$ is a weight zero Rota--Baxter operator. Indeed, let
$$
    F(x):=I[f](x)=\int_0^x f(t)\,dt,\quad G(x):=I[g](x)=\int_0^x g(t)\, dt.
$$
Then the integration-by-parts formula for the Riemann integral
states that
$$
    \int_0^x F(t)G'(t) dt = F(x)G(x) - \int_0^x F'(t)G(t) dt,
$$
that is,
$$
    I\big[I[f] g\big](x) = I[f](x)I[g](x) - I\big[fI[g]\big](x).
$$
Thus the
weight $\theta$ Rota--Baxter relation (\ref{def:RBR}) may be
interpreted as a generalization of the integration-by-parts rule
for integral-like operators on suitable function spaces.

Such a generalization of Riemann's integral map away from weight
zero is given by the $q$-{\it{analog}} of the Riemann integral,
also known as {\it{Jackson's integral}}. By this we mean the
following operator on a well-chosen function algebra
$\mathcal{F}$, defined for $0 < q < 1 $ \allowdisplaybreaks{
\begin{eqnarray*}
    J[f](x) := \int_{0}^{x} f(y) d_qy = (1-q)\:\sum_{n \ge 0} f(xq^n) xq^n.
\end{eqnarray*}}
This may be written in a more algebraic way, using the operator
\begin{equation*}
    P_q[f] := \sum_{n>0} E_q^{n}[f],
\end{equation*}
where the algebra endomorphism ($q$-{\it{dilatation}})
$E_q[f](x):=f(qx)$, $f \in \calf$. The map $P_q$ is a Rota--Baxter
operator of weight $-1$ and hence $\id_{\calf} + P_q =: \hat{P}_q$
is of weight $+1$.

Jackson's integral is given in terms of the above operators $P_q$
and the multiplication operator $M_{\id}(f)(x):=xf(x)$, $f \in
\calf$ as follows
\begin{equation*}
     J[f](x) = (1-q) \hat{P}_q\:M_{\id} \:[f](x)
\end{equation*}
satisfying the following `mixed' Rota--Baxter relation of weight
$(1-q)$,
\begin{equation*}
    J[f]\: J[g] + (1-q)J\:M_{\id}[f \: g] = J \Big[J [f] \: g + f \: J[g]
    \Big],\qquad f,g \in \calf,
\end{equation*}
replacing the integration-by-parts rule for the Riemann
integral.\\


\subsubsection{Summation}

On a suitable class of functions, we define
$$
    S(f)(x) := \sum_{n\geq 1} f(x+n).
$$
Then $S$ is a Rota--Baxter operator of weight $-1$, since
\begin{align*}
    &\Big(\sum_{n\geq 1} f(x+n)\Big)\Big(\sum_{m\geq 1} g(x+m)\Big)
                = \sum_{n\geq 1, m\geq 1} f(x+n)g(x+m)\\
    &= \Big (\sum_{n>m\geq 1} +\sum_{m>n\geq 1}+ \sum_{m=n\geq 1}\Big) f(x+n) g(x+m) \\
    &= \sum_{m\geq 1} \big( \sum_{k\geq 1}f(x+\underbrace{k+m}_{=n})\big) g(x+m)
        + \sum_{n\geq 1} \big (\sum_{k\geq 1} g(x+\underbrace{k+n}_{=m})\big) f(x+n)\\
    &\hspace{8cm}+ \sum_{n\geq 1} f(x+n)g(x+n)\\
    &= S\big(S(f) g\big)(x) + S\big(fS(g)\big)(x)+S(fg)(x).
\end{align*}


\subsubsection{Partial sums}

Let $A$ be the set of sequences $\{a_n\}$ with values in
$\mathbb{K}$. Then $A$ is a $\mathbb{K}$-algebra with termwise
sum, product and scalar product. Define
$$
    R:A \to A,\quad R(a_1,a_2,\cdots):=(a_1,a_1+a_2,\cdots).
$$
Then $R$ is a Rota--Baxter operator of weight one.

Let $A$ be a unitary algebra and let
$\mathbb{N}:=\mathbb{N}_{>0}$. Define
$\cala:={\rm{map}}(\mathbb{N}, A) = A^\mathbb{N}$ to be the
algebra of maps $f: \mathbb{N} \to A$ with point-wise operations.
Define the linear operator
\begin{equation}
    Z := Z_A: \cala \to \cala, \quad Z[f](k):= \begin{cases}
                                                      \sum_{i=1}^{k-1} f(i),\ k>1, \\
                                                                                0,\ k=1.
                                               \end{cases}
    \label{def:Zsum}
\end{equation}
The map $Z$ is a Rota--Baxter operator on $\cala$ of weight $-1$.
Furthermore, we have, for $f_1, \cdots, f_n \in \cala$, the
iteration
\begin{equation*}
    Z\Big[f_1 Z\big[f_2 \cdots Z[f_n]\cdots \big]\Big](k)
                                    = \sum_{k>i_1>\cdots>i_n>0} f_1(i_1) \cdots f_n(i_n)
\end{equation*}
and thus in the limit
\begin{equation}
    \label{eq:limitZ}
    \lim_{k\to \infty} Z\Big[f_1 Z\big[f_2 \cdots Z[f_n]\cdots \big]\Big](k) =
                                                            \sum_{i_1 > \cdots > i_n > 0} f_1(i_1) \cdots f_n(i_n)
\end{equation}
if the nested infinite sum on the righthand side exists. For the
particular choice of functions $f_i(x) \in \cala':=\{
f_s(x):=x^{-s} \ |\ s \in \mathbb{N}\} \subset \cala$, we obtain
from the summation map $Z$ in (\ref{def:Zsum}) respectively its
limit (\ref{eq:limitZ}) the multiple-zeta-value
\begin{equation}
    \label{eq:mzv}
    \zeta(s_1,\cdots,s_k) := \sum_{s_1>s_2>\cdots >s_k \geq 1} \frac{1}{n_1^{s_1} \cdots n_k^{s_k}}
                           = \lim_{l \to \infty} Z\big[{x^{-s_1}}Z\big[{x^{-s_2}}\cdots
                                                                     Z[{x^{-s_k}}]\cdots \big]\big](l).
\end{equation}
of length $\ell(\zeta(s_1,\cdots,s_k)):=k$ and weight
$w(\zeta(s_1,\cdots,s_k)):=\sum_{i=1}^k s_i$. The reader may take
a glimpse into the proceedings contribution by Weinzierl for the
link to multiple-zeta-values and their generalizations, see also
\cite{EGmzv05,MUW02}. The {\it{shuffle}} and {\it{quasi-shuffle}}
relation mentioned for these numbers~\cite{Hoffman05} are simply a
consequence of the Rota--Baxter relation for the Riemann integral
operator (\ref{def:Riemann}) and the summation operator $Z$ in
(\ref{def:Zsum}). More details about these generalized shuffle
products and its relation to free commutative Rota--Baxter algebra
\cite{GK2000} can be found in \cite{EGmixShuf}.\\


\subsubsection{Matrix Rota--Baxter maps} \label{ex:Miller}

The next example was introduced in \cite{Miller69} and is related
to Rota's original (free) Rota--Baxter algebra in \cite{Rota1}.
Let $A$ be a finite dimensional $\mathbb{K}$-vector space with
basis $e_1,\dots, e_n$. We make it into a commutative
$\mathbb{K}$-algebra by defining the product componentwise on the
column vectors of \makebox{$n=s+t$} components. Then the following
matrix $R \in \mathcal{M}_n(\mathbb{K})$ defines a Rota--Baxter
operator of weight $\theta=1$ on $A$
\begin{equation}
 R:=\left( \begin{matrix}
                    S_s & 0 \cr
                    0 & T_t \cr
           \end{matrix} \right),
\end{equation}
where the matrices $S,T$ of size $s$ respectively $t$ are given by
\begin{equation}
  S_s:={\small{ \left( \begin{matrix}
                    1 & 1  & \ldots & 1      \cr
                    0 & 1  & \ldots & 1      \cr
                    \vdots & \ddots & \ddots & \vdots\cr
                    0 & \ldots & 0&  1 \cr
                              \end{matrix}\right)}}_{s \times s},
                             \;\;\;
   T_t:={\small{ \left( \begin{matrix}
                     0 &  0  & \ldots & 0    \cr
                    -1 &  0  & \ldots & 0    \cr
                \vdots & \ddots & \ddots & \vdots \cr
                    -1 & \ldots & -1 & 0  \cr
                              \end{matrix}\right)}}_{t \times t}.
\end{equation}


\subsubsection{Triangular matrices}
On the algebra of upper triangular matrices
$\mathcal{M}^u_n(\mathbb{K})$, define the linear operator $D$
$$
    \big(D(\alpha_{kl})\big)_{ij} := \theta \delta_{ij} \sum_{k \geq i} \alpha_{ik},
                                \qquad \alpha_{ik} \in \mathcal{M}^u_n(\mathbb{K}).
$$
Then $D$ is a Rota--Baxter operator on
$\mathcal{M}^u_n(\mathbb{K})$ of weight $\theta \in \mathbb{K}$
\cite{Leroux04}.\\


\subsubsection{Idempotent Rota--Baxter maps}

If the $\mathbb{K}$-algebra $A$, for instance Lie or associative,
decomposes directly into subalgebras $A_1$ and $A_2$ , $A = A_1
\oplus A_2$, then the projection to $A_1$, $R: A \to A$,
$R(a_1,a_2)=a_1$, is an idempotent Rota--Baxter operator. Let us
verify this briefly for $a,b \in A= A_1 \oplus A_2$
\begin{align*}
    R(a)b + aR(b) - ab &= R(a)\big(R(b) + (\id_A-R)(b)\big)\\
                       & \qquad \quad - \big(R(a) + (\id_A-R)(a)\big)(\id_A-R)(b)\\
                       &= R(a)R(b) - (\id_A-R)(a)(\id_A-R)(b)
\end{align*}
such that applying $R$ on both sides gives
$$
    R\big( R(a)b + aR(b) - ab \big) = R(a)R(b),
$$
since it kills the term $(\id_A-R)(a)(\id_A-R)(b)$  without
changing the term $R(a)R(b)$, as $R (\id_A-R)(a)=0$ and $A_1,A_2$
are subalgebras.

As an example we saw already the {\it{minimal subtraction scheme}}
in {\it{dimensional regularization}}, where
$A=\CC[\varepsilon^{-1},\varepsilon]]=\varepsilon^{-1}\CC[\varepsilon^{-1}]
\oplus \CC[[\varepsilon]]$. So $R\big(\sum_{n \geq -N} a_n
\varepsilon^n\big):=\sum_{n= -N}^{-1} a_n \varepsilon^n$ is a
Rota--Baxter operator. In general if $R$ is multiplicative and
idempotent, then $R$ is a Rota--Baxter operator.


\subsection{Properties of Rota--Baxter algebras}
\label{ssect:properties}

Let $(A,R)$, $(B,P)$ be two Rota--Baxter algebras of the same
weight. We call an algebra morphism $f$ between $A$ and $B$ a
{\it{Rota--Baxter homomorphism}}, if $f \circ R = P \circ f$. A
{\it{Rota--Baxter ideal}} of a Rota--Baxter algebra $(A,R)$ is an
ideal $I$ of $A$ such that $R(I)\subseteq I$. The following basic
properties of a Rota--Baxter algebra $(A,R)$ are readily verified.
If $R$ has weight $\theta$, then $\theta^{-1} R$ has weight one.
We already remarked that $\tilde{R}:=\theta \id_A - R$ is a
Rota-Baxter operator. The images $A_{-}:=R(A)$ as well as
$A_{+}:=\tilde{R}(A)$ are subalgebras in $A$. The so-called
{\it{double Rota--Baxter product}}
\begin{equation}
 \label{def:doubleRBprod}
    x *_R y := xR(y)+R(x)y- \theta xy, \qquad x,y \in A,
\end{equation}
equips the vector space underlying $A$ with another Rota--Baxter
structure denoted by $(A, \ast_R, R)$ where $R$ satisfies the
Rota--Baxter relation for this product. One readily verifies that
\begin{equation}
 \label{eqs:doubleRBhom}
    R(x *_R y) = R(x)R(y) \quad {\rm{ and }} \quad
    \tilde{R}(x *_R y) = -\tilde{R}(x)\tilde{R}(y),\qquad x,y \in A.
\end{equation}
This construction might be continued giving a hierarchy of
Rota--Baxter algebras. Later, we will need the following useful
identity
\begin{equation}
     a_1  *_R \dots  *_R a_n = \frac{1}{\theta}\Big( \prod_{i=1}^{n}R(a_i)
                                                    - (-1)^n\prod_{i=1}^{n} \tilde{R}(a_i) \Big),
     \qquad a_i \in A.
           \label{eq:Kingman}
\end{equation}

If $(A,R)$ is a Rota-Baxter algebra with idempotent Rota--Baxter
map ($R^2=R$), then $A = A_- \oplus A_+$. More generally, we have
that a linear operator $R: A \to A$ is a Rota--Baxter operator if
and only if the following is true, first, $A_+$ and $A_-$ are both
closed under multiplication, secondly, if for $x,y \in A$ we have
that for a $z \in A$, $R(x)R(y)=R(z)$ implies
$\tilde{R}(x)\tilde{R}(y)=-\tilde{R}(z)$. We find moreover that
$K_{+} :=\ker(R)$ and $K_{-} :=\ker(\tilde{R})$ are ideals in
$A_R$ and therefore $A_{\pm} = A_R / K_{\pm}$. They are also
ideals in $A_{\mp}$ and the map $ \xi: A_{+} / K_{+} \to A_{-} /
K_{-}$, defined by $\xi \big( R(x)\big) := \tilde{R}(x)$ is an
algebra isomorphism.\smallskip

From a mathematical as well as physics point of view it is
interesting to mention that the Rota--Baxter relation was
independently (re)discovered in the 1980s by several Russian
physicists, including Semenov-Tian-Shansky, and Belavin and
Drinfeld \cite{BelavinDrinfeld1,STS83} (see \cite{BBT} for more
details). In the particular context of Lie algebras they studied
$r$-matrix solutions to the {\it{classical Yang--Baxter equation}}
\begin{equation}
   [r_{13},r_{12}] + [r_{23},r_{12}] + [r_{23},r_{13}]=0
   \label{eq:CYB}
\end{equation}
named after the physicists C.N. Yang from China and the Australian
Rodney Baxter. Under suitable circumstances a related operator $R$
satisfies
\begin{eqnarray*}
     [R(x),R(y)] + R([x,y]) = R\big([R(x),y] + [x,R(y)]\big).
\end{eqnarray*}
Especially Semenov-Tian-Shansky found a link between the modified
classical Yang--Baxter equation and the Riemann--Hilbert
problem~\cite{STS83,STS00}. His results are a generalization of
the Kostant--Adler scheme.


\subsection{Spitzer's identity for commutative Rota--Baxter algebras}
\label{ssect:Spitzer}

We mentioned Spitzer's paper \cite{Spitzer56} playing an important
role in Baxter's work. A transparent way to understand this
identity may be seen by following Baxter's original approach. Let
$A={\bf Cont}(\mathbb{R})$ be the ring of continuous functions on
$\mathbb{R}$ with the Riemann integral map $I(f)(x)=\int_0^x
f(u)du$ (\ref{def:Riemann}). It is well-known that the initial
value problem
\begin{equation}
    \label{eq:IVP}
    \frac{d}{dt}y(t)=a(t)y(t),\quad y(0)=1, \qquad  a \in A
\end{equation}
has a unique solution, $y(t)=\exp\left (\int_0^t a(u)du\right )$.
As we may transform the differential equation into an integral
equation by applying $I$ to (\ref{eq:IVP})
\begin{equation}
    \label{eq:integralIVP1}
    y(t) = 1 + I(ay)(t)
\end{equation}
we arrive naturally at the non-trivial identity
\begin{equation}
    \label{exp-sol}
   \exp \left (\int_0^t a(u)du\right)
            = 1 + \sum_{n=1}^\infty  \underbrace{I\big(I(I( \cdots I}_{n-times}(a)a)\dots a)a\big)(t)
\end{equation}
which follows from the identity
\begin{equation}
    \label{eq:BSzero}
    \big(I(a)(t)\big)^n =  n!\underbrace{I\big(I(I( \cdots I}_{n-times}(a)a)\dots a)a\big)(t).
\end{equation}
This relation is the weight zero case of the Bohnenblust--Spitzer
identity, see Eq.~(\ref{eq:BS}) below. Baxter's simple question
was to replace within the {\it{commutative}} setting the weight zero
Rota--Baxter map $I$ in (\ref{eq:integralIVP1}) by an arbitrary
Rota--Baxter map $R$ of weight $\theta \neq 0$ and to find the
analog of the exponential solution respectively the corresponding
relation (\ref{exp-sol}). The result is the {\it{classical Spitzer
identity}} for a commutative Rota--Baxter algebra $(A,R)$ of
weight $\theta \neq 0$, \allowdisplaybreaks{
\begin{eqnarray}
  \exp\Big(- R\Big(\frac{\log(1_A - \theta a x)}{\theta}\Big) \Big)
         &=&\sum_{n=0}^\infty  x^n\underbrace{R\big(R(R( \cdots R}_{n-times}(a)a)\dots
         a)a\big),
  \label{SpitzerId-theta1}
\end{eqnarray}}
in the ring of power series $A[[x]]$. We may interpret this
identity as an infinite set of identities in $A$ followed by
comparison of coefficients of powers of $x$ \cite{RotaSmith72}.
Observe that $\frac{\log(1_A - \theta a x)}{\theta}$ reduces to
$-ax$ in the limit $\theta \to 0$, which gives back the classical
Riemann integral identity (\ref{exp-sol}).

To get ourselves acquainted with this identity let us consider a
simple, but interesting example of its use. We go back to the
$Z$-summation Rota--Baxter map in (\ref{def:Zsum}) and its link to
multiple-zeta-values (\ref{eq:mzv}). Observe that a simple
calculation gives for the function $f_k:=1/x^k$
\begin{eqnarray*}
    Z\big[\log(1 + f_k t)\big](m) \!\!\!&=&\!\!\! Z \left[\sum_{i=1}^{\infty} \frac{(-1)^{i-1}}{i}
                                                           \left(\frac{t}{x^k} \right)^i \right](m)
                = \sum_{i=1}^\infty \frac{(-1)^{i-1} t^i}{i} Z\left[\frac{1}{x^{ki}}\right](m).
\end{eqnarray*}
Hence, in the limit
$$
  \lim_{m \to \infty} Z\big[\log (1 + f_k t)\big](m) = \sum_{i=1}^\infty \frac{(-1)^{i-1} t^i}{i} \zeta(ki).
$$
Therefore, Spitzer's classical identity for the $Z$-summation map
gives
\begin{eqnarray*}
 \exp\left(\sum_{i=1}^\infty (-1)^{i-1} \zeta(ik)
 \frac{t^{i}}{i}\right)&=&
   1 + \sum_{n>0}^\infty t^i
       \underbrace{Z\big[ Z[ \cdots Z[Z}_{n\mbox{\rm -} {\rm times}}[f_k] f_k ]f_k \cdots ]f_k \big]\\
       &=& 1 + \sum_{n>0}^\infty t^{i} \zeta (\underbrace{k,\cdots, k}_{n-times}).
\end{eqnarray*}

Another example is given for the case where $R$ is the --trivial
weight one Rota--Baxter map-- $\id_A$. Then the left hand side of
Eq.~(\ref{SpitzerId-theta1}) becomes the power series
$$
    \exp\big( -\log(1_A - ax)\big)=\exp(\log(1_A-ax)^{-1}) = \frac{1}{1_A-ax}
$$
and the right hand side is $\sum_{n=0}^\infty x^n a^n
=\sum_{n=0}^\infty (ax)^n$. So we have the familiar geometric
expansion.

Let us remark here, that there exist several proofs of Spitzer's
classical identity in the commutative setting. Other than the
original proof of Spitzer in the case of probability
distributions, there are the combinatorial proofs of Baxter and
Kingman~\cite{Kingman62}, the analytic proofs of
Atkinson~\cite{Atkinson63} and Wendel~\cite{Wendel}, as well as
the algebraic proofs of Cartier and Rota--Smith working with free
commutative Rota--Baxter algebras and making use of Waring's
formula.

The Rota--Baxter point of view opens simple ways to recognize
particular identities, e.g. for multiple-zeta-values
(\ref{eq:mzv}) \cite{EGmzv05}, where Hoffman found the so-called
{\it{Partition Identity}} \cite{Hoffman05}, which is just a
particular example of the Bohnenblust--Spitzer formula for
commutative Rota--Baxter algebras.

Let us briefly recall the exact form of the
{\it{Bohnenblust--Spitzer formula}} \cite{Gioev02,RotaSmith72} of
weight $\theta$. Let $(A,R)$ be a commutative Rota--Baxter algebra
of weight $\theta$ and fix $s_1, \dots, s_n \in A$, $n>0$. Let
$S_n$ be the set of permutations of $\{1,\cdots, n\}$. Then
\begin{equation}
 \sum_{\sigma \in S_n} R\Big(s_{\sigma(1)}
        R\big(s_{\sigma(2)}\cdots R(s_{\sigma(n)})\cdots \big)\Big) = \sum_{\mathcal{T}}
        \theta^{n-|\mathcal{T}|} \prod_{T \in \mathcal{T}}(|T|-1)! \:R( \prod_{j \in T} s_j).
    \label{eq:BS}
\end{equation}
Here $\mathcal{T}$ runs through all unordered set partitions of
$\{1,\cdots, n\}$. The weight $\theta=0$ case reduces the sum over
$\mathcal{T}$ to $|\mathcal{T}|=n$. The Rota--Baxter relation
itself appears as a particular case for $n=2$. \medskip

In the next part, we will see that the {\it{noncommutative}}
version of Spitzer's identity naturally follows from
Theorem~\ref{thm:bch} in the context of a filtered associative
Rota--Baxter algebra of weight $\theta$. The limit $\theta \to 0$
retrieves Magnus' solution to the initial value problem
(\ref{eq:IVP}) in a noncommutative setting, say, for matrix valued
functions. The results presented here can be found in
\cite{EGMbch06} and were achieved together with D.~Manchon.


\subsection{Spitzer's identity for non-commutative Rota--Baxter algebras}

We will derive the noncommutative version of Spitzer's identity
from Theorem~\ref{thm:bch} by comparison with the following
fundamental result for associative Rota--Baxter algebras due to
F.~V.~Atkinson. His comprehensive 1963 paper~\cite{Atkinson63} is
important in the understanding of the role played by Rota--Baxter
algebras in the description of renormalization as a factorization
problem. For convenience we put ourselves into the realm of
complete filtered associative unital Rota--Baxter algebras
$(\cala,R)$ with filtration preserving Rota--Baxter map $R$ of
weight $\theta \neq 0$ over the field $\mathbb{K}$ of
characteristic zero.

For instance, recall Example~\ref{ex:matrix} of triangular
matrices. More generally, let $(A,R)$ be a Rota--Baxter algebra of
weight $\theta$. Define a Rota--Baxter map $\mathrm{R}$ on
$\mathcal{M}^\ell_n(A)$ by extending the Rota--Baxter map $R$
entrywise,
$$
    \mathrm{R}(\alpha) =  \big ( R(\alpha_{ij})\big).
$$
We have the following important, though simple theorem
\begin{theorem} \cite{EG,EGGV}\label{thm:matrixRB}
The triple
$\left(\mathcal{M}^\ell_n(A),\mathrm{R},\{\mathcal{M}^\ell_n(A)_k\}_{k\geq
1}\right)$ forms a complete filtered Rota--Baxter algebra of
weight $\theta$.
\end{theorem}

Atkinson observed in \cite{Atkinson63} a multiplicative
decomposition theorem for associative unital Rota--Baxter algebras
of weight $\theta$. Let $a \in \cala\fil{1}$. For elements $X$ and $Y$
in $\cala$ such that
\begin{equation}
    X = 1_\cala - R(X\ a)
        \quad {\mathrm{ respectively }} \quad
    Y = 1_\cala - \tilde{R}(a\ Y)
    \label{eq:recursion1}
\end{equation}
we find by simple Rota--Baxter gymnastic, that
\begin{equation}
    X (1_\cala + \theta a) Y = 1_\cala.
    \label{eq:atkinson1}
\end{equation}
We therefore obtain the factorization
\begin{equation}
    (1_\cala + \theta a) = X^{-1}Y{}^{-1}
    \label{eq:atkinson2}
\end{equation}
for elements in $1_\cala + \theta \cala\fil{1}$. If $R$ is
idempotent, then this is the unique decomposition of $1_\cala +
\theta a$ into a product of an element in
$\mathfrak{A}_{-}:=1_\cala + R(\mathcal{A}\fil{1})$ with an
element in $\mathfrak{A}_{+}:=1_\cala+
\tilde{R}(\mathcal{A}\fil{1})$. Both, $\mathfrak{A}_{\pm}$ are
subgroups, due to the Rota--Baxter relation. The inverses
$X':=X^{-1}$ and $Y':=Y^{-1}$ satisfy the equations
\begin{equation}
    X' = 1_\cala - R(\check{a}\ X')
        \quad {\mathrm{ respectively }} \quad
    Y' = 1_\cala - \tilde{R}(Y'\ \check{a}),
  \label{eq:recursionInverse}
\end{equation}
where $\check{a}:= (1_\cala+\theta a)^{-1}-1_\cala$ in $\cala\fil{1}$.\smallskip

Now, recall Theorem~\ref{thm:bch}, which we generalize to linear
filtration preserving maps $P$ on $\cala$, such that $P +
\tilde{P}=\theta \id_\cala$. This is easily achieved upon the
generalization of the $BC\!H$-recursion from $\chi$ to $\chi_\theta$.

\begin{theorem} \label{thm:BCHtheta}
Let $\cala$ be a complete filtered $\mathbb{K}$-algebra and $P$ a
linear filtration preserving map, such that $P + \tilde{P}=\theta
\id_\cala$. The map $\chi$ in factorization (\ref{BCHrecursion1})
generalizes to
\begin{equation}
    \chi_{\theta}(u) = u -
    \frac{1}{\theta}{\BCH}\Big(P\big(\chi_{\theta}(u)\big),\tilde{P}\big(\chi_{\theta}(u)\big)\Big).
    \label{BCH-recursion3}
\end{equation}
Similarly the recursion in Eq.~(\ref{BCHrecursion2}) transposes
into
\begin{equation}
    \label{BCH-recursion4}
    \chi_{\theta}(u)= u + \frac{1}{\theta}\BCH\Big(-P\big(\chi_{\theta}(u)\big),\theta u\Big),
    \qquad u \in \cala\fil{1}.
\end{equation}
 Such that for all $\exp(\theta u) \in 1_\cala + \theta \cala\fil{1}$, $u \in \cala\fil{1}$, we have the decomposition
\begin{equation}
    \exp(\theta u) = \exp\big(P(\chi_\theta (u))\big) \exp\big(\tilde{P}(\chi_\theta (u))\big).
    \label{eq:exptheta}
\end{equation}
\end{theorem}

\noindent We call $\chi_{\theta}$ the $BC\!H$-recursion of weight
$\theta \in \mathbb{K}$, or simply $\theta$-$BC\!H$-recursion. In
the light of Atkinson's decomposition (\ref{eq:atkinson2}) and the
factorization in Eq.~(\ref{eq:exptheta}) of
Theorem~\ref{thm:BCHtheta} we observe by taking into account
Eqs.~(\ref{eqs:doubleRBhom}) that
\begin{equation}
 \label{eq:SpitzerA}
    x:=\exp\big(-R(\chi_\theta (u))\big)=1_\cala + R\big(\exp^{*_R}\big(-\chi_\theta (u)\big)-1_\cala \big),
\end{equation}
for $1_\cala + \theta a=\exp(\theta u)$, $u \in
\cala\fil{1}$. One readily verifies using the identity in
Eq.~(\ref{eq:Kingman}) that \allowdisplaybreaks{
\begin{eqnarray}
 \exp^{*_R}\big(-\chi_\theta (u)\big)-1_A &=& \sum_{n>0} \frac{(-\chi_\theta (u))^{*_Rn}}{n!}\\
        &=& \sum_{n>0} \frac{(-1)^n}{n!\theta}\big(R(\chi_\theta (u))^n - (-\tilde{R}(\chi_\theta (u))^n)\big)
            \nonumber\\
        &=& \frac{1}{\theta} \exp\big(-R(\chi_\theta (u)\big) - \frac{1}{\theta} \exp\big(\tilde{R}(\chi_\theta (u)\big)
            \nonumber\\
        &=& - \exp\big(-R(\chi_\theta (u)\big) a. \nonumber
\end{eqnarray}}
In the last equality we used (\ref{eq:exptheta}). Hence we arrive
at Spitzer's identity for a complete filtered noncommutative
Rota--Baxter algebras $(\cala,R)$ of weight $\theta \neq 0$
\allowdisplaybreaks{
\begin{eqnarray}
  \exp\Big(- R\Big(\chi_{\theta}\Big(\frac{\log(1_\cala + \theta a)}{\theta}\Big)\Big) \Big)
         &=&\sum_{n=0}^\infty  (-1)^n \underbrace{R\big(R(R( \cdots R}_{n-times}(a)a)\dots
         a)a\big),
  \label{id:NC-SpitzerId-theta}
\end{eqnarray}}
for $a \in \cala_1$. It is obvious that $\chi_{\theta}$ reduces to
the identity for commutative algebras, giving back Spitzer's
classical identity (\ref{SpitzerId-theta1}).

We clearly see that in Theorem~\ref{thm:BCHtheta} no Rota--Baxter
property is needed. Instead, in identity
(\ref{id:NC-SpitzerId-theta}) respectively (\ref{eq:SpitzerA}) it
becomes evident that in the presence of a Rota--Baxter map we may
express the exponential factors in Eq.(\ref{eq:exptheta}) in terms
of Atkinson's recursions (\ref{eq:recursion1})
$$
    X=\exp\Big(- R\Big(\chi_{\theta}\Big(\frac{\log(1_\cala + \theta a)}{\theta}\Big)\Big) \Big)
    \qquad
    Y=\exp\Big(- \tilde{R}\Big(\chi_{\theta}\Big(\frac{\log(1_\cala + \theta a)}{\theta}\Big)\Big)\Big).
$$

As we will see in the following part, the particular appearance of
the weight $\theta$ in
Eqs.~(\ref{BCH-recursion3},\ref{BCH-recursion4}) reflects the fact
that in the case of weight $\theta=0$ we find $\tilde{R}=-R$ such
that Atkinson's factorization formula (\ref{eq:atkinson1})
collapses to
\begin{equation}
    \label{AtkinsonZero}
     XY=\big(1_\cala - R(X\ a)\big)\big(1_\cala + R(a\ Y)\big)=1_\cala
\end{equation}
for any $a \in \cala_1$ which is compatible with the $\theta =0$
limit in Eq.~(\ref{eq:exptheta}).


\subsection{Magnus' expansion and the weight zero $BC\!H$-recursion}
\label{ssect:Magnus}

In the light of Baxter's approach to Spitzer's classical identity
and its generalization to noncom\-mu\-ta\-tive associative
Rota--Baxter algebras of weight $\theta$ it seems to be natural to
look at the initial value problem in Section~\ref{ssect:Spitzer}
in a noncommutative setting, say, for matrix valued functions, $
\frac{d}{dt}\alpha(t)= \beta(t) \alpha(t)$, $\alpha(0)=1$, where
$\alpha(t)$ and $\beta(t)$ take values in the matrices
$\calm_n(\mathbb{K})$ of size $n \times n$, $1<n<\infty$. Again,
we can traverse the differential equation of the initial value
problem into an integral equation, which we write more generally
like
\begin{equation}
    \label{eq:integralIVP}
    \alpha(t) = {\bf{1}} + I(\beta\, \alpha)(t).
\end{equation}
But, under this assumption we can not write its solution as a
simple exponential.

Instead, recall the seminal work of Magnus~\cite{Magnus54} on
initial value problems of the above type. He proposed an
exponential solution
$$
    \alpha(t) = \exp\big(\Omega[\beta](t)\big)
$$
with $\Omega[\beta](0)=0$. For $\Omega[\beta](t)$ we assume an
expansion, $\Omega[\beta](t) := \sum_{n>0}\Omega^{(n)}[\beta](t)$,
in terms of multiple Riemann integrals of nested commutators of
the matrix $\beta(t)$. Magnus established a recursive equation for
the terms $\Omega^{(n)}[\beta](t)$ defined in terms of the
differential equation
\begin{equation}
    \label{pre-Magnus}
    \frac{d}{dt}\Omega[\beta](t) = \frac{\hbox{ad}\,
                \Omega[\beta]}{{\rm{e}}^{\hbox{\eightrm ad}\,\Omega[\beta]}-1}(\beta)(t).
\end{equation}

Let us go back to Eq.~(\ref{id:NC-SpitzerId-theta}) for a moment.
It happens to be useful to write Eq.(\ref{eq:fullBCH}) as a sum
~\cite{Reutenauer}
$$
  C(a,b)=a + b + \BCH(a,b)=\sum_{n \geq 0} H_n(a,b),
$$
where each $H_n(a,b)$ is homogenous of degree $n$ with respect to
$b$. Especially, $H_0(a,b)=a$. For $n=1$ we have
$$
    H_1(a,b)=\frac{\hbox{ad}\,a}{1-{\rm{e}}^{-\hbox{\eightrm ad}\,a}}(b).
$$
Hence, for Eq.(\ref{BCH-recursion4}) we get in the limit $\theta
\to 0$ a non-linear map $\chi_0$ inductively defined on the
pro-nilpotent Lie algebra $\cala_1$ by the formula
\allowdisplaybreaks{
\begin{eqnarray}
    \chi_0(a)&=&-\frac{\hbox{ad}R\big(\chi_0(a)\big)}{\id_\cala-{\rm{e}}^{\hbox{\eightrm ad}R(\chi_0(a))}}(a)
               = \bigg(\id_\cala + \sum_{n>0} b_n \Big[\hbox{\rm ad}R\big(\chi_0(a)\big)\Big]^{n}\bigg)(a)
                \label{BCH-recur5b}
\end{eqnarray}}
where $R$ is now a weight zero Rota--Baxter operator. We call this
the weight zero $BC\!H$-recursion. The coefficients
$b_n:=\frac{B_n}{n!}$ where $B_n$ are the Bernoulli numbers. For
$n=1,2,3,4$ we find the numbers $b_1=-1/2$, $b_2=1/12$, $b_3=0$
and $b_4=-1/720$. The first three terms in (\ref{BCH-recur5b}) are
\begin{equation}
    \chi_0(a)= a - \frac 12[R(a),\,a]+\Big(\frac 14\big[R\big([R(a),\,a]\big),\,a\big]
    +\frac{1}{12}\big[R(a),\,[R(a),\,a]\big]\Big)+\cdots
\end{equation}

\begin{lemma}\label{thm:0-ncSpitzer}
Let $(\cala,R)$ be a complete filtered Rota--Baxter algebra of
weight zero. For $a \in \cala_1$ the weight zero $BC\!H$-recursion
$\chi_0 :\cala_1 \to \cala_1$ is given by the recursion in
Eq.~(\ref{BCH-recur5b})
\begin{equation*}
 \chi_0(a)=-\frac{\hbox{ad}R\big(\chi_0(a)\big)}{\id_\cala-{\rm{e}}^{\hbox{\eightrm ad}R(\chi_0(a))}}(a).
\end{equation*}
\begin{enumerate}
\item\label{eq:exp1o} The equation $x=1_\cala-R(x\ a)$ has a
                      unique solution $x = \exp\big(-R(\chi_0(a))\big)$.
\item\label{eq:exp2o} The equation $y=1_\cala + R(a\ y)$ has a
unique
                      solution $y=\exp\big(R(\chi_0(a))\big).$
\end{enumerate}
\end{lemma}
Atkinson's factorization for the weight zero case,
Eq.~(\ref{AtkinsonZero}), follows immediately from the preceding
lemma.

Comparison of (\ref{BCH-recur5b}) with (\ref{pre-Magnus}) reveals
the link between Magnus' recursion and the $BC\!H$-recursion in
the context of a vanishing Rota--Baxter weight, namely

\begin{coro} Let $A$ be a function algebra over $\mathbb{R}$ with
values in an operator algebra. Let $I$ denote the indefinite
Riemann integral operator. Magnus' $\Omega$ expansion is given by
the formula
\begin{equation}
    \label{magnus-link}
    \Omega[a](x)=I\big(\chi_0(a)\big)(x).
\end{equation}
\end{coro}

Hence, the $\theta$-$BC\!H$-recursion (\ref{BCH-recursion3})
generalizes Magnus' expansion to general weight $\theta \neq 0$
Rota--Baxter operators $R$ by replacing the weight zero Riemann
integral in $F= 1 + I\{aF\}$.

The following commutative diagram captures the picture
established. We observe on the right wing of (\ref{diag:general})
that starting from a complete filtered associative Rota--Baxter
algebra $(\cala,R)$ of weight $\theta \neq 0$, we arrive at
Spitzer's classical identity assuming a commutative algebra.
Taking the limit $\theta \to 0$ gives the classical expression for
integral-like equations. Instead, following the left wing
(\ref{diag:general}) and first taking the limit $\theta \to 0$
while maintaining noncommutativity we retrieve Magnus' expansion.
It reduces to the classical exponential solution for the initial
value problem in (\ref{eq:IVP}) when the underlying algebra is
commutative.
\begin{equation}
    \label{diag:general}
    \xymatrix{
                        & {\hbox{{\eightrm{$\exp\!\Big(\!\!-\!\! R\Big(\chi_{\theta}
                                                    \Big(\frac{\log(1_\cala - \theta
                                                    b)}{\theta}\Big)\!\!\Big)\!\Big)$}}}
                                                    \atop \theta \neq 0,\ non-com. }
                                                     \ar[dd]^{{com.\atop \theta \to 0}}
                                                     \ar[rd]_{\theta \neq 0 \atop com.}
                                                     \ar[ld]^{\theta \to 0 \atop {non-com.}}
                                                     &\\
   {\hbox{\eightrm{$\exp\!\big(R\big(\chi_0(b)\big)\!\big)$}}
    \atop {\rm{Magnus}}}
     \ar[rd]^{com.}   &
                                                    & {\hbox{{\eightrm{$\exp\!\Big(\!\!-\!\!R\Big(
                                                     \!\frac{\log(1_\cala - \theta b)}{\theta}
                                                     \!\Big)\!\Big)$}}}
                                                      \atop {\rm{cl.\ Spitzer}}}
                                                      \ar[ld]_{\theta \to 0}\\
                      & {\hbox{\eightrm{$\exp\!\big(R(b)\big)$}}
                        \atop \theta = 0,\ com.}&
                }
\end{equation}


\section{The Rota--Baxter structure of renormalization in \textsf{pQFT}}
\label{sect:RBmatrixCK}

Let us briefly recapitulate Section~\ref{sect:HopfCK}. We assume a
renormalizable \textsf{QFT} treated perturbatively. Recall that
for a proper --Hopf algebraic-- renormalization picture in
\textsf{pQFT} we need the combinatorial Hopf algebra of 1PI
Feynman graphs, $\calh_\calf$, as well as Feynman rules $\phi$
which are identified as particular multiplicative maps on
$\calh_\calf$ into the base field $\mathbb{K}$. Both, the set of
1PI Feynman graphs $F$ and Feynman rules are dictated by the
\textsf{QFT} itself upon physical constraints. Renormalization
enters with the choice of a regularization prescription rendering
ill-defined Feynman amplitudes formally finite, together with a
renormalization scheme, the latter may reflect some input from
physics (physical renormalization scheme). The last two objects
enter the setting via the regularization target space $A \neq
\mathbb{K}$ in $\mathrm{Hom}(\calh_\calf,A)$ additionally equipped
with a particular linear map, denoted by $R$, which extracts
divergent parts of regularized Feynman amplitudes. The choice of
$A$, that is, the regularization prescription for the Feynman
rules, is mainly guided by practical reasons and severe demands
for conserving as many  as possible of the physical properties of
the unregulated \textsf{QFT}.\smallskip

Motivated by the example of the minimal subtraction scheme in
dimensional regularization, leading to regularized Hopf algebra
characters with values in
$A=\CC[\varepsilon^{-1},\varepsilon]]=\varepsilon^{-1}\CC[\varepsilon^{-1}]
\oplus \CC[[\varepsilon]]$ forming a commutative unital
Rota--Baxter algebra $(A,R)$, where $R$ is the projection of a
Laurent series onto its pole part, we assume in general that
regularization prescriptions and the related renormalization
schemes form a commutative unital Rota--Baxter algebra denoted by
the tuple $(A,R)$ with the renormalization scheme given by the
Rota--Baxter  map $R$.


\subsection{Birkhoff decomposition in Rota--Baxter algebras}

We will now see how in the light of the last sections this purely
algebraic {\it{Rota--Baxter}} picture naturally leads to an
algebraic Birkhoff type decomposition of Feynman rules capturing
the process of renormalization. Minimal subtraction in dimensional
regularization as renormalization prescription simply corresponds
to the choice of a particular Rota--Baxter algebra with an
idempotent Rota--Baxter map. In principle any other `sensible'
choice of such an algebra would be as good, though dimensional
regularization has proven itself to be a veritable setting
reflecting the state of the art in several areas of high energy
particle physics.

Eventually we will cast the process of renormalization into a
simple calculus on upper or lower triangular matrices with entries
in the regularization algebra $(A,R)$. This gives rise to a
linear representation of Connes--Kreimer's group $G_A$ of
$A$-valued characters and its Birkhoff decomposition, which is
recovered in terms of a factorization of such matrices implied by
the underlying Rota--Baxter structure. \medskip

The following theorem describes the Birkhoff decomposition of
Connes and Kreimer in Theorem \ref{thm:ck} using the algebraic
setting developed in the earlier sections, to wit, the
Rota--Baxter structure on the target space implied by the choice
of the renormalization scheme.

Remember that the Hopf algebra of Feynman graphs $\calh_\calf$,
which is the polynomial algebra generated by the set $F$ of 1PI
Feynman graphs, has the coproduct defined in (\ref{def:coprod})
which is of combinatorial type, i.e., for $\Gamma \in F$ the
cograph $\Gamma/\gamma$ in (\ref{def:coprod}) always lies in $F$.
Recall $\calf:=\mathbb{K}F$. We have
\begin{equation}
        \calf \xrightarrow{\Delta} \calh_{\calf} \otimes \calf.
\end{equation}
Remember that we called $\mathrm{ker}(\epsilon)$ the augmentation
ideal of $\mathcal{H}_\calf$ and that we denoted by $P$ the
projection $\calh_\calf \rightarrow \mathrm{ker}(\epsilon)$ onto
the augmentation ideal, $P := \id - \eta \circ \epsilon.$

Recall from Proposition~\ref{pp:filter} that the increasing
filtration on $\mathcal{H}_\mathcal{F}$ by its loop number grading
implies in duality a complete decreasing filtration on the
associative $\mathbb{K}$-algebra
$\cala:=(\mathrm{Hom}(\calh_\calf,A),\star,e_A)$ with unit
$e_A:=\eta_A \circ \epsilon$. Here $(A,R,\eta_A)$ is a unital
commutative Rota--Baxter algebra with idempotent weight one
Rota--Baxter map $R$ dictated by the regularization prescription.
In $\cala$ we have the group
$G_A:={\mchar{(\mathcal{H}_\mathcal{F},A)}}$ of multiplicative
maps and its corresponding Lie algebra of infinitesimal characters
$\call_A$, related bijectively via the exponential map
$\exp^{\star}$.

The crucial point is the observation that the linearity of $R$ on
$A$ allows for a lifting of the Rota--Baxter structure to $\cala$,
making it into a unital associative non-commutative Rota--Baxter
algebra $(\cala,\calr)$ of weight one. Having such a Rota--Baxter
algebra on $\cala$ at hand we arrive at the following theorem.

\begin{theorem}\label{thm:BCH-RB-QFT}{\rm{\cite{E-G-K2,E-G-K3}}}
$(\Hom(\mathcal{H}_\mathcal{F},A),\mathcal{R})$ is a complete
filtered Rota--Baxter algebra with Rota--Baxter operator
$\mathcal{R}(\phi):= R \circ \phi$ and filtration from
$\mathcal{H}_\calf$. We denote its unit by $e_A:= \eta_A \circ
\epsilon$. For an $A$-valued character $\phi=\exp^{\star}(b) \in
\mchar(\mathcal{H}_\mathcal{F},A)$ take $a:=\phi - e_A$. Then $a$
is in $\cala\fil{1}$ and Theorem \ref{thm:BCHtheta} implies that
\begin{enumerate}
\item \label{it:BogoRecursions}
    the equations in (\ref{eq:recursion1}) are the recursive formulae for
    $X =: \phi_- \in G_A$ and $Y =:\phi_+^{-1} \in G_A$ in Theorem~\ref{thm:ck} of
    Connes--Kreimer. They take the following form for $\Gamma \in \ker(\epsilon)$
 \allowdisplaybreaks{
\begin{eqnarray}
    \phi_-(\Gamma) &=& e_A(\Gamma) - \calr\big( \phi_{-} \star a \big)(\Gamma) \label{eq:special1}\\
                   &=& -R \big( \phi(\Gamma) + \sum_{(\gamma \subset \Gamma)} \phi_-(\gamma) \phi(\Gamma / \gamma) \big),
                   \nonumber\\
    \phi_+(\Gamma) &=& e_A(\Gamma) - \tilde{\calr}( \phi_{+} \star (\phi^{-1} - e_A))(\Gamma) \label{eq:special2}\\
                   &=& \tilde{R}\big( \phi(\Gamma) + \sum_{(\gamma \subset \Gamma)}\phi_-(\gamma)\phi(\Gamma / \gamma) \big).
                   \nonumber
\end{eqnarray}}

\item
    Spitzer's identity for non-commutative complete filtered Rota--Baxter
    algebras (\ref{id:NC-SpitzerId-theta}) implies that the above recursions for the
    linear maps $\phi_-$ and $\phi_+$ can be written as the exponential factors in Eq.~(\ref{eq:bch})
    giving the unique explicit formulae for $X=\phi_-$ and $Y^{-1}=\phi_+$. They are algebra homomorphisms,
    i.e., $A$-valued characters in $G^-_A:=e_A + \calr(\cala\fil{1})$
    respectively in $G^+_A:=e_A + \tilde{\calr}(\cala\fil{1})$
\begin{equation}
     \label{eq:SpitzerCKexp}
    \phi_- = \exp^{\star}\Big( \calr \big( \chi(b) \big)\Big) \ \mathrm{ resp.} \quad
    \phi_+ = \exp^{\star}\Big( \tilde{\calr}\big( \chi(b) \big)\Big).
\end{equation}

\item \label{it:CK-birkhofBCH}
    equation (\ref{eq:bch}) gives the unique Birkhoff
    decomposition of  $\phi = \phi_-^{-1} \star \phi_+$
    found in Theorem~\ref{thm:ck} of Connes--Kreimer;

\item \label{it:BogoMap}
    Bogoliubov's $\bar{\rm{R}}$-map, $\bar{\rm{R}}[\phi]: \mathcal{H}_\calf \to
    A$,  is given by $\bar{{\rm{R}}}[\phi] = \exp^{\star_\mathcal{R}}\big(-\chi(\log^{\star}(\phi))\big)$,
    for $\phi \in G_A$, such that $\tilde{\mathcal{R}}(\bar{{\rm{R}}}[\phi])
    =2e_A-\phi_{+}$ and $\mathcal{R}(\bar{{\rm{R}}}[\phi])
    =\phi_{-} - e_A$.
    Here, the double Rota--Baxter product (\ref{def:doubleRBprod})
    is defined in terms of $\calr$
    $$
        \phi_1{\star_\mathcal{R}}\phi_2:=\mathcal{R}(\phi_1)\star\phi_2 + \phi_1\star\mathcal{R}(\phi_2) - \phi_1 \star
        \phi_2,\quad \phi_1,\phi_2\in \cala.
    $$
\end{enumerate}
\end{theorem}

Observe that Eq.~(\ref{eq:special2}) for $Y^{-1}=\phi_+$ in
item~(\ref{it:BogoRecursions}) of Theorem~\ref{thm:BCH-RB-QFT}
does not contain the counterterm $\phi_{-}$ explicitly
$$
    \phi_{+} = e_A - \mathcal{\tilde{R}}\big(\phi_{+}\star(\phi^{-1} - e_A)\big).
$$
\noindent Recall that the inverse of $\phi \in G_A$ is given by
the composition with the antipode of $\calh_\calf$,
$\phi^{-1}=\phi \circ S$. The reader will easily check
that $\phi_{+}\star(\phi^{-1} - e_A) = -\phi_{-}\star(\phi - e_A)$
in accordance with the classical expression in terms of
Bogoliubov's $\bar{\mathrm{R}}$-operation (\ref{eq:BogoClassic}).

In the light of the last remark we may emphasize the observation
in item~(\ref{it:BogoMap}) of the above theorem, where we see that
Bogoliubov's $\bar{\mathrm{R}}$-operation (\ref{eq:BogoClassic})
(see also Eq.~(\ref{eq:Bogoclassic})) can be written as an
exponential, defined with respect to the double Rota--Baxter
convolution product $\star_{\calr}$ defined on $(\cala,\calr)$,
and the $BC\!H$-recursion $\chi$ in (\ref{BCHrecursion1})
$$
  \bar{\mathrm{R}}[\phi] = \phi_{-}\star (\phi-e_A)
                         = -\exp^{\star_{\calr}}\big(-\chi(\log^{\star}(\phi)) \big)
$$
for elements in the augmentation ideal of $\calh_\calf$.\\

The above theorem presents a purely algebraic setting for the
formulation of renormalization as a factorization problem in the
group of regularized Hopf algebra characters, situated in the
theory of non-commutative Rota--Baxter algebras with idempotent
Rota--Baxter map. The formulae for the counterterm
(\ref{eq:special1}) and renormalized character (\ref{eq:special2})
are completely dictated by a general decomposition structure,
which characterizes Rota--Baxter algebras. The additional property
of $R$ being a projector implies a direct decomposition of the
algebra $A$, hence the uniqueness of the factorization in
item~(\ref{it:CK-birkhofBCH}). We would like to emphasis the
necessary freedom in the choice of the regularization
prescription, encoded in the particular structure of the
commutative Rota--Baxter algebra $A$ as target space of linear
Hopf algebra functionals in
${\rm{Hom}}(\calh_{\calf},A)$.\\

{\bf{Geometric picture for dimensional regularization:}}
Specializing the target space Rota--Baxter algebra $A$ to the
field of Laurent series, i.e., using dimensional regularization
together with minimal subtraction scheme, the above theorem
amounts to the decomposition of the Laurent series
$\phi(\Gamma)(\varepsilon)$, which has poles of finite order in
the regulator parameter $\varepsilon$, into a part holomorphic at
the origin and a part holomorphic at complex infinity. This has a
geometric interpretation upon considering the Birkhoff
decomposition of a loop around the origin, which is central in the
work of Connes and Kreimer \cite{CK2,CK3}.

In this special case the theorem opens a hitherto hidden geometric
viewpoint on perturbative renormalization and puts the mathematics
in the theory of renormalization in a broader context of the
Riemann--Hilbert correspondence which has its origin in Hilbert's
21st problem and has been extended into several areas of analysis,
geometry, mathematical physics~\cite{Its03}.\medskip

Finally, let us mention that the above notion of complete
Rota--Baxter algebra and the statements in Theorem
\ref{thm:BCH-RB-QFT} become very transparent for (pro) uni-
respectively nilpotent upper (or lower) triangular matrices with
entries in a unital commutative Rota--Baxter algebra. This will be
our last point in the following final section. It is extracted
from our recent work~\cite{EGGV,EG}, which was conducted together
with J.~M.~Gracia-Bond\'{i}a and J.~C.~V\'{a}rilly. It shows how
the combinatorics of perturbative renormalization can be
represented by matrix factorization of unipotent triangular
matrices with entries in a commutative Rota--Baxter algebra. As we
have seen above such triangular matrices provide a simple example
of a complete filtered Rota--Baxter algebra.


\subsection{Matrix calculus for the renormalization process in \textsf{pQFT}}
\label{ssect:Matrep}

Recall that $\mathcal{H}_\mathcal{F}$ is defined to be the graded
polynomial algebra generated by $F$. Connes and Kreimer's coproduct $\Delta:
\mathcal{H}_\mathcal{F} \to \mathcal{H}_\mathcal{F} \otimes
\mathcal{H}_\mathcal{F}$ on $\mathcal{H}_\mathcal{F}$, making it
into a combinatorial Hopf algebra, is simply given by
(\ref{def:coprod}) which we repeat here for $\Gamma \in
\calh_\calf$
\begin{equation}
    \Delta(\Gamma) = \Gamma \otimes 1_\calf  + 1_\calf \otimes \Gamma +
                            \sum_{\gamma{\subset}\Gamma} \gamma \otimes \Gamma/\gamma,
    \label{eq:coprod1}
\end{equation}
where the sum is over superficially divergent subgraphs $\gamma$
of $\Gamma$ and $\Gamma/\gamma \in \calf$ was the corresponding
1PI cograph. Recall the Remark~\ref{rmk:forestcoprod} after
(\ref{def:coprod}) on the spinney structure of graphs and the
terms in the coproduct.

Let $A$ be a commutative Rota--Baxter algebra with Rota--Baxter
operator $R$ and define
$\mathcal{A}:={\rm{Hom}}(\mathcal{H}_\mathcal{F},A)$ which is
associative with respect to convolution and has a unit
$e_A:=\eta_A \circ \epsilon$. The grading of $\calh_{\calf}$ and
the Rota--Baxter structure on $A$ imply that $\cala$ is a complete
filtered non-commutative unital Rota--Baxter algebra.

In this final section we will establish a representation of
$\cala$ in terms of lower triangular matrices with entries in $A$.
The group $G_A := e_A + \cala\fil{1} \in \cala$ and its
corresponding Lie algebra $\call_A$ are associated to (pro)
unipotent and nilpotent lower triangular matrices, respectively.

Let us choose a subset $F' \subseteq F$ of graphs and order them,
$\Gamma_1:=1_\calf, \Gamma_2, \Gamma_3,\dots$ in accordance to the
grading by degree. For graphs having the same degree we choose an
arbitrary order. By the definition of the coproduct $\Delta$ in
(\ref{eq:coprod1}) we have
\begin{equation}
 \label{coprod}
 \Delta(\Gamma_i) = \Gamma_i \otimes 1_\calf + 1_\calf \otimes \Gamma_i +
 \sum_{j=2}^{i-1} \Gamma_{ij} \otimes \Gamma_j
 = \sum_{j=1}^i \Gamma_{ij}\otimes \Gamma_j,
\end{equation}
where $\Gamma_{ij} \in \mathcal{H}_\mathcal{F}$ and
$\mathrm{deg}(\Gamma_j) < \deg(\Gamma_i)$ for $j<i$. Here we see
that to assure that the restriction of the coproduct becomes a
good comodule map, $\Delta|_{\calf'}: \calf' \to \calh_\calf
\otimes \calf'$ where $\calf'=\mathbb{K}\,F'$, we must have that
for any $\Gamma_i \in F'$ all corresponding cographs in
(\ref{coprod}) appear in $F'$, too.

Define the following possibly infinite lower triangular matrix
$$
  M_{\mathcal{H}_\calf}(F')=(\Gamma_{ij})_{1\leq j \leq i\leq |F'|}
$$
to be the {\it{coproduct matrix}} of $\mathcal{H}_\mathcal{F}$
corresponding to $F'$. $M_{\mathcal{H}_\calf}(F')$ is an $ |F'|
\times |F'|$ lower triangular matrix ($\infty \times \infty$ if
$F' \subseteq F$ contains infinitely many graphs) with entries in
$\mathcal{H}_\mathcal{F}$ and with unit on the diagonal.

For example, take the following simple set of electron self-energy
graphs borrowed from $\textsf{QED}$ in four space-time dimensions
\begin{equation}
F':= \left\{\Gamma_1:= 1_\calf,\
\Gamma_2:= \scalebox{0.8}{\tw1},\
\Gamma_3:= \scalebox{0.8}{\sw1},\
\Gamma_4:=\!\!\scalebox{0.8}{\uw1},\
\Gamma_5:=\!\scalebox{0.8}{\vw1}\ \right\} \label{eq:F}
\end{equation}
and with the prescribed order. Then $\Delta(\Gamma_i) =
\sum_{j\leq i} \Gamma_{ij} \otimes \Gamma_j$. So, that by a slight
abuse of the tensor product notation as matrix product we have
$$
    \Delta \left( {\small{\begin{array}{l} 1_\calf\\ \Gamma_2\\ \Gamma_3\\ \Gamma_4\\ \Gamma_5\end{array}}} \right )
    = \big( \Gamma_{ij} \big) \otimes
      \left( {\small{\begin{array}{l} 1_\calf\\ \Gamma_2\\ \Gamma_3\\ \Gamma_4\\ \Gamma_5\end{array}}} \right )
    = \left( {\small{\begin{array}{lllll}
          1_\calf & 0                & 0         & 0       & 0 \\
         \Gamma_2 & 1_\calf          & 0         & 0       & 0\\
         \Gamma_3 & \Gamma_2         & 1_\calf   & 0       & 0\\
         \Gamma_4 & \Gamma_3         & \Gamma_2  & 1_\calf & 0\\
         \Gamma_5 & \Gamma_2\Gamma_2 & 2\Gamma_2 & 0       & 1_\calf \end{array}}} \right )
        \otimes
    \left({\small{\begin{array}{l} 1_\calf\\\Gamma_2\\ \Gamma_3\\\Gamma_4\\ \Gamma_5 \end{array}}} \right),
$$
where the coproduct matrix $(\Gamma_{ij}) \in
\mathcal{M}_{\calh_\calf}(F')$ is lower $5 \times 5$ triangular
with unit diagonal and entries in ${\calh_\calf}$. In the last row
we used that
\begin{align*}
    \Delta\big({\vw1}\big) &= {\vw1} \otimes 1_\calf + 1_\calf \otimes {\vw1} \\
                    &\qquad\qquad + {\tw1}\ {\tw1} \otimes {\tw1} + 2 {\tw1} \otimes {\sw1}
\end{align*}
As a remark we should underline that the entries in the above
coproduct matrix can also be seen directly from the spinneys in
the woods $\bar{\mathfrak{W}}(\Gamma_i)$, $i=1,2,3,4,5$, see
Eqs.~(\ref{graphs:forest3}-\ref{graphs:forest6}) from
Section~\ref{ssect:BogoRbar}. Indeed, this is in accordance with
the remark after the definition of the coproduct on Feynman graphs
in Eq.~(\ref{def:coprod}) of Section~\ref{sect:HopfCK}, since we
observe that the $k$-th row of $(\Gamma_{ij})$ corresponding to
the graph $\Gamma_k$ simply contains the spinneys in
$\bar{\mathfrak{W}}(\Gamma_k)$. More precisely, the entry in
$(\Gamma_{kj})$ for $j=1,2,3,4,5$ is given by the particular
spinney in $\bar{\mathfrak{W}}(\Gamma_k)$ which produces the
cograph $\Gamma_j$ when it is contracted in $\Gamma_k$. Recall
that $\bar{\mathfrak{W}}(\Gamma_k)$ contains the spinneys
$\{\Gamma_k\}$ and $\{\emptyset\}$ with cographs
$\Gamma_k/\{\emptyset\}=\Gamma_k$ and
$\Gamma_k/\{\Gamma_k\}=\emptyset=1_\calf$, leading to the entries
$(\Gamma_{k1})$ and $(\Gamma_{kk})$, respectively.\smallskip

We must put some emphasis on the fact that $F'$ can be an infinite
subset of $F$, or even $F$ itself, in the latter case we denote
the coproduct matrix by $M_{\calh_\calf}$. Both cases would imply
that the matrix $M_{\calh_\calf}(F')$ is triangular of infinite
size. Such matrices form a very well-behaved subalgebra.\smallskip

Let us define $n:=|F'|$, $n\leq \infty$. Now let $f:
\mathcal{H}_\mathcal{F} \to A$ denote an arbitrary regularized
linear functional on $\calh_\calf$. Simply applying $f$ to either
$M_{\calh_\calf}(F')$ or, in the full case, to
$M_{\mathcal{H}_\calf}$ entry by entry gives a lower triangular
matrix
$$
  \widehat{f}:=\widehat{f}_{F'}:= f\big(M_{\mathcal{H}_\calf}(F')\big)
              = \big( f(\Gamma_{ij}) \big)_{1\leq j \leq i \leq
              |F'|} \in \mathcal{M}^{\ell}_{|F'|}(A)
$$
with $f(\Gamma_1),f(\Gamma_2),f(\Gamma_3),\cdots,f(\Gamma_{|F'|})$
for all graphs $\Gamma_i$, $i=1,\dots,n$ (possibly infinite), as
the first column. The matrix $\widehat{f}$ is now in the algebra
of lower triangular matrices of size $|F'|$ with entries in the
commutative Rota--Baxter algebra $A$.

For a Feynman rules character $\phi \in G_A$ we obtain the
unipotent {\it{Feynman rules matrix}} $\widehat{\phi}$ with unit
diagonal. For example, with our $F'$ in (\ref{eq:F}), we have
\begin{equation}
    \label{ex:QEDmatrix}
\widehat{\phi} = \left( {\small{\begin{array}{lllll}
                    1_A & 0                            & 0               & 0   & 0\\
         \phi(\Gamma_2) & 1_A                          & 0               & 0   & 0\\
         \phi(\Gamma_3) & \phi(\Gamma_2)               & 1_A             & 0   & 0\\
         \phi(\Gamma_4) & \phi(\Gamma_3)               & \phi(\Gamma_2)  & 1_A & 0\\
         \phi(\Gamma_5) & \phi(\Gamma_2)\phi(\Gamma_2) & 2\phi(\Gamma_2) & 0   & 1_A \end{array}}} \right )
\end{equation}
The unit $e_A \in \cala$ obviously maps to the unit matrix
$\widehat{e_A} = {\bf{1}}$, where
$     {\bf 1}
    :=(\delta_{ij}1_A)_{1\leq i,j\leq n}
$
is the identity matrix.
Compare this with the matrix representation of the identity map
$\id_{\calh_\calf}$, which is just the coproduct matrix
$$
    \widehat{\id_{\calh_\calf}} = M_{\mathcal{H}_\calf}(F').
$$
Since $A$ is assumed to be commutative the space of those
triangular matrices with unit diagonal forms the Lie group
$\widehat{G_A} := {\bf{1}} + \mathcal{M}^\ell_{n}(A)\fil{1}
\subset \mathcal{M}^{\ell}_{|F'|}(A)$.

On the other hand, for elements in the corresponding Lie algebra
$\call_A$ of infinitesimal characters we find that $Z \in \call_A$
applied to $M_{\calh_\calf}(F')$ maps the unit diagonal and
non-linear entries to zero due to relation
(\ref{def:inifiniChar}). For $F'$ in (\ref{eq:F}), this gives
$$
\widehat{Z} = \left( {\small{\begin{array}{lllll}
                   0 & 0           & 0             & 0 & 0\\
         Z(\Gamma_2) & 0           & 0             & 0 & 0\\
         Z(\Gamma_3) & Z(\Gamma_2) & 0             & 0 & 0\\
         Z(\Gamma_4) & Z(\Gamma_3) & Z(\Gamma_2)   & 0 & 0\\
         Z(\Gamma_5) & 0           & 2 Z(\Gamma_2) & 0 & 0 \end{array}}} \right)
$$
Hence, derivations correspond to nilpotent matrices forming the
matrix Lie algebra $\widehat{\call_A} \subset
\mathcal{M}^{\ell}_{|F'|}(A)$ bijectively corresponding to
$\widehat{G_A}$ via the matrix exponential map respectively
logarithmic map. Let us underline that we are just in the setting
described in Example~\ref{ex:matrix}, i.e., the complete filtered
(non-commutative) algebra of lower triangular matrices with
entries in a commutative unital algebra $A$. It contains the group
$\widehat{G_A}$ of unit diagonal triangular matrices as well as
its Lie algebra $\widehat{\call_A}$. In fact, this picture
represents most transparently the pro-nilpotent structure of the
Connes--Kreimer group of combinatorial Hopf algebra characters.
For any finite subset $F' \subseteq F$, $n=|F'| < \infty$ we see
that finite powers of derivations $\widehat{Z}^m=0$, $Z \in
\widehat{\call_A}$ and $m>n$.

Observe that for the Feynman rules character $\phi$ the first
column of the associated matrix $\widehat{\phi}$ contains the
Feynman amplitudes corresponding to the elements in $F'$ (resp.
$\calf$). Therefore it is natural to introduce a {\it{bra-ket
notation}}, i.e., we identify the elements in $\Gamma_i \in
\calf'$, ${1\leq i \leq n}$ as column vectors $\{ |\Gamma_i
\rangle \}_{1\leq i \leq |F'|}$ with unit entry at position $i$
and zero else. Such that for ${1\leq i \leq |F'|}$
$$
    \langle \Gamma_i| \ \widehat{\phi}\  |\Gamma_i\rangle = \phi(\Gamma_i) \in A.
$$

An intriguing simple observation is the following. Let $\phi \in
G_A$ respectively $\log^{\star}(\phi) \in \call_A$. Then, in terms
of matrices we have immediately
$$
\widehat{\phi}=\exp\big( \log(\widehat{\phi}) \big)
              = \exp\Big(\phi\big(\log\big(\widehat{\id_{\calh_\calf}}\big)\big) \Big),
$$
since $\phi$ is multiplicative. The matrix
$\log\big(\widehat{\id_{\calh_\calf}}\big) \in
\mathcal{M}_{|F'|}(\calh_\calf)$ is easily calculated
$$
 \log\big(\widehat{\id_{\calh_\calf}}\big)=\log\big({\bf{1}} + (\widehat{\id_{\calh_\calf}} -{\bf{1}}) \big)
                                          = - \sum_{k>0} \frac{\big(-(\widehat{\id_{\calh_\calf}} -{\bf{1}})\big)^k}{k}.
$$
In the above \textsf{QED} example $F'$ in (\ref{eq:F}) we find
the following concrete matrix
$$
\log\big(\widehat{\id_{\calh_\calf}}\big) = \left(
{\small{\begin{array}{lllll}
                0                                             & 0                                  & 0          &0  & 0\\
         \Gamma_2                                             & 0                                  & 0          &0  & 0\\
         \Gamma_3 - \frac{1}{2}\Gamma_2{}^2                   & \Gamma_2                           & 0          & 0 & 0\\
         \Gamma_4 - \Gamma_2\Gamma_3 +\frac{1}{3}\Gamma_2{}^3 & \Gamma_3 - \frac{1}{2}\Gamma_2{}^2 & \Gamma_2   & 0 & 0\\
         \Gamma_5 - \Gamma_2\Gamma_3 +\frac{1}{6}\Gamma_2{}^3 &   0                                & 2 \Gamma_2 & 0 & 0
         \end{array}}} \right )
$$
This matrix just contains the {\it{normal coordinates}} used in
the exponential representation of $\phi$, see
\cite{Mexicans02,FG05}. In the same spirit we may calculate the
inverse of an element in $\widehat{G_A}$. Recall that for $\phi
\in G_A$ the inverse was given by composition with the Hopf
algebra antipode, $\phi^{-1}:=\phi \circ S$. Since $S$ is
supposed to be the convolution inverse of the identity on
$\calh_\calf$, we readily see that
$$
    \widehat{\phi^{-1}} = \phi \circ {\widehat{\id_{\calh_\calf}^{-1}}}
                        = \phi \circ \Big(\sum_{k \geq 0} \big(-(\widehat{\id_{\calh_\calf}} -{\bf{1}})\big)^k\Big)
$$
where again in our \textsf{QED} example we find
$$
    {\bf{1}} + \sum_{k > 0} \big(-(\widehat{\id_{\calh_\calf}} -{\bf{1}})\big)^k
    =
    \left(
{\small{\begin{array}{lllll}
           1_\calf                                     & 0                       &       0    & 0       & 0\\
         -\Gamma_2                                     & 1_\calf                 &       0    & 0       & 0\\
         -\Gamma_3 - \Gamma_2{}^2                      & - \Gamma_2              & 1_\calf    & 0       & 0\\
         -\Gamma_4 + 2 \Gamma_2\Gamma_3 - \Gamma_2{}^3 & \Gamma_3 - \Gamma_2{}^2 & -\Gamma_2  & 1_\calf & 0\\
         -\Gamma_5 + 2 \Gamma_2\Gamma_3 - \Gamma_2{}^3 & \Gamma_2{}^2            & - \Gamma_2 & 0       & 1_\calf
         \end{array}}} \right )
$$
One confirms without difficulty that
$\widehat{\phi^{-1}}=\widehat{\phi}{}^{-1}$. In the light of
Theorem~\ref{thm:matrixRB} and Theorem~\ref{thm:BCH-RB-QFT} the
reader may guess the next step. Indeed, we lift the Rota--Baxter
structure from $(A,R)$ to $\mathcal{M}^{\ell}_{|\calf'|}(A)$ in
the usual way by defining for $\alpha \in
\mathcal{M}_{|\calf'|}^\ell(A)$
$$
    \mathrm{R}(\alpha) := \big( R(\alpha_{ij}) \big)_{1\leq j \leq i \leq |F'|}.
$$

All this is readily summarized in \cite{EG,EGGV}. In fact, denote
by $A\,\mathcal{F}':= A \otimes \calf'$ the free $A$-module with
basis $F'$. Recall that
$\cala:=\mathrm{Hom}(\mathcal{H}_\mathcal{F},A)$ is a complete
filtered unital non-commutative Rota--Baxter algebra of weight one
with idempotent Rota--Baxter map $\calr(f) := R \circ f$. Define a
map
$$
    \Psi:=\Psi_{A,\mathcal{F}'}: \Hom(\calh_\calf,A)\to {\rm End}(A\otimes \calf')
$$
by the composition
\allowdisplaybreaks{
\begin{eqnarray*}
 \Psi[f]: A \otimes \calf'
                \xrightarrow{\id_A \otimes \Delta} A \otimes \mathcal{H}_\calf \otimes \calf'
                \xrightarrow{\id_A \otimes f \otimes \id_{\calf'}} A \otimes A \otimes \calf'
                \xrightarrow{m_A \otimes \id_{\calf'}} A \otimes
                \calf'.
\end{eqnarray*}}
So that for $\Psi[f]$, $f \in \cala$ applied to $\Gamma_i \in F'$
we find
$$
    \Psi[f](\Gamma_i) = f \star \id_{\calf}(\Gamma_i)
                      = \sum_{j=1}^{i} f(\Gamma_{ij})\Gamma_j \in A\,\calf'.
$$
It was shown in~\cite{EG} that $\Psi$ is an anti-homomorphism of
algebras. We then obtain an algebra homomorphism
$\Hom(\calh_\calf,A)\to \mathcal{M}^\ell_{|F'|}(A)$, still denoted
by $\Psi$, when composing this with the anti-homomorphism
$$ {\rm End}(A\otimes \calf') \to \mathcal{M}^\ell_{|F'|}(A) $$
sending $P: A\otimes \calf' \to A\otimes \calf'$ to its standard
matrix $M=(m_{ij})$ with respect to the ordered basis $F'$:
$P(\Gamma_i)=\sum_{j} m_{ij} \Gamma_j$. Altogether, we have the
algebra morphism
$$
  \Psi_{A,\calf'}: \big({\rm{Hom}}(\mathcal{H}_\mathcal{F},A),\calr\big)
                    \to \big(\mathcal{M}_{|F'|}^\ell(A),\mathrm{R}\big),\
   \quad \Psi_{A,\calf'}[f]=\widehat{f}
$$
of complete filtered Rota--Baxter algebras. We omit the indices
now for notational transparency. So in particular for $f,g \in
\cala$
$$
  \Psi[f \star g]=\Psi[f]\ \Psi[g]=\widehat{f}\ \widehat{g},
  \qquad
  \Psi[\mathcal{R}(f)]=\mathrm{R}(\Psi[f])=\mathrm{R}(\widehat{f}).
$$
The map $\Psi$ gives a matrix representation of
$\Hom(\mathcal{H}_\calf,A)$ in terms of lower triangular matrices.
In a more abstract context, $\Psi$ gives a morphism between the
affine group scheme of $\calh_\calf$ and the matrix group scheme.

Let us go back to
Theorem~\ref{thm:BCH-RB-QFT}. Applying $\Psi$ to
item~(\ref{it:CK-birkhofBCH}) we have for the matrix
representation of the character $\phi$ respectively its Birkhoff
decomposition
\begin{equation}
    \widehat{\phi} = \widehat{\phi_-}{}^{-1}\ \widehat{\phi_+}{},
    \label{eq:matrixCK}
\end{equation}
where $\widehat{\phi_-} = \Psi[\phi_-]$ and
$\widehat{\phi_+}=\Psi[\phi_+]$. Hence, we recover
$\widehat{\phi_+}$ and $\widehat{\phi_-}$ for each element in
$F'$ from the first column of the corresponding matrices.

Further the two matrix factors, $ \widehat{\phi_+}{}^{-1}$,
$\widehat{\phi_-}$, are unique solutions to the matrix Spitzer
identities, i.e., Bogoliubov's formulae for the counterterm and
renormalized Feynman rules, respectively \allowdisplaybreaks{
\begin{eqnarray}
    \widehat{\phi_-} \!&=& {\bf{1}} - \mathrm{R}\Big(\widehat{\phi_-} \ (\widehat{\phi}-{\bf{1}})\Big),
    \label{eq:mphi-}\\
    \widehat{\phi_+}{}^{-1} \!&=& {\bf{1}} - \tilde{\mathrm{R}}\Big( (\widehat{\phi}-{\bf{1}}) \ \widehat{\phi_+}{}^{-1} \Big)
    \quad \mathrm{ and }\quad
    \widehat{\phi_+} \!= {\bf{1}} - \tilde{\mathrm{R}}\Big( \widehat{\phi_+} \ (\widehat{\phi}^{-1}-{\bf{1}}) \Big)
    \label{eq:mphi+}
\end{eqnarray}}
In \cite{EG,EGGV} we worked with an upper triangular
representation which makes $\Psi$ is an anti-morphism, exchanging
the order of products when comparing with the `scalar' formulae in
the Hopf algebra formalism. The matrix entries can be calculated
without recursions using $\alpha = \widehat{\phi}$ from the
equations \allowdisplaybreaks{
\begin{eqnarray}
    (\widehat{\phi_-})_{ij} \!\!&\!\!\!=\!\!\!\!&\!\! -R(\alpha_{ij})\!
                                +\! \sum_{k=2}^{j-i}\: \sum_{i>l_1> \cdots >l_{k-1}>j}\!\!\!\!\!\!\!\!\!\!\!\! (-1)^{k}\!
       {R}\big({R}(\cdots {R}(\alpha_{i l_{1}})\alpha_{l_{1} l_{2}}) \cdots \alpha_{ l_{k-1} j}\big)
                                             \label{eq:one}\\
    (\widehat{\phi_+}^{-1})_{ij} \!\!&\!\!\!=\!\!\!\!&\!\!  -\tilde{R}(\alpha_{ij})
                                \!+ \!\sum_{k=2}^{j-i}\: \sum_{i>l_1> \cdots > l_{k-1} > j}\!\!\!\!\!\!\!\!\!\!\!\! (-1)^{k}\!
        \tilde{R}\big(\alpha_{il_1}\tilde{R}(\alpha_{l_1l_2} \cdots \tilde{R}(\alpha_{l_{k-1}j})\cdots)\big).
                                              \label{eq:two}
\end{eqnarray}}

Let us finish this section and the paper with the simple example
from \textsf{QED}. We should emphasize that the calculations of
the matrix Birkhoff factorization are just meant to point up the
combinatorial structure involved in the renormalization of matrix
Feynman rules characters.

The Feynman rules matrix (\ref{ex:QEDmatrix}) of the graphs up to
$3$ loops from the \textsf{QED} example decomposes into the
counterterm
 \allowdisplaybreaks{
\begin{eqnarray}
  \widehat{\phi_-}:= \left( {\small{
        \begin{array}{ccccc}
                          1_A & 0                      & 0                     &   0 & 0 \\
         \phi_-\big(\tw1\big) & 1_A                    & 0                     &   0 & 0 \\
         \phi_-\big(\sw1\big) & \phi_-\big(\tw1\big)   & 1_A                   &   0 & 0 \\
         \phi_-\big(\uw1\big) & \phi_-\big(\sw1\big)   & \phi_-\big(\tw1\big)  & 1_A & 0 \\
         \phi_-\big(\vw1\big) & \phi_-\big(\tw1\big)^2 & 2\phi_-\big(\tw1\big) & 0   & 1_A
        \end{array} }}\right)
\end{eqnarray}}
Of course, this follows immediately from general reasoning by
applying the abstract Hopf algebra character $\phi_-$ to the
coproduct matrix $M_{\calh_\calf}(F')$. Instead, the goal is
to calculate the matrix entries directly either by use of the
matrix Equation~(\ref{eq:mphi-}), which terminates after 3
iterations, or entrywise with help of Eq.~(\ref{eq:one}).

In this example the only nontrivial counterterm matrix entry we
need to calculate is position $(5,1)$ in $\widehat{\phi}_{-}$.
Applying formula (\ref{eq:one}) we find \allowdisplaybreaks{
\begin{eqnarray*}
(\widehat{\phi_{-}})_{51} &=&-R\bigg(\phi\Big( \vw1 \Big)\bigg)
                                    +R\bigg( R\Big(\phi\big(\tw1\big)^2\Big)
                                            \phi\big(\tw1\big) \bigg)\\
           & & \hspace{2cm}         +R\bigg( R\Big(2\phi\big(\tw1\big)\Big)
                                            \phi\big(\sw1\big)\bigg)\\
           & & \hspace{3cm} - R\bigg(R \Big( R \big(
                2\phi\big(\tw1\big) \big)
                 \phi\big(\tw1\big) \Big)
                 \phi\big(\tw1\big) \bigg).\nonumber
\end{eqnarray*}}

The identity $R(a)^2=2R(aR(a))-R(a^2)$ which follows from the
Rota--Baxter identity, and which is true only for commutative
algebras, immediately implies \allowdisplaybreaks{
\begin{eqnarray*}
    (\widehat{\phi_{-}})_{51} &=&-R\bigg(\phi\Big(\vw1\Big)\bigg)
                                    - R\bigg( R\Big(\phi\big(\tw1\big) \Big)^2
                                                    \phi\big(\tw1\big) \bigg)\\
           & & \hspace{5cm}
                                    +2R\bigg( R\Big(\phi\big(\tw1\big) \Big)
                                                    \phi\big(\sw1\big) \bigg).
\end{eqnarray*}}
Likewise for the corresponding entry in the renormalized matrix
$\widehat{\phi_+}= \widehat{ \phi_-} \ \widehat{ \phi}$ we find
\allowdisplaybreaks{
\begin{eqnarray*}
  (\widehat{ \phi_+})_{51} &=& \phi\Big(\vw1\Big)
                                    +\phi\big(\tw1\big)
                                     R\Big(\phi\big(\tw1\big)\Big)^2\\
                                  & &
                                    -2\phi\big(\sw1\big)
                                     R\Big(\phi\big(\tw1\big)\Big)
                                    -R\bigg(\phi\Big(\vw1\Big)\\
                                  & &
                                               +\phi\big(\tw1\big)
                                               R\Big(\phi\big(\tw1\big)\Big)^2
                                              -2\phi\big(\tw1\big)
                                               R\Big(\phi\big(\tw1\big)\Big)\bigg).\nonumber
\end{eqnarray*}}
More examples can be found in \cite{EGGV,EG}.


\section{Conclusion and outlook}
\label{sect:conclusionoutlook}

The algebraic-combinatorial structure of renormalization in
perturbative quantum field theory has found a concise formulation
in terms of Hopf algebras of Feynman graphs. The process of
renormalization is captured by an algebraic Birkhoff decomposition
of regularized Feynman characters discovered by Connes and
Kreimer. Associative Rota--Baxter algebras naturally provide a
suitable general underpinning for such factorizations in terms of
Atkinson's theorem and Spitzer's identity. This approach gives a
different perspective on the original result due to Connes and
Kreimer in the context of other renormalization schemes. It
enabled us to establish a simple matrix calculus for
renormalization. However, a general characterization of the
algebraic structure underlying the notion of renormalization
schemes needs to be explored in future work. We should underline
that our results presented here go beyond the particular
application in the context of perturbative renormalization as we
presented a general factorization theorem for filtered algebras in
terms of a recursion defined using the Baker--Campbell--Hausdorff
formula for which we found a closed formula under suitable
circumstances. Moreover, we showed its natural link to a classical
result of Magnus known from matrix differential equations. The
examples of Rota--Baxter algebras, especially its appearance in
classical integrable systems makes it a rich field to be further
explored.

\vspace{1cm}

\noindent {\bf{Acknowledgements}.} We would like to thank the
organizers for giving us the opportunity to present parts of our
research results which we achieved in joint work with D.~Kreimer,
and partly with J.~M.~Gracia-Bond\'ia, J.~V\'arilly and more
recently D.~Manchon. The stimulating atmosphere we found at this
memorable workshop affected us considerably and helped a lot in
shaping our understanding of the concept and applications of
renormalization. The generous support, relaxing atmosphere and
warm hospitality we experienced at the Fields Institute is greatly
acknowledged. Thanks goes especially to its staff for the
efficient organization. The first author acknowledges greatly the
support by the European Post-Doctoral Institute (E.P.D.I.) and
Institute des Hautes \'Etudes Scientifiques (I.H.\'E.S.), and last
but not least D.~Kreimer's ongoing collaboration. He would also
like to thank the department of theoretical physics at the
University of Zaragoza (Spain) where this work was finished for
warm hospitality and support. The second author thanks NSF and
Rutgers University Research Council for support. Dirk Kreimer is
thanked for his corrections, comments and remarks.\smallskip

URL of KEF: http://www.th.physik.uni-bonn.de/th/People/fard/}


\end{document}